\documentclass[final]{siamart0516}

\usepackage{amsfonts}
\usepackage{graphicx}
\usepackage{algorithmic}
\ifpdf
  \DeclareGraphicsExtensions{.eps,.pdf,.png,.jpg}
\else
  \DeclareGraphicsExtensions{.eps}
\fi

\newcommand{\TheTitle}{An Optimized, Parallel Computation of the\\ Ghost Layer
                       for Adaptive Hybrid Forest Meshes}
\newcommand{\TheShortTitle}{Optimized ghost algorithm for hybrid forest AMR}
\newcommand{\TheAuthors}{J. Holke and D. Knapp and C. Burstedde}

\headers{\TheShortTitle}{\TheAuthors}

\title{{\TheTitle}%
}
\author{
  Johannes Holke%
  \thanks{German Aerospace Center (DLR), Cologne, Germany
  (\email{johannes.holke@dlr.de})}
  \and
  David Knapp\thanks{Rheinische Friedrich-Wilhelms-Universit{\"a}t Bonn, Germany, 
  and DLR, Cologne, Germany}
  \and
  Carsten Burstedde\thanks{Institut f{\"u}r Numerische Simulation (INS)
    and Hausdorff Center for Mathematics (HCM),
    Rheinische Friedrich-Wilhelms-Universit{\"a}t Bonn, Germany}
}

\ifpdf
\hypersetup{
  pdftitle={\TheTitle},
  pdfauthor={\TheAuthors}
}
\fi

\usepackage{amssymb}
\usepackage{cite} %
\usepackage{caption} %
\usepackage{mathrsfs} %
\usepackage{multirow}	  %
\usepackage[usenames,dvipsnames]{xcolor} 	%
\usepackage{enumerate}
\usepackage[ruled,linesnumbered,algosection,vlined]{algorithm2e}
\usepackage{hhline}	  %
\usepackage{subcaption} %
\usepackage{longtable}
\usepackage{tabularx}
\usepackage{verbatim}

\newif\ifANMan
\ANMantrue %
\newif\ifDEBUG

\newcommand{\set}[1]{\left\lbrace\, #1 \,\right\rbrace}

\newcommand{\abst}[1]{\, #1 \,} %
\newcommand{\mytabvspace}{\vphantom{${X^X}^X$}} %
\newcommand{\myhugetabvspace}{{\huge\vphantom{${X^X}^X$}}} %
\newcommand{\forest}[1]{{\mathscr{#1}}} %

\newcommand{\Comment}[1]{\tcc*[r]{#1}}   %
\newcommand{\IfComment}[1]{\tcc*[f]{#1}} %
\let\oldnl\nl%
\newcommand{\nonl}{\renewcommand{\nl}{\let\nl\oldnl}}%
\newcommand{\algoand}{\textbf{ and }} %
\newcommand{\algor}{\textbf{ or }} %
\newcommand{\algoresult}{\nonl\textbf{Result:\quad }} %
\newcommand{\algofor}[1]{\For{\upshape #1}} %
\newcommand{\algoforcom}[2]{\For(#1){\upshape #2}} %
\newcommand{\algoif}[1]{\If{\upshape #1}} %
\newcommand{\algoeifcom}[2]{\eIf(#1){\upshape #2}} %

\newcommand{\e}{\text{e}} %
\newcommand{\confer}{cf.\xspace}

\newcommand{\pforest}{\texttt{p4est}\xspace} %
\newcommand{\tetcode}{\texttt{t8code}\xspace} %

\newcommand{\ghostb}{\texttt{Ghost\_v1}\xspace} %
\newcommand{\ghostn}{\texttt{Ghost}\xspace} %
\newcommand{\ghostmatch}{\texttt{ghost\_match}\xspace}
\newcommand{\juqueen}{JUQUEEN\xspace}
\newcommand{\juwels}{JUWELS\xspace}

\ifANMan
\newcommand{\ANM}[1]{\footnote{ANMERKUNG: #1}}  %
\newcommand{\ANMtext}[1]{\footnotetext{ANMERKUNG: #1}}
\newcommand{\ANMmarkn}[1]{\footnotemark[#1]}           %
\newcommand{\ANMtextn}[2]{\footnotetext[#1]{ANMERKUNG: #2}}
\else
\newcommand{\ANM}[1]{}  %
\newcommand{\ANMtext}[1]{}
\newcommand{\ANMmarkn}[1]{}          %
\newcommand{\ANMtextn}[2]{}
\fi

\ifDEBUG
\newcommand{\figlabel}[1]{\caption*{\color{red} #1}\label{#1}}
\else
\newcommand{\figlabel}[1]{\label{#1}}
\fi
\newcommand{\figref}[1]{Figure~\ref{fig:#1}}
\newcommand{\eqnlab}[1]{\label{eq:#1}}
\newcommand{\eqnref}[1]{\eqref{eq:#1}}
\newcommand{\seclab}[1]{\label{sec:#1}}
\newcommand{\secref}[1]{Section~\ref{sec:#1}}

\definecolor{mygreen}{rgb}{0,0.7,0}

\DeclareMathOperator{\orient}{orientation}
\DeclareMathOperator{\firstd}{fd}
\DeclareMathOperator{\lastd}{ld}
\DeclareMathOperator{\type}{type}				%
\DeclareMathOperator{\sign}{sign}				%

\newcommand{\rfirst}{\text{first}}
\newcommand{\rlast}{\text{last}}
\newcommand{\microsec}{$\mu\mathrm{s}$\xspace}

\newcommand{\lowlevel}[2]{
\vspace{\baselineskip}%
\begin{minipage}{\dimexpr\textwidth-8ex}
\texttt{#1}

#2
\end{minipage}
\vspace{\baselineskip}%
}%

\newcommand{\lowleveloneline}[3]{
\vspace{\baselineskip}%
\begin{minipage}{\dimexpr\textwidth-8ex}
\texttt{#1}
\newline\mbox{}\hfill
\texttt{#2}{#3}
\end{minipage}%
\vspace{\baselineskip}%
}%

\newcolumntype{L}[1]{>{\raggedright\arraybackslash}p{#1}}
\newcolumntype{C}[1]{>{\centering\arraybackslash}p{#1}}
\newcolumntype{R}[1]{>{\raggedleft\arraybackslash}p{#1}}

\theoremstyle{plain}

\newtheorem{remark}[theorem]{Remark}
\newtheorem{rationale}[theorem]{Rationale}

\graphicspath{{../pictures/trianglemorton/},{../pictures/tetrahedra/},%
        {../pictures/trianglemorton/TheCube/},{../timings/adapt/},%
        {../timings/partition/},{../pictures/Notes/},%
        {./pictures/amr/ghost/},{./pictures/amr/coarse_mesh/},%
        {./pictures/amr/balance/},%
        {../pictures/partition/},{./pictures/},{./pictures/png/},%
        {./pictures/timings/},{./pictures/amr/tets/}}

\begin{document}

\maketitle

\begin{abstract}%
We discuss parallel algorithms to gather topological information about
off-process mesh neighbor elements.
This information is commonly called the ghost layer, whose creation is a
fundamental, necessary task in executing most parallel, element-based computer
simulations.
Approaches differ in that the ghost layer may either be inherently part of
the mesh data structure that is maintained and modified, or kept separate and
constructed/deleted as needed.

In this work, we present an updated design following the latter approach, which
we favor for its modularity of algorithms and data structures.
We target arbitrary adaptive, non-conforming forest-of-(oc)trees meshes of
mixed element shapes, such as cubes, prisms, and tetrahedra, and restrict
ourselves to face-ghosts.
Our algorithm has low complexity and redundancy since we reduce it to generic
codimension-1 subalgorithms that can be flexibly combined.
We cover several existing solutions as special cases and optimize further
using recursive, amortized tree searches and traversals.
\end{abstract}

\begin{keywords}
  Adaptive mesh refinement,
  parallel algorithms,
  forest of octrees,
  ghost layer
\end{keywords}

\begin{AMS}
  65M50, %
  68W10, %
  65Y05, %
  65D18  %
\end{AMS}

\section{Introduction}
\seclab{intro}

In the parallel mesh-based numerical solution of partial differential
equations, the notion of a ghost or halo layer is
ubiquitous.
It refers to connectivity information about all elements owned by any remote
process and directly adjacent to at least one process-local element.
As such, it is implemented in many general purpose software packages; see
for example~\cite{BangerthBursteddeHeisterEtAl11, DednerKlofkornNolte14,
                  CoupezSilvaDigonnet16, RichardsonWells16}.
If the numerical method only couples directly adjacent elements, which applies
to most finite element and finite volume methods, the combined set of variables
on local and ghost elements suffices to complete a basic global step of the
method, be it the assembly of a system matrix or an explicit or implicit solve.
In particular with adaptive refinement, the ghost layer aides in globally
numbering the degrees of freedom and in computing refinement and coarsening
indicators.

The concept of the ghost layer is widely applied due to several benefits it
provides, such as the locality of parallel communication, the transparency to
the discretization code, and the overlap of communication and computation it
encourages.
If the mesh structure is replicated in parallel, information on the individual
process partition and the ghost elements is replicated, too, providing a global
view of the partition data for every process.
If, on the other hand, the mesh is distributed in parallel, constructing the
ghost layer becomes a parallel algorithm in its own right.

When using unstructured meshes, the ghost layer is often part of the
graph-based encoding of the mesh.
Graph partitioners
\cite{KarypisKumar98, DevineBomanHeaphyEtAl02, ChevalierPellegrini08}
can be executed and the result queried for both ghost and local elements, often
encoded by lookup tables or other convenient data structures; see e.g.\ \cite{%
TautgesMeyersMerkleyEtAl04,
LawlorChakravortyWilmarthEtAl06,
TautgesKraftcheckBertramEtAl12,
RasquinSmithChitaleEtAl14,
LiuZhaoChengEtAl16}.
Tree-based meshes, on the other hand, often allow to build the ghost
information without referring to external software, but using the
hierarchy and coordinate information inherent in the tree structure;
see for example
\cite{SbalzariniWaltherBergdorfEtAl06,
      TeunissenKeppens19
}.

In some approaches, the ghost layer is inherently part of the mesh data
structure, which removes the need for its explicit computation
\cite{TuOHallaronGhattas05,
      BangerthBursteddeHeisterEtAl11, SchornbaumRuede16}.
On the other hand, every mesh update requires to update the ghost layer
alongside to maintain consistency of the mesh data.

This work focuses on the alternative that the ghost layer is not considered
first-class mesh data, but may be constructed when needed by executing a
suitable algorithm \cite{BursteddeWilcoxGhattas11}.
Using this extra algorithm adds a cost, but simplifies the core mesh data
structure as a benefit.
In addition, it improves modularity and permits to optimize the ghost
layer computation independently of other meshing algorithms
\cite{IsaacBursteddeWilcoxEtAl15}, and it allows to omit the ghost construction
altogether when it is not required by the numerical method \cite{Burstedde18}.

The frame for our algorithm development is set by the forest-of-(oc-)trees
approach to meshing \cite{StewartEdwards04, BangerthHartmannKanschat07,
BursteddeWilcoxGhattas11}, and the implementation is provided within the
\tetcode software library \cite{BursteddeHolke16, tetcodeweb19}.
The unique property of this set of algorithms is that it works with hybrid
meshes, that is, the same mesh can mix shapes such as triangles and
quadrilaterals in 2D or tetrahedra, prisms, and hexahedra in 3D.
Refinement is tree-based and allows for hanging faces, which lends many of the
benefits of hexahedral forest/tree-structures to the hybrid case.

\subsection{Contributions}
\seclab{contributions}

In this paper we present two novel contributions.
Firstly, we extend the computation of a ghost layer for forest-based AMR to meshes
with arbitrary element shapes, and in particular hybrid meshes.
Secondly, we optimize the proposed algorithm to obtain optimal runtime.

To achieve the first goal, the most important step is the construction of (same-level)
face-neighbors across tree boundaries.
This is of particular interest if the connected trees have different shapes.
The challenging part here is to perform the necessary transformations to
account for tree-to-tree coordinate changes.
Since the underlying low-level implementations for the element shapes should be
exchangeable, it is crucial to avoid dependencies between these implementations.
Such dependencies would for example arise if we directly transform the
coordinates of one element (for example a hexahedron) into coordinates of the
neighbor element (for example a prism).
Instead, our proposed approach is to
construct the $(d-1)$-dimensional face element as an intermediate object.
We then perform the necessary coordinate transformation in $d-1$ dimensions and
extrude the resulting element into the desired $d$-dimensional face-neighbor.

To achieve the second goal, the optimization of runtime, we utilize recent
developments of tree-based search routines~\cite{IsaacBursteddeWilcoxEtAl15}
to exclude locally surrounded portions of the mesh and thus limit our
computational effort to the partition boundary elements.

For further technical details and background, as well as the in-depth
discussion of triangular and tetrahedral space-filling curves, we refer to H.'s
thesis \cite{Holke18}.

Perspectively, our algorithms enable applications beyond their direct use in
ele\-ment-based numerics, such as the efficient design of further meshing
sub-algorithms.
One example is the implementation of a 2:1 balance algorithm
by ripple propagation \cite{TuOHallaronGhattas05},
and another the efficient, globally consistent numbering of degrees of freedom
\cite{IsaacBursteddeWilcoxEtAl15}.
We leave these developments to future work and close with numerical examples
that prove the parallel scalability of the proposed ghost algorithms on their
own.

\subsection{Fundamental concepts}
\label{ch:ghost}
\seclab{defs}

Throughout this document, we assume a forest-of-trees mesh structure.
The tree roots can be of any shape as long as their faces conform to all
neighbor trees.
For example, a hexahedron and a tetrahedron tree may both connect to a prism
tree but not to each other.
The trees are refined recursively, and the length of the path from the root to
an element is called its level.
Thus, two elements may be a descendant or ancestor of each other
(in fact both if they are equal) or unrelated.
Given this generality, the number of child elements $n$ may be a
constant 4 (triangles/quadrilaterals) or 8 (tetrahedra/cubes/prisms), but also
a different number, and even varying within the tree.
For example, we might consider the one-dimensional line, $n = 2$,
a Peano-style $1:n = 3^d$ refinement of cubes~\cite{WeinzierlMehl11,Peano90},
or use a Peano refinement on even and Morton refinement on odd levels.
All elements are derived by recursively and adaptively subdividing the tree roots.
Only the leaf elements of the forest are maintained in memory to make up the
mesh, which is often described as linear tree storage
\cite{SundarSampathBiros08}.
The ghost layer will be assembled as a linear array as well, augmented with
offset arrays to encode the owner process of each ghost element.

For each element, we assume that sub-algorithms exist to count and index its
faces, to construct its parent or any of its children, et cetera.
We consider these sub-algorithms an opaque, low-level functionality:
They will vary by implementation and by shape, and we do not wish to depend on
their internal mechanisms.
Instead we impose abstract consistency requirements between the refinements of
volumes, faces, and edges, within and between trees.
Specifically, if we consider the faces of a tree as separate
$(d-1)$-dimensional refinement trees, then the refinement of the volume cells
restricted to a tree face must be a possible face refinement.
A similar rule holds for face-neighbor elements in the same tree.
This approach enables modularity and extensibility and keeps the technical
complexity low \cite{BursteddeHolke16}.

If the forest mesh is partitioned among multiple processes, then a neighbor
leaf of a leaf element owned by process $p$ may be owned and primarily
stored by a different process $q\neq p$.
In this paper, we discuss face-neighbors exclusively, believing our algorithms
to be extensible to edge- and corner-neighbors by somewhat tedious, yet
manageable work.
When referring to the forest mesh, we speak of its topological properties,
excluding geometry maps and numerical data structures.

The connectivity between trees across tree-faces is in fact a mesh of its own.
This ``genesis mesh'' \cite{StewartEdwards04} or coarse mesh
\cite{BangerthHartmannKanschat07}
is conforming even though the elements may be
arbitrarily non-conforming by adaptive refinement.
Throughout this document, we assume that we can access the coarse mesh
information of each neighbor tree of a local tree.
This is ensured since the
coarse mesh is either replicated on all processes \cite{BursteddeWilcoxGhattas11}
or stores a layer of ghost trees, where a ghost tree is understood as the
topological shell without regarding the elements in it.
We have previously proposed sharp parallel algorithms to gather the ghost
trees~\cite{BursteddeHolke17}.

\begin{definition} %
 A ghost element (or just ghost for short) of a process
 $p$ in a forest $\forest F$ is a leaf element $G$ of a process
 $q\neq p$, such that there exists a face-neighbor $E$ of $G$ that is a
 local leaf element of $p$.
\end{definition}

\begin{definition} %
\label{def:boundary_element}
 We call a local leaf element $E$ of a process $p$ a partition boundary element
 if it has at least one face-neighbor that is a ghost element
 (the term ``mirror element'' has been used as well
 \cite{GuittetIsaacBursteddeEtAl16}).
 The remote processes to $E$ are all processes $q\neq p$ that own
 ghost elements of $E$.
 The union of all remote processes over all local elements of $p$ are the
 remote processes of $p$.
\end{definition}

\begin{definition}
\label{def:loc_sur_element}
  We say that an %
  element is a locally surrounded
  element of
  process $p$ if all of its leaf descendants and all of its leaf face-neighbors
  are owned by $p$.
\end{definition}

\begin{definition} %
 By $R_p^q$ we denote the set of partition boundary elements of process $p$ that
 have process $q$ as a remote process.
 The ghost layer for process $p$ is thus
 \begin{equation}
   \eqnlab{ghostdef}%
   \forest G _p = \cup_{q} R_p^q .
 \end{equation}
\end{definition}
By construction, we have the following symmetry:
\begin{equation}
  \eqnlab{symmcomm}%
  R_p^q \ne \emptyset
  \qquad \Leftrightarrow \qquad
  R_q^p \ne \emptyset .
\end{equation}

Adaptation of the element mesh proceeds recursively from the root, which
assigns a unique level $\ell \ge 0$ to each element.
For completeness, we give the definition of a 2:1-balanced forest since
older algorithms depend on it.
\begin{definition}
  \label{def:balance}
We call a forest $\forest F$ balanced if each pair $E$, $E'$ of
face-neighboring leaf elements of $\forest F$ satisfies
  \begin{equation}
    \ell(E) - 1 \leq \ell(E') \leq \ell(E) + 1.
  \end{equation}
Here, $E$ and $E'$ may belong to different processes. Thus, any two
face-neighbors differ by at most one in their refinement levels.
If the condition is not fulfilled, we say that $\forest F$ is unbalanced.
\end{definition}
\begin{remark}
  In some publications \emph{graded} is used instead of \emph{balanced}
  \cite{CohenKaberMuellerEtAl03, MuellerStiriba07}.
\end{remark}
Note that for unbalanced forests the number of neighbors of an element
$E$ that are ghosts can be arbitrarily large.  It is only bounded by the number
of elements at maximum refinement level that can touch the faces of $E$.
Therefore, the number of remote processes is not easily bounded from above.

\subsection{Technical procedure}
\seclab{technique}

We will eventually describe three variants of %
constructing the ghost layer.
\ghostb is
the simplest version in that it only works on balanced forests.
\texttt{Ghost\_v2} %
works on arbitrary forests, while in the present paper we optimize its runtime
leading to the latest hybrid version \texttt{Ghost}.
For the first two versions we refer to existing \pforest implementations for
quadrilateral/hexa\-hedral meshes
\cite{BursteddeWilcoxGhattas11, IsaacBursteddeWilcoxEtAl15}.
However, we discuss them in our new, element-shape independent formalism, which
not only extends to simplicial meshes but also to hybrid meshes consisting of
multiple element shapes.

When refering to the shape of an element, we refer to the associated low-level
operations at the same time.
In our reference implementation \tetcode~\cite{tetcodeweb19} we provide line,
quadrilateral, and hexahedral elements ordered by the Morton index, as well as
triangular and tetrahedral elements using the tetrahedral Morton (TM-)index.
This also provides us with an implementation of prism elements, since we can
model these as the cross product of a line and a triangle \cite{Knapp17}.

The basic idea of \ghostb and \texttt{Ghost\_v2} is to first identify all partition boundary
elements and their remote processes, thus building the sets $R_p^q$ and
identifying the non-empty ones. In a second step, each process $p$ sends all
elements in $R_p^q$ to $q$.
The senders are known to the receivers due to the symmetry of the
communication pattern \eqnref{symmcomm}.
In the first step we iterate over all local leaves and for each over
all of its faces. We then have to decide for each face $F$ of a leaf $E$ which
processes own leaves that touch this face. The difference between \ghostb and
\texttt{Ghost\_v2} lies in this decision process.

In \pforest, the runtime is optimized by performing a so called $(3\times
3)$-neighborhood check of an element~\cite{IsaacBursteddeWilcoxEtAl15}
inspired by the ``insulation layer'' concept \cite{SundarSampathBiros08}.
For a local hexahedral/quadrilateral element, it is tested whether all possible
same-level face- (or edge-/vertex-) neighbors would also be process-local and
if so, the element is excluded from further iteration.
Since this check makes explicit use of the classical Morton code and its
properties, it is difficult to generalize for arbitrary element shapes and
hybrid meshes.

To improve generality, we present a new variant
that replaces the iteration over all leaves with a top-down forest traversal.
While traversing, we exclude locally surrounded elements from the iteration,
which is equivalent to an early pruning of the search tree.
This approach supersedes the ($3 \times 3$) test and improves the overall
runtime over any naive hybrid version of \texttt{Ghost\_v2}.
We design this algorithm shape-independent from the start.
To this end, we require a minimal set of element sub-algorithms
that we develop in \secref{faceneighbors}.
We discuss fast algorithms of finding the owner process of neighbor elements in
\secref{ghost-findowner}
and expose the high-level forest traversal algorithms to construct $\forest
G_p$ beginning with \secref{theghostalgorithms}.

\section{Low-level element functions}
\seclab{faceneighbors}

Our goal is to rebuild and advance the history of ghost algorithms
and to present it as a consistent combination of methods.
We will adhere throughout to the abstraction of low-level, per-element
algorithms on the one hand and high-level, global parallel algorithms on the
other.
Introducing for example a new element shape like the pyramid, or an alternative
space filling curve such as the Hilbert curve, will be implemented on the
low-level side without requiring a change in the high-level algorithms.
Conversely, improving the high-level algorithms further will be possible with
or without defining new low-level interface functions, depending on the
algorithmic idea.

We will use this section to add several low-level functions that we require for
the high-level algorithms formulated later in this document.
We will motivate and discuss the abstract interface first and then
propose reference implementations for the (T\mbox{-)}Morton curves currently
available in \pforest and \tetcode, which will enable the reader to recreate
our algorithms in new code and, more importantly, to substitute their favorite
element implementation if so desired.

An important part of any \texttt{Ghost} routine is to construct the same-level
neighbor of a given element $E$ across a face $f$.
To support older ghost algorithms, we will also construct refined neighbors.
Here, to construct means to compute information defining this neighbor as a
possible (hypothetical) element in a mesh, not necessarily as a leaf that
exists on this or another process.
The hypothetical element can than be compared with the existing local leaves
(which may be descesdants or ancestors) or the partition boundaries (which
are encoded using deepest-level hypothetical elements).

As long as such a face-neighbor is inside the same tree as $E$, this problem is
addressed by the corresponding low-level function
\texttt{t8\_element\_face\-\_neigh\-bor\_inside}.
We describe its version for the TM-index
in~\cite[Algorithm~4.6]{BursteddeHolke16}; additionally, see
\cite{BursteddeWilcoxGhattas11} for an implementation for the classical Morton
index and~\cite{Knapp17} for an implementation for prism elements.

It is more challenging to find element face-neighbors across tree boundaries.
One reason is that neighbor trees may be rotated against each other.
For hybrid meshes, a new challenge occurs in that multiple shapes of trees exist
in the same forest;
this is for example the case if a hexahedron tree is neighbor of a
prism tree.
In this section, we develop a step-by-step procedure that fits our needs.
The core idea we propose is to explicitly build the face as a lower dimensional
element.

\begin{figure}
   \center
    \includegraphics{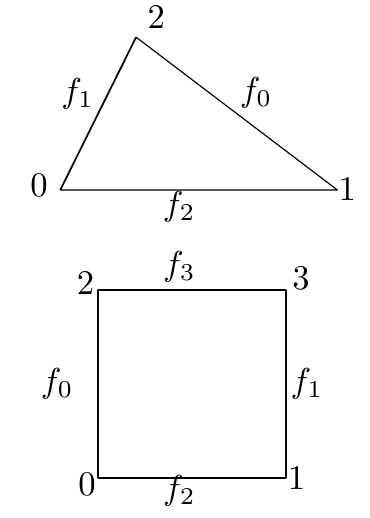}
    \includegraphics{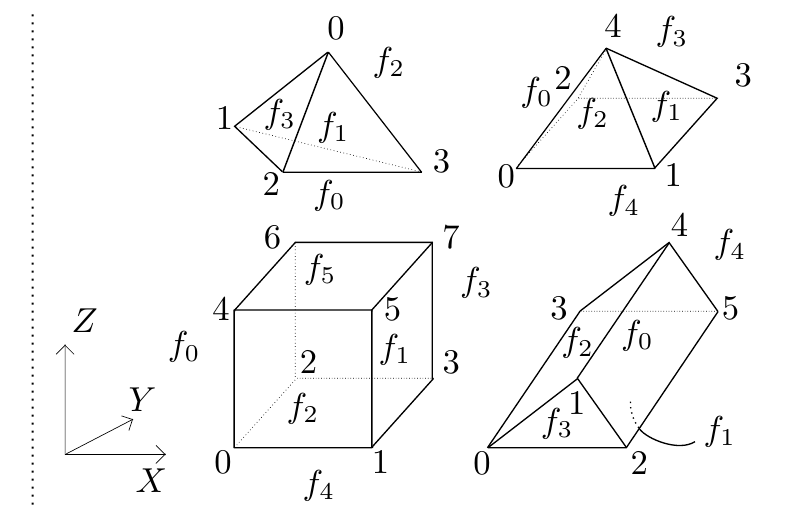}
   \caption{The vertex and face labels of the 2D (left) and 3D (right) tree shapes.
            While the low-level functions for the
            pyramid are not yet implemented in \tetcode, this element shape is
            covered by our new algorithm exactly like
            the others.}%
   \label{fig:vertexface}%
\end{figure}%
For reference, we display our numbering convention in Figure~\ref{fig:vertexface}.
\begin{figure}
\center
\def\svgwidth{0.5\textwidth}
\begingroup%
  \makeatletter%
  \providecommand\color[2][]{%
    \errmessage{(Inkscape) Color is used for the text in Inkscape, but the package 'color.sty' is not loaded}%
    \renewcommand\color[2][]{}%
  }%
  \providecommand\transparent[1]{%
    \errmessage{(Inkscape) Transparency is used (non-zero) for the text in Inkscape, but the package 'transparent.sty' is not loaded}%
    \renewcommand\transparent[1]{}%
  }%
  \providecommand\rotatebox[2]{#2}%
  \ifx\svgwidth\undefined%
    \setlength{\unitlength}{458.05712891bp}%
    \ifx\svgscale\undefined%
      \relax%
    \else%
      \setlength{\unitlength}{\unitlength * \real{\svgscale}}%
    \fi%
  \else%
    \setlength{\unitlength}{\svgwidth}%
  \fi%
  \global\let\svgwidth\undefined%
  \global\let\svgscale\undefined%
  \makeatother%
  \begin{picture}(1,0.79044063)%
    \put(0,0){\includegraphics[width=\unitlength]{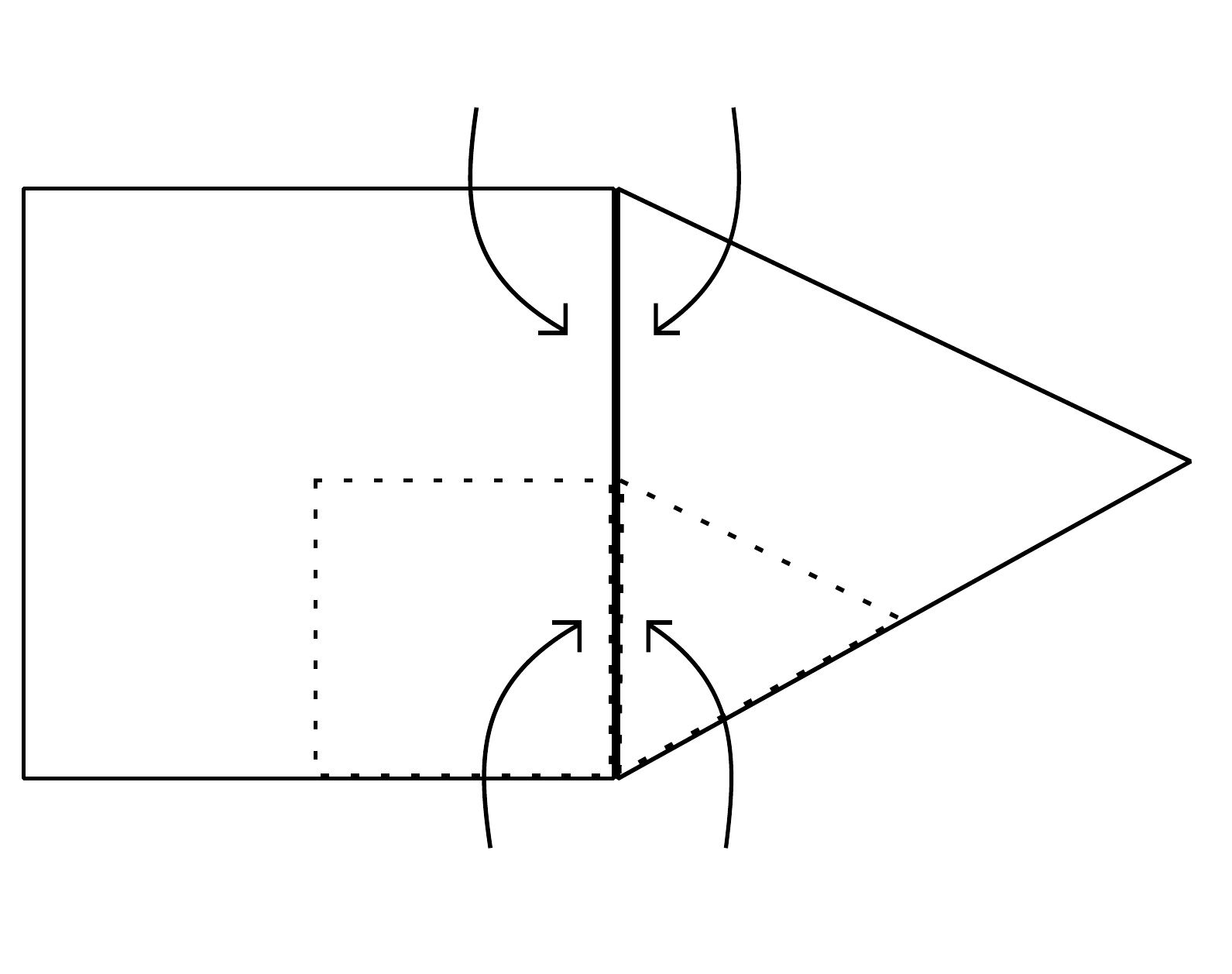}}%
    \put(0.10029941,0.50622876){\color[rgb]{0,0,0}\makebox(0,0)[lb]{\smash{$K$}}}%
    \put(0.71154646,0.41801682){\color[rgb]{0,0,0}\makebox(0,0)[lb]{\smash{$K'$}}}%
    \put(0.37475053,0.28167788){\color[rgb]{0,0,0}\makebox(0,0)[lb]{\smash{$E$}}}%
    \put(0.55164924,0.28167788){\color[rgb]{0,0,0}\makebox(0,0)[lb]{\smash{$E'$}}}%
    \put(0.37325352,0.72204715){\color[rgb]{0,0,0}\makebox(0,0)[lb]{\smash{$G$}}}%
    \put(0.57584834,0.72204715){\color[rgb]{0,0,0}\makebox(0,0)[lb]{\smash{$G'$}}}%
    \put(0.36726547,0.0184543){\color[rgb]{0,0,0}\makebox(0,0)[lb]{\smash{$F$}}}%
    \put(0.56561873,0.0184543){\color[rgb]{0,0,0}\makebox(0,0)[lb]{\smash{$F'$}}}%
  \end{picture}%
\endgroup%
 \caption[Element face-neighbor]{We show a tree $K$, an element $E$, and a face
$F$ of $E$ that is a subface of a tree face. The task is to construct the
face-neighbor element $E'$.
A subtask is to identify the tree
faces $G$ and $G'$, taking into account the coordinate systems of both trees,
and the face $F'$.
}%
\figlabel{fig:facesitu}%
\end{figure}
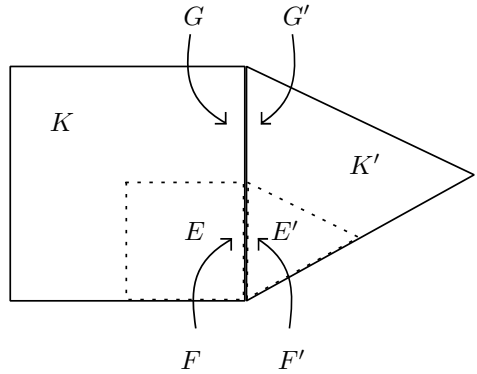%
We use capital letters ($K$, $E$, $F$, $G$) for entities
such as trees, elements, and faces, and use lower case for indices.
We draw a quadrilateral-triangle tree connection in Figure~\ref{fig:facesitu}.

The coordinate systems of neighbor trees may
not be aligned.
We must properly transform the $(d-1)$-dimensional coordinates of the faces
between the two systems.
To decouple neighbor trees of different shapes,
we consider the face
$G$ of the tree $K$ as a $(d-1)$-dimensional root tree
and explicitly construct the face $F$ of $E$ as a $(d-1)$-dimensional element
descendant of $G$.
Thus, we identify four major substeps in the computation of face-neighbors across
tree boundaries (Algorithm~\ref{alg:forestfaceneighbor} and Figure~\ref{fig:face-neighbor-hex}):
\begin{enumerate}[(i)]
\item From an element and its face number,
      identify the number $g$ of the tree face.
\item Construct the  $(d-1)$-dimensional face element $F$.
\item Transform the coordinates of $F$ to obtain the neighbor face
      element $F'$.
\item Extrude $F'$ to the $d$-dimensional neighbor element $E'$.
\end{enumerate}

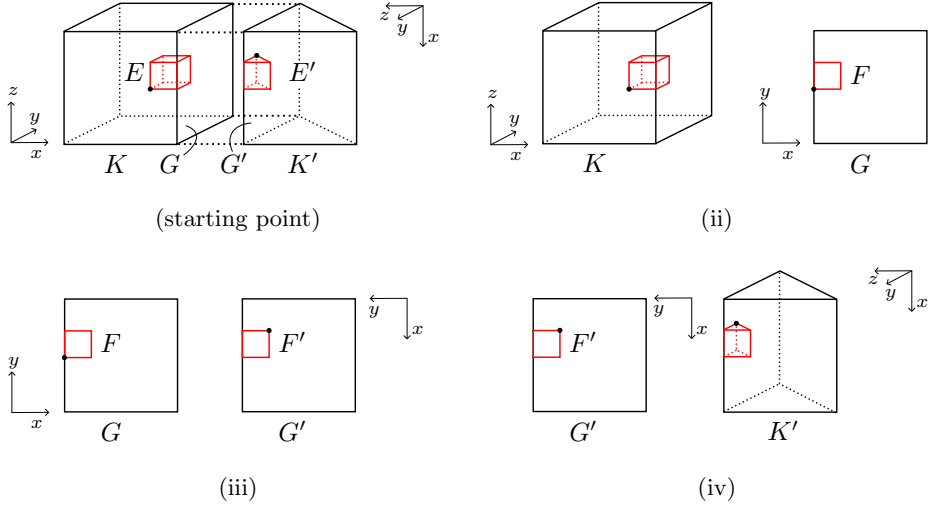
\begin{figure}
\begin{subfigure}[t]{0.48\textwidth}
 \def\svgwidth{\textwidth}
\begingroup%
  \makeatletter%
  \providecommand\color[2][]{%
    \errmessage{(Inkscape) Color is used for the text in Inkscape, but the package 'color.sty' is not loaded}%
    \renewcommand\color[2][]{}%
  }%
  \providecommand\transparent[1]{%
    \errmessage{(Inkscape) Transparency is used (non-zero) for the text in Inkscape, but the package 'transparent.sty' is not loaded}%
    \renewcommand\transparent[1]{}%
  }%
  \providecommand\rotatebox[2]{#2}%
  \ifx\svgwidth\undefined%
    \setlength{\unitlength}{799.10239258bp}%
    \ifx\svgscale\undefined%
      \relax%
    \else%
      \setlength{\unitlength}{\unitlength * \real{\svgscale}}%
    \fi%
  \else%
    \setlength{\unitlength}{\svgwidth}%
  \fi%
  \global\let\svgwidth\undefined%
  \global\let\svgscale\undefined%
  \makeatother%
  \begin{picture}(1,0.41401537)%
    \put(0,0){\includegraphics[width=\unitlength]{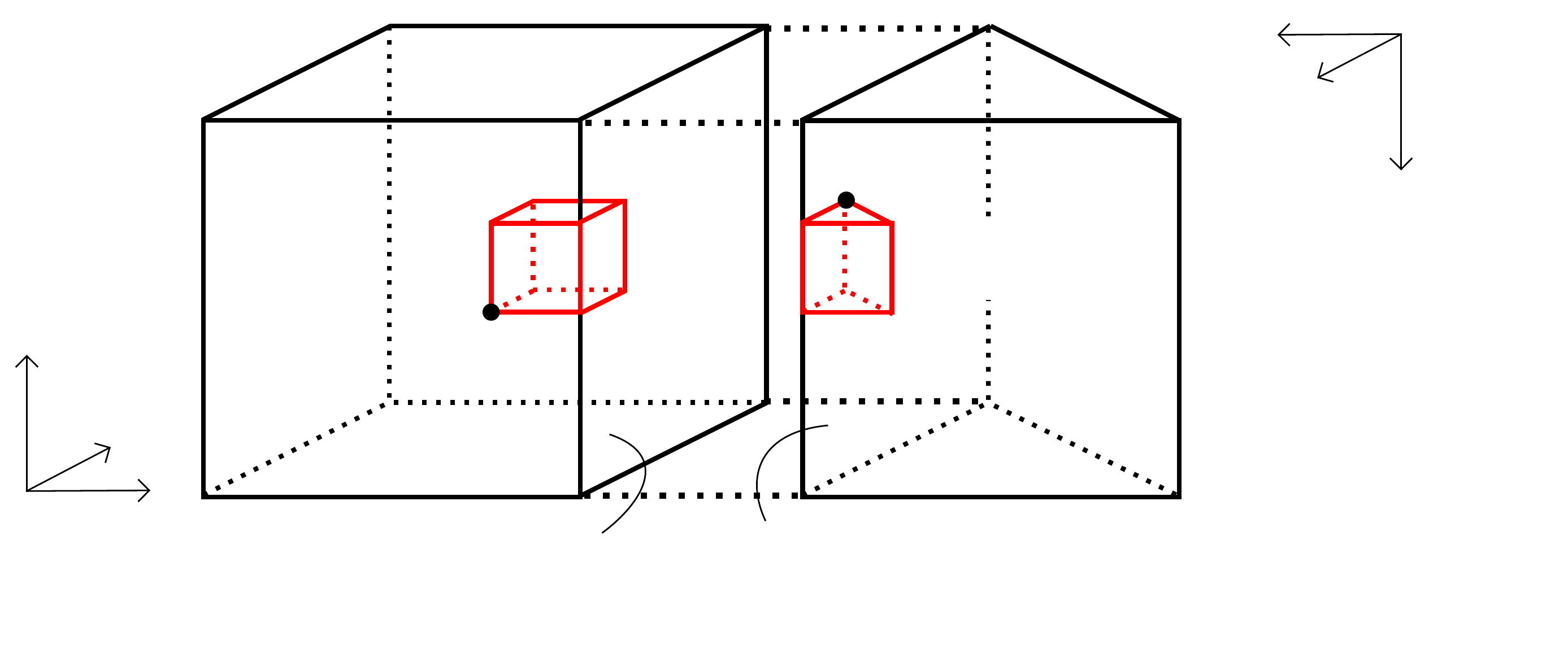}}%
    \put(0.21437256,0.03658345){\color[rgb]{0,0,0}\makebox(0,0)[lb]{\smash{$K$}}}%
    \put(0.60470357,0.03658345){\color[rgb]{0,0,0}\makebox(0,0)[lb]{\smash{$K'$}}}%
    \put(0.33266427,0.03658345){\color[rgb]{0,0,0}\makebox(0,0)[lb]{\smash{$G$}}}%
    \put(0.46158235,0.03658345){\color[rgb]{0,0,0}\makebox(0,0)[lb]{\smash{$G'$}}}%
    \put(0.25670404,0.23115977){\color[rgb]{0,0,0}\makebox(0,0)[lb]{\smash{$E$}}}%
    \put(0.60728489,0.23115977){\color[rgb]{0,0,0}\makebox(0,0)[lb]{\smash{$E'$}}}%
    \put(0.05828918,0.0758545){\color[rgb]{0,0,0}\makebox(0,0)[lb]{\smash{$_x$}}}%
    \put(0.0555752,0.15506053){\color[rgb]{0,0,0}\makebox(0,0)[lb]{\smash{$_y$}}}%
    \put(0.0098054,0.20967906){\color[rgb]{0,0,0}\makebox(0,0)[lb]{\smash{$_z$}}}%
    \put(0.90390524,0.32307219){\color[rgb]{0,0,0}\makebox(0,0)[lb]{\smash{$_x$}}}%
    \put(0.83749876,0.34477421){\color[rgb]{0,0,0}\makebox(0,0)[lb]{\smash{$_y$}}}%
    \put(0.80025501,0.37132049){\color[rgb]{0,0,0}\makebox(0,0)[lb]{\smash{$_z$}}}%
  \end{picture}%
\endgroup%
\hfill
 \caption*{(starting point)}
\end{subfigure}
\begin{subfigure}[t]{0.48\textwidth}
 \def\svgwidth{\textwidth}
\begingroup%
  \makeatletter%
  \providecommand\color[2][]{%
    \errmessage{(Inkscape) Color is used for the text in Inkscape, but the package 'color.sty' is not loaded}%
    \renewcommand\color[2][]{}%
  }%
  \providecommand\transparent[1]{%
    \errmessage{(Inkscape) Transparency is used (non-zero) for the text in Inkscape, but the package 'transparent.sty' is not loaded}%
    \renewcommand\transparent[1]{}%
  }%
  \providecommand\rotatebox[2]{#2}%
  \ifx\svgwidth\undefined%
    \setlength{\unitlength}{799.10400391bp}%
    \ifx\svgscale\undefined%
      \relax%
    \else%
      \setlength{\unitlength}{\unitlength * \real{\svgscale}}%
    \fi%
  \else%
    \setlength{\unitlength}{\svgwidth}%
  \fi%
  \global\let\svgwidth\undefined%
  \global\let\svgscale\undefined%
  \makeatother%
  \begin{picture}(1,0.41401368)%
    \put(0,0){\includegraphics[width=\unitlength]{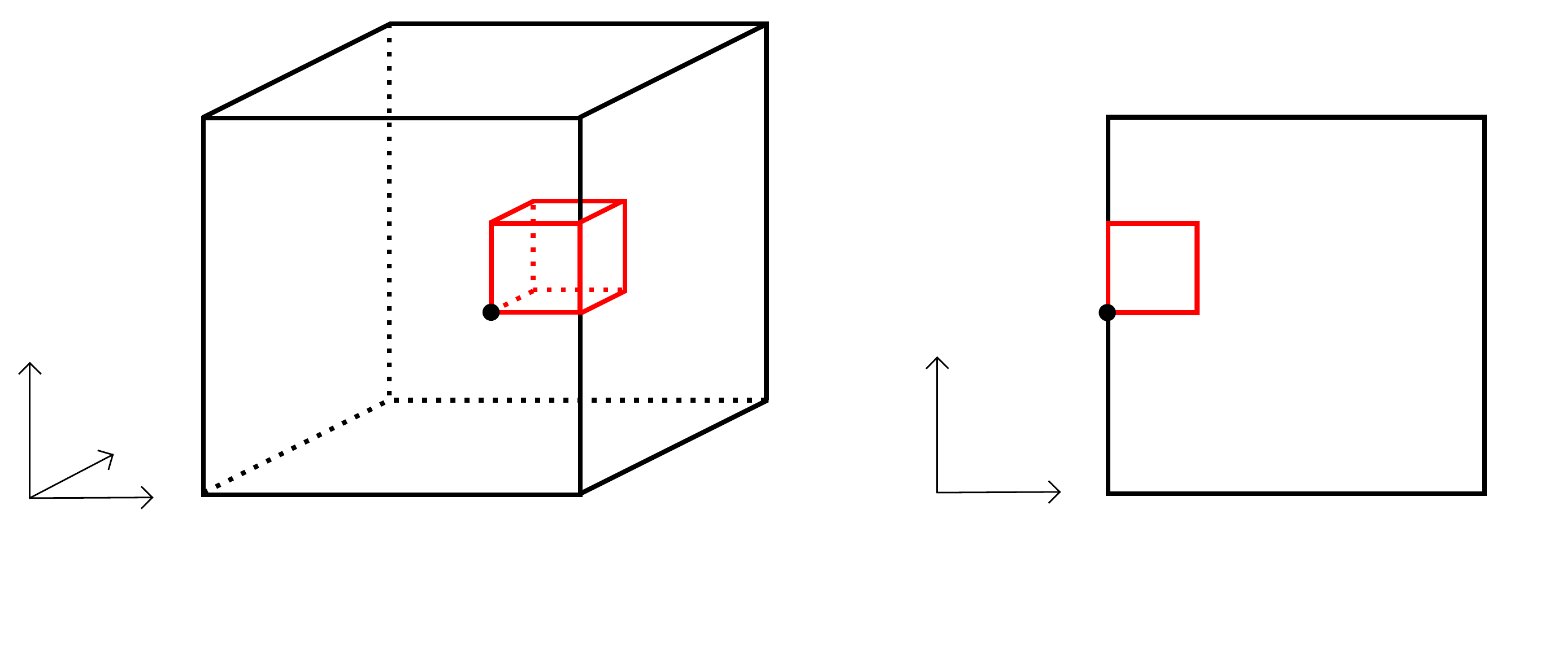}}%
    \put(0.21442341,0.03802052){\color[rgb]{0,0,0}\makebox(0,0)[lb]{\smash{$K$}}}%
    \put(0.78363383,0.03670257){\color[rgb]{0,0,0}\makebox(0,0)[lb]{\smash{$G$}}}%
    \put(0.78360257,0.22622664){\color[rgb]{0,0,0}\makebox(0,0)[lb]{\smash{$F$}}}%
    \put(0.06004851,0.07120157){\color[rgb]{0,0,0}\makebox(0,0)[lb]{\smash{$_x$}}}%
    \put(0.05733453,0.15040741){\color[rgb]{0,0,0}\makebox(0,0)[lb]{\smash{$_y$}}}%
    \put(0.01156483,0.20502583){\color[rgb]{0,0,0}\makebox(0,0)[lb]{\smash{$_z$}}}%
    \put(0.63894708,0.07523844){\color[rgb]{0,0,0}\makebox(0,0)[lb]{\smash{$_x$}}}%
    \put(0.5904634,0.2090627){\color[rgb]{0,0,0}\makebox(0,0)[lb]{\smash{$_y$}}}%
  \end{picture}%
\endgroup%
  \caption*{(ii)}
\end{subfigure}

\begin{subfigure}[t]{0.48\textwidth}
 \def\svgwidth{\textwidth}
\begingroup%
  \makeatletter%
  \providecommand\color[2][]{%
    \errmessage{(Inkscape) Color is used for the text in Inkscape, but the package 'color.sty' is not loaded}%
    \renewcommand\color[2][]{}%
  }%
  \providecommand\transparent[1]{%
    \errmessage{(Inkscape) Transparency is used (non-zero) for the text in Inkscape, but the package 'transparent.sty' is not loaded}%
    \renewcommand\transparent[1]{}%
  }%
  \providecommand\rotatebox[2]{#2}%
  \ifx\svgwidth\undefined%
    \setlength{\unitlength}{799.10400391bp}%
    \ifx\svgscale\undefined%
      \relax%
    \else%
      \setlength{\unitlength}{\unitlength * \real{\svgscale}}%
    \fi%
  \else%
    \setlength{\unitlength}{\svgwidth}%
  \fi%
  \global\let\svgwidth\undefined%
  \global\let\svgscale\undefined%
  \makeatother%
  \begin{picture}(1,0.41401368)%
    \put(0,0){\includegraphics[width=\unitlength]{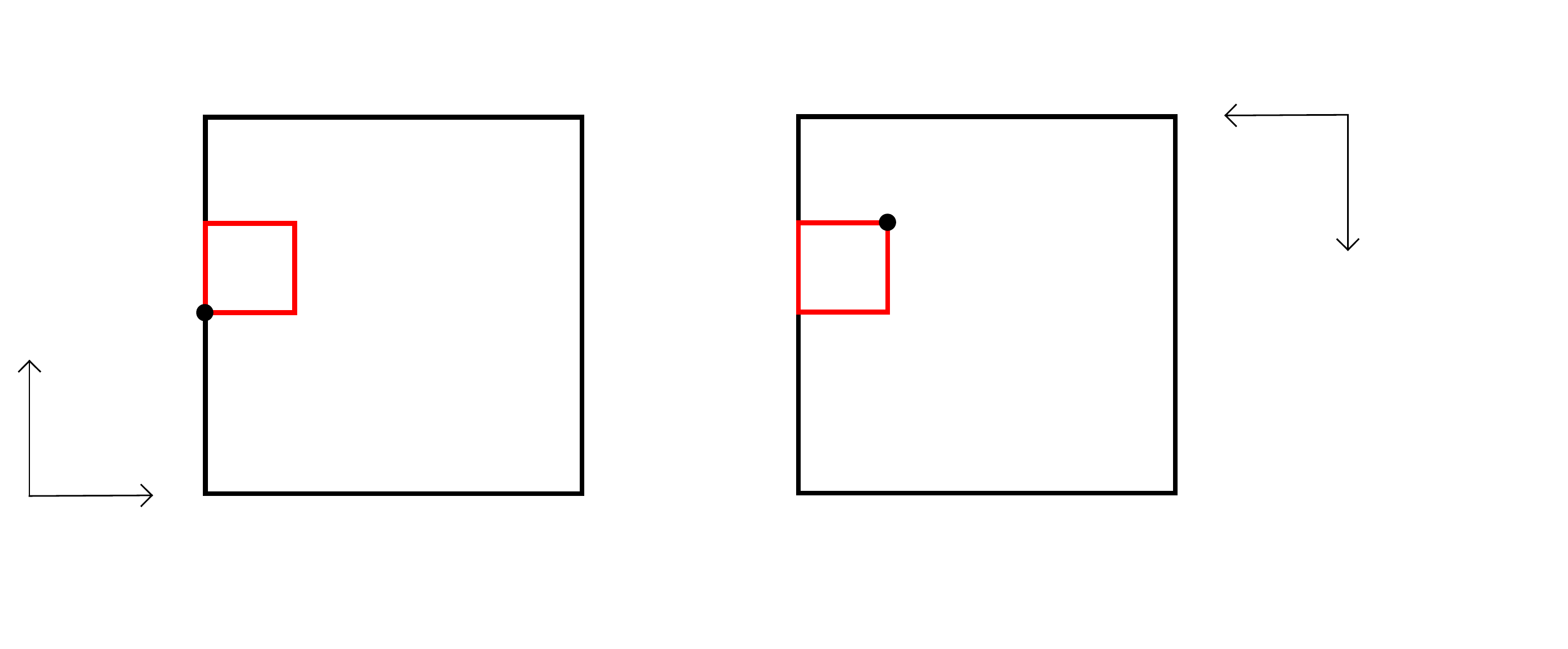}}%
    \put(0.20798911,0.03670257){\color[rgb]{0,0,0}\makebox(0,0)[lb]{\smash{$G$}}}%
    \put(0.58625273,0.03708453){\color[rgb]{0,0,0}\makebox(0,0)[lb]{\smash{$G'$}}}%
    \put(0.20761742,0.22645079){\color[rgb]{0,0,0}\makebox(0,0)[lb]{\smash{$F$}}}%
    \put(0.58574337,0.22660859){\color[rgb]{0,0,0}\makebox(0,0)[lb]{\smash{$F'$}}}%
    \put(0.0592312,0.07354792){\color[rgb]{0,0,0}\makebox(0,0)[lb]{\smash{$_x$}}}%
    \put(0.01074751,0.20737218){\color[rgb]{0,0,0}\makebox(0,0)[lb]{\smash{$_y$}}}%
    \put(0.86950839,0.27228219){\color[rgb]{0,0,0}\makebox(0,0)[lb]{\smash{$_x$}}}%
    \put(0.77787182,0.31652589){\color[rgb]{0,0,0}\makebox(0,0)[lb]{\smash{$_y$}}}%
  \end{picture}%
\endgroup%
\hfill
 \caption*{(iii)}
\end{subfigure}
\begin{subfigure}[t]{0.48\textwidth}
 \def\svgwidth{\textwidth}
\begingroup%
  \makeatletter%
  \providecommand\color[2][]{%
    \errmessage{(Inkscape) Color is used for the text in Inkscape, but the package 'color.sty' is not loaded}%
    \renewcommand\color[2][]{}%
  }%
  \providecommand\transparent[1]{%
    \errmessage{(Inkscape) Transparency is used (non-zero) for the text in Inkscape, but the package 'transparent.sty' is not loaded}%
    \renewcommand\transparent[1]{}%
  }%
  \providecommand\rotatebox[2]{#2}%
  \ifx\svgwidth\undefined%
    \setlength{\unitlength}{799.10400391bp}%
    \ifx\svgscale\undefined%
      \relax%
    \else%
      \setlength{\unitlength}{\unitlength * \real{\svgscale}}%
    \fi%
  \else%
    \setlength{\unitlength}{\svgwidth}%
  \fi%
  \global\let\svgwidth\undefined%
  \global\let\svgscale\undefined%
  \makeatother%
  \begin{picture}(1,0.41401368)%
    \put(0,0){\includegraphics[width=\unitlength]{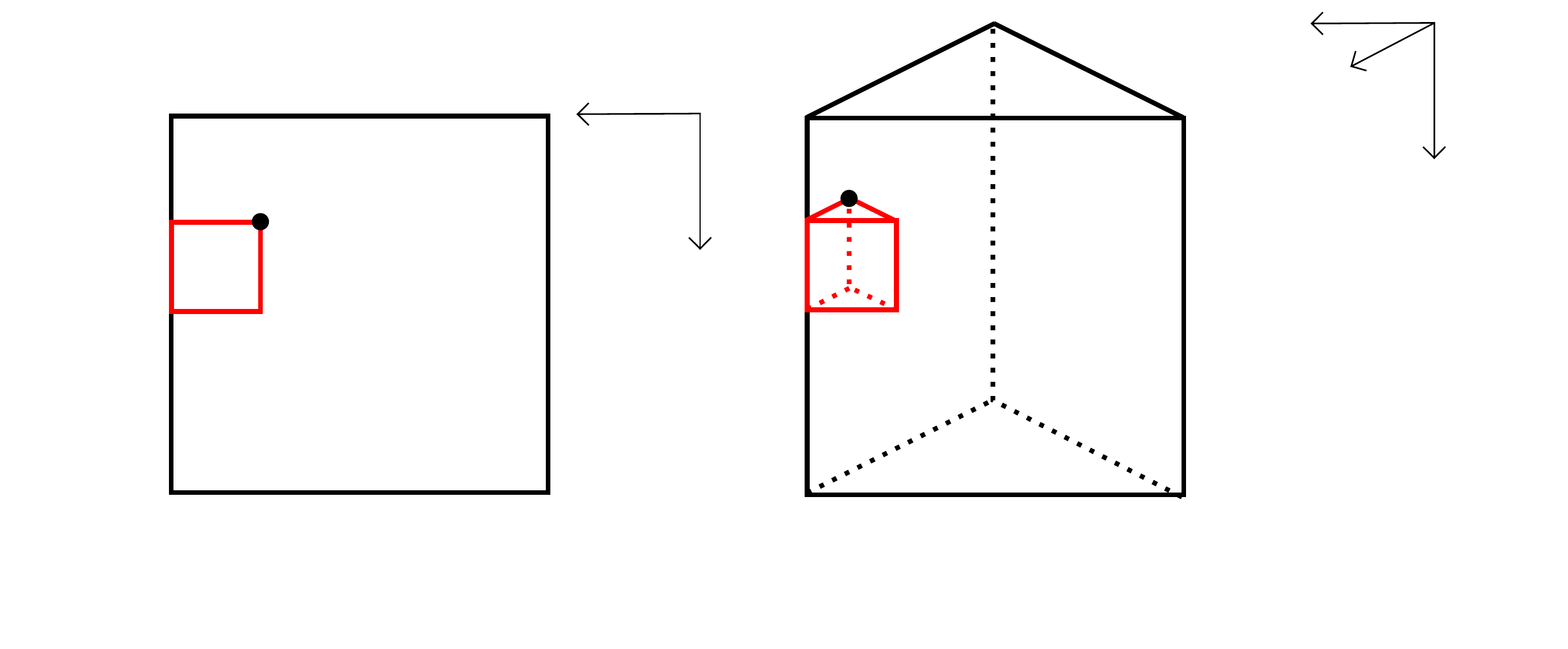}}%
    \put(0.18627798,0.0374569){\color[rgb]{0,0,0}\makebox(0,0)[lb]{\smash{$G'$}}}%
    \put(0.60757599,0.03822802){\color[rgb]{0,0,0}\makebox(0,0)[lb]{\smash{$K'$}}}%
    \put(0.18560335,0.22696489){\color[rgb]{0,0,0}\makebox(0,0)[lb]{\smash{$F'$}}}%
    \put(0.45584682,0.27293441){\color[rgb]{0,0,0}\makebox(0,0)[lb]{\smash{$_x$}}}%
    \put(0.36421025,0.31717811){\color[rgb]{0,0,0}\makebox(0,0)[lb]{\smash{$_y$}}}%
    \put(0.92464063,0.33055057){\color[rgb]{0,0,0}\makebox(0,0)[lb]{\smash{$_x$}}}%
    \put(0.85823435,0.35225255){\color[rgb]{0,0,0}\makebox(0,0)[lb]{\smash{$_y$}}}%
    \put(0.82099067,0.37879878){\color[rgb]{0,0,0}\makebox(0,0)[lb]{\smash{$_z$}}}%
  \end{picture}%
\endgroup%
  \caption*{(iv)}
\end{subfigure}
\caption
{
Example: a hexahedron and a prism element that are face-neighbors across tree
boundaries.
Constructing the face-neighbor $E'$ of $E$ across
the face amounts to computing its anchor node (black) from the anchor node of
$E$ and the coarse mesh connectivity information about the two neighbor trees.
Here, the coordinate systems of the two trees are rotated against each other.
In step (ii) we construct the face element $F$ from the element $E$.
The coordinate system of the face root is inferred from that of the left tree.
In step (iii) we transform the face element $F$ to the neighbor
face element $F'$.
In the last step (iv) we extrude the face-neighbor $E'$ from the face element
$F'$.}%
\figlabel{fig:face-neighbor-hex}%
\end{figure}%

\begin{rationale}
  We deliberately choose this method of using lower dimensional entities over
  directly transforming the tree coordinates from one tree to the other---as it
  is done for example in \cite{BursteddeWilcoxGhattas11}---since
  our approach allows for maximum flexibility of the implementations of the different
  element shapes and SFC choices.
  This holds since all intermediate operations are either local to one element
  or change the dimension (i.e.\ hexahedra to quadrilaterals, tetrahedra to
  triangles, and back), but not both.
  Therefore, even if, for example, a hexahedron tree is neighbor to a
  prism tree, no function in the implementation of the hexahedral
  elements relies on knowledge about the implementation of the prism
  elements.
  Hence, it is possible to exchange the implementation of the SFC for one
  element shape without changing the others.
\end{rationale}

We construct refined neighbors in a modular, two-step process:
First, we construct the children of an element touching a given face, which is
a tree-local operation.
Then, we construct the same-size neighbor for each child using the above steps,
which encapsulates all operations connecting two trees.

\begin{algorithm}
\DontPrintSemicolon
\caption{$E' \gets \texttt{t8\_forest\_face\_neighbor}$\newline
         \mbox{}\hfill
         (forest $\forest F$, tree $K$, element $E$, face number $f$)}%
\label{alg:forestfaceneighbor}%
\algoresult{The same-level face-neighbor $E'$ of $E$ across face $f$}\;
\algoif{\texttt{element\_neighbor\_inside\_root ($E$, $f$)}}
 { %
  $E'\gets$ \texttt{t8\_element\_face\_neighbor\_inside ($E$, $f$)}\;
 }%
\Else{
$g\gets$ \texttt{t8\_element\_tree\_face ($E$, $f$)}
\Comment{(i) Tree face no.\ and}
$\mathrm{o}\gets$ \texttt{face\_orientation ($\forest F$, $K$, $g$)}
\Comment{relative orientation}
$F \gets$ \texttt{t8\_element\_boundary\_face ($E$, $f$)}
\Comment{(ii) Face element}
$F'\gets$ \texttt{t8\_element\_transform\_face ($F$, $o$)}
\Comment{(iii) Neighb.\ fc.\mbox{}}
$g'\gets$ \texttt{tree\_neighbor\_face ($\forest F$, $K$, $g$)}
\Comment{Neighbor tree face}
$E' \gets$ \texttt{t8\_element\_extrude\_face ($F'$, $g'$)}
\Comment{(iv) and element}%
}%
\end{algorithm}

\subsection{(i) Identifying the tree face}

The first subproblem is to identify the tree face $G$, respectively its
face index $g$, from $E$, $f$, and the tree $K$.
For this task we define a new low-level function:

\lowlevel{$g \leftarrow$
          t8\_element\_tree\_face (element $E$, face index $f$)}{%
  The element face index $f$ designates a subface of a tree face.
  Return the face index $g$ of this root tree face.
  Only valid if face $f$ of $E$ is on a tree boundary.
}

For lines, quadrilaterals,
and hexahedra with the Morton index, the root tree face indices are the same as
the element's face indices \cite{BursteddeWilcoxGhattas11} and thus
\texttt{t8\_element\_tree\_face} always returns $g = f$.

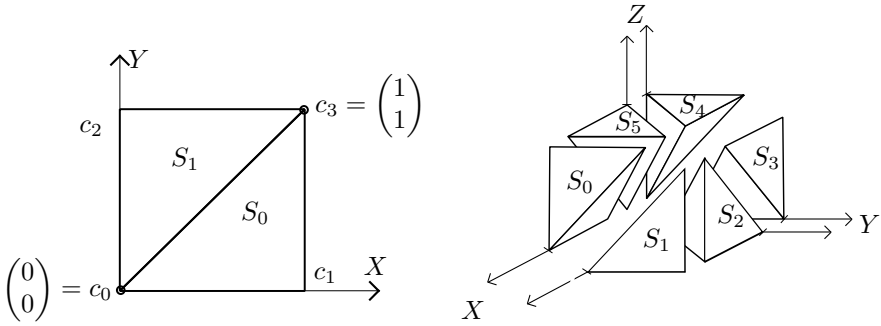
\begin{figure}
   \def\svgwidth{0.48\textwidth}%
\begingroup%
  \makeatletter%
  \providecommand\color[2][]{%
    \errmessage{(Inkscape) Color is used for the text in Inkscape, but the package 'color.sty' is not loaded}%
    \renewcommand\color[2][]{}%
  }%
  \providecommand\transparent[1]{%
    \errmessage{(Inkscape) Transparency is used (non-zero) for the text in Inkscape, but the package 'transparent.sty' is not loaded}%
    \renewcommand\transparent[1]{}%
  }%
  \providecommand\rotatebox[2]{#2}%
  \ifx\svgwidth\undefined%
    \setlength{\unitlength}{380.95061035bp}%
    \ifx\svgscale\undefined%
      \relax%
    \else%
      \setlength{\unitlength}{\unitlength * \real{\svgscale}}%
    \fi%
  \else%
    \setlength{\unitlength}{\svgwidth}%
  \fi%
  \global\let\svgwidth\undefined%
  \global\let\svgscale\undefined%
  \makeatother%
  \begin{picture}(1,0.63546393)%
    \put(0,0){\includegraphics[width=\unitlength]{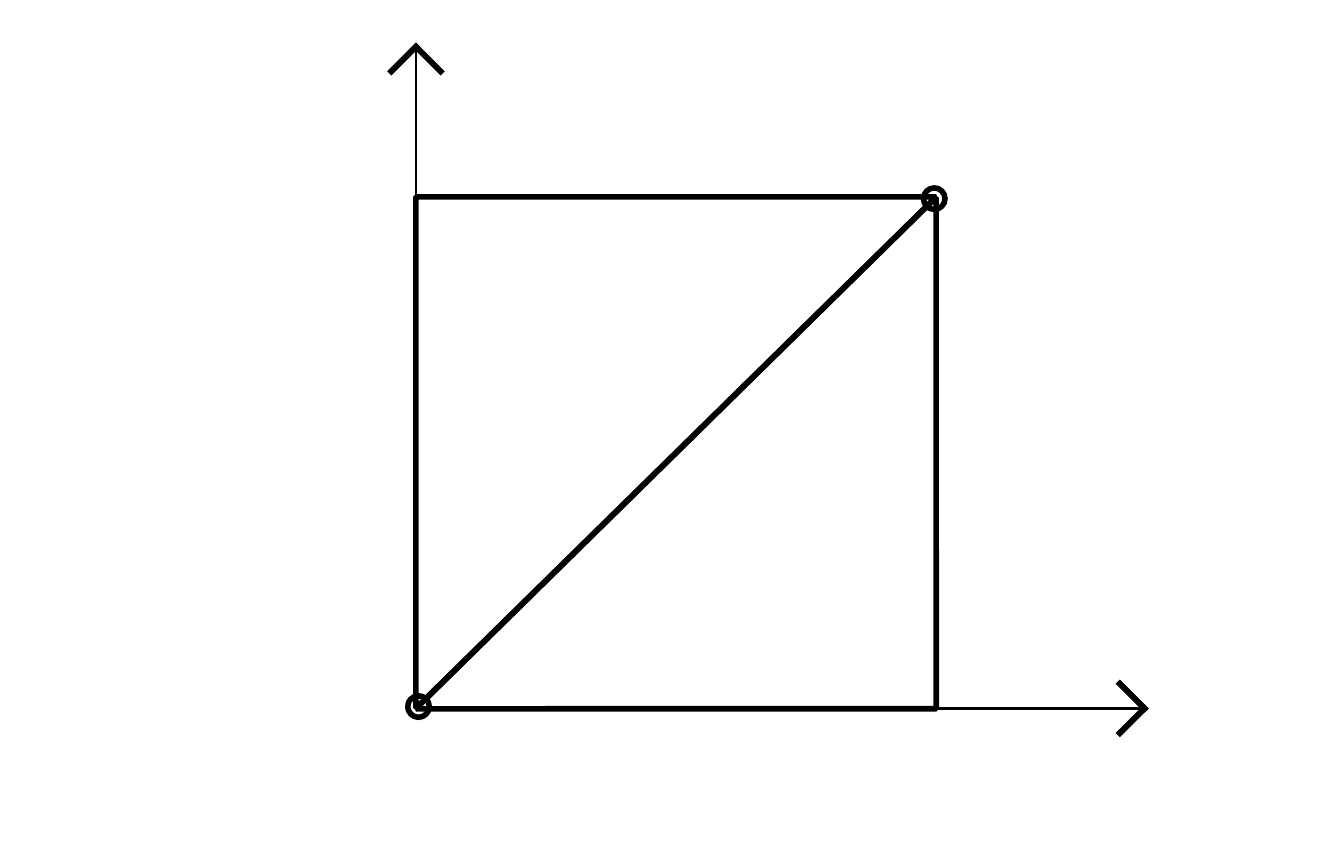}}%
    \put(0.83323174,0.13373889){\color[rgb]{0,0,0}\makebox(0,0)[lb]{\smash{$X$}}}%
    \put(0.33187745,0.57393166){\color[rgb]{0,0,0}\makebox(0,0)[lb]{\smash{$Y$}}}%
    \put(0.42372984,0.36473997){\color[rgb]{0,0,0}\makebox(0,0)[lb]{\smash{$S_1$}}}%
    \put(0.57283053,0.25028943){\color[rgb]{0,0,0}\makebox(0,0)[lb]{\smash{$S_0$}}}%
    \put(0.73033127,0.48024049){\color[rgb]{0,0,0}\makebox(0,0)[lb]{\smash{$c_3=\begin{pmatrix}
1\\1
\end{pmatrix}$}}}%
    \put(0.22886495,0.43750411){\color[rgb]{0,0,0}\makebox(0,0)[lb]{\smash{$c_2$}}}%
    \put(0.72709512,0.12591086){\color[rgb]{0,0,0}\makebox(0,0)[lb]{\smash{$c_1$}}}%
    \put(0.06300029,0.08895305){\color[rgb]{0,0,0}\makebox(0,0)[lb]{\smash{$\begin{pmatrix}
0\\0
\end{pmatrix}=c_0$}}}%
  \end{picture}%
\endgroup%
   \def\svgwidth{0.5\textwidth}%
\begingroup%
  \makeatletter%
  \providecommand\color[2][]{%
    \errmessage{(Inkscape) Color is used for the text in Inkscape, but the package 'color.sty' is not loaded}%
    \renewcommand\color[2][]{}%
  }%
  \providecommand\transparent[1]{%
    \errmessage{(Inkscape) Transparency is used (non-zero) for the text in Inkscape, but the package 'transparent.sty' is not loaded}%
    \renewcommand\transparent[1]{}%
  }%
  \providecommand\rotatebox[2]{#2}%
  \ifx\svgwidth\undefined%
    \setlength{\unitlength}{550.32397461bp}%
    \ifx\svgscale\undefined%
      \relax%
    \else%
      \setlength{\unitlength}{\unitlength * \real{\svgscale}}%
    \fi%
  \else%
    \setlength{\unitlength}{\svgwidth}%
  \fi%
  \global\let\svgwidth\undefined%
  \global\let\svgscale\undefined%
  \makeatother%
  \begin{picture}(1,0.71689905)%
    \put(0,0){\includegraphics[width=\unitlength]{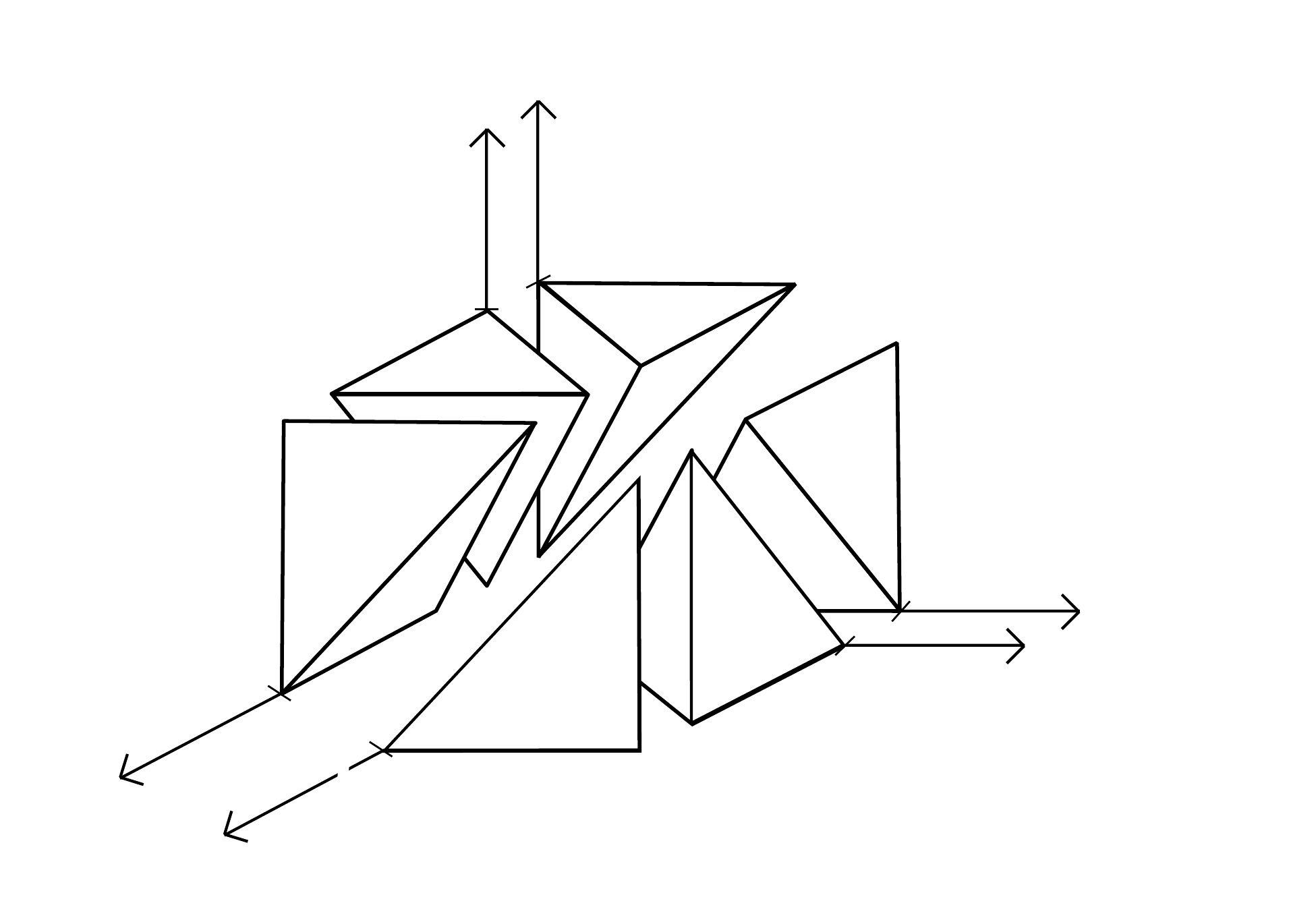}}%
    \put(0.25338018,0.30385227){\color[rgb]{0,0,0}\makebox(0,0)[lb]{\smash{$S_0$}}}%
    \put(0.41026491,0.18572463){\color[rgb]{0,0,0}\makebox(0,0)[lb]{\smash{$S_1$}}}%
    \put(0.56197011,0.23375293){\color[rgb]{0,0,0}\makebox(0,0)[lb]{\smash{$S_2$}}}%
    \put(0.63363954,0.35339003){\color[rgb]{0,0,0}\makebox(0,0)[lb]{\smash{$S_3$}}}%
    \put(0.352629,0.42566703){\color[rgb]{0,0,0}\makebox(0,0)[lb]{\smash{$S_5$}}}%
    \put(0.03878397,0.03761989){\color[rgb]{0,0,0}\makebox(0,0)[lb]{\smash{$X$}}}%
    \put(0.85164988,0.21446232){\color[rgb]{0,0,0}\makebox(0,0)[lb]{\smash{$Y$}}}%
    \put(0.37631668,0.64364674){\color[rgb]{0,0,0}\makebox(0,0)[lb]{\smash{$Z$}}}%
    \put(0.48367022,0.45951426){\color[rgb]{0,0,0}\makebox(0,0)[lb]{\smash{$S_4$}}}%
  \end{picture}%
\endgroup%
  \caption{%
Implementation of the TM-index:
The basic types $i$ for triangles (2D) and tetrahedra (3D) $S_i$ obtained by
dividing $[0,1]^d$ into simplices.
Left:
The unit square can be divided into two triangles sharing the
edge from $(0,0)^T$ to $(1,1)^T$.
The four corners of the square are numbered $c_0,\ldots,c_3$ in $yx$-order.
Right (exploded view):
In three dimensions the unit cube can be divided
into six tetrahedra, all sharing the edge from the origin to $(1,1,1)^T$.
The eight corners of the cube are numbered $c_0,\ldots,c_7$ in $zyx$-order
(redrawn and modified with permission \cite{Bey92}).}
\label{fig:sechstetras}
\end{figure}

For simplices with the TM index \cite{BursteddeHolke16},
the enumeration of their faces depends on their simplex type.
By convention, the face number $i$ refers to the unique face that does
not contain the vertex $\vec{x}_i$ (see \figref{vertexface}), and the vertex
numbering relative to the surrounding cube corners differs by type
(\figref{sechstetras}).

We observe that for triangles of type $0$, the face number is the same as the
face number of the root tree (since triangles of type $0$ are scaled copies of
the root tree).  Triangles of type $1$ cannot lie on the boundary of the root
tree and thus we never call \texttt{t8\_element\_tree\_face} with a type $1$
triangle.

For tetrahedra of type $0$ the same reasoning holds as for type
$0$ triangles, and \texttt{t8\_element\_tree\_face} returns $f$.
Tetrahedra of type $3$ cannot lie on the boundary of the root tree. For each
of the remaining four types there is exactly one face that can lie
on the root tree boundary. Face $0$ of type $1$ tetrahedra is a descendant of the
root face $0$; face 2 of type $2$ tetrahedra is a descendant of the root face $1$;
face 1 of type $4$ tetrahedra is a descendant of the root face $2$.
Finally, face $3$ of type $5$ tetrahedra is a descendant of the root face $3$.
We list these indices in Table~\ref{tab:elementtreeface}.

Note that for face indices $f$ of faces that cannot lie on the root boundary,
calling
\texttt{t8\_element\_tree\_face} is
illegal.
This behavior is well-defined, since we ensure in
Algorithm~\ref{alg:forestfaceneighbor} that the function is only called if the
face $f$ lies on the root boundary.

\begin{table}
\center
\begin{tabular}{|c|c|c||c|c|c|}
\hline
\multicolumn{6}{|c|}{Tetrahedron}\\ \hline
  $\type(T)$ & $f$ & $g$ & $\type(T)$ & $f$ & $g$\\ \hline
0 & $i$ & $i$ & 3 & $-$ & $-$ \\\hline
1 & $0$ & $0$ & 4 & $1$ & $2$ \\\hline
2 & $2$ & $1$ & 5 & $3$ & $3$ \\\hline
\end{tabular}
\caption[Face number and type for tetrahedral subfaces]
  {\texttt{$g =$ t8\_element\_tree\_face ($T$, $f$)} for a tetrahedron $T$ and a face $f$
  of $T$ that lies on a tree face.
  Depending on $T$'s type, all, exactly one, or none of its faces can be a
  subface of a face of the root tetrahedron tree.
  We show the tetrahedron's face number $f$ and the corresponding face number
  $g$ in the root tetrahedron.%
}
\figlabel{tab:elementtreeface}
\end{table}

\subsection{(ii) Constructing the face element}
\seclab{faceelement}%

As a next step, we build the face $F$ as a $(d-1)$-dimensional element.
We do this via the low-level function:

\lowlevel{$F \leftarrow$
         t8\_element\_boundary\_face (element $E$, face index $f$)}{
   Return the $(d-1)$-dimensional face element $F$ of element $E$
   specified by the face index $f$.
   Required for all elements of positive dimension.}

In other words, the lower dimensional face element $F$ is created from $E$.
For the Morton index this is equivalent to computing the coordinates of its
anchor node and additionally its type for the TM-index.
Hereby we interpret the tree face
$G$ as a $(d-1)$-dimensional root tree of which $F$ is a descendant element;
see also Figure~\ref{fig:construct-face}.

\begin{figure}
\center
\def\svgwidth{0.6\textwidth}
\begingroup%
  \makeatletter%
  \providecommand\color[2][]{%
    \errmessage{(Inkscape) Color is used for the text in Inkscape, but the package 'color.sty' is not loaded}%
    \renewcommand\color[2][]{}%
  }%
  \providecommand\transparent[1]{%
    \errmessage{(Inkscape) Transparency is used (non-zero) for the text in Inkscape, but the package 'transparent.sty' is not loaded}%
    \renewcommand\transparent[1]{}%
  }%
  \providecommand\rotatebox[2]{#2}%
  \ifx\svgwidth\undefined%
    \setlength{\unitlength}{658.38232422bp}%
    \ifx\svgscale\undefined%
      \relax%
    \else%
      \setlength{\unitlength}{\unitlength * \real{\svgscale}}%
    \fi%
  \else%
    \setlength{\unitlength}{\svgwidth}%
  \fi%
  \global\let\svgwidth\undefined%
  \global\let\svgscale\undefined%
  \makeatother%
  \begin{picture}(1,0.47149704)%
    \put(0,0){\includegraphics[width=\unitlength,page=1]{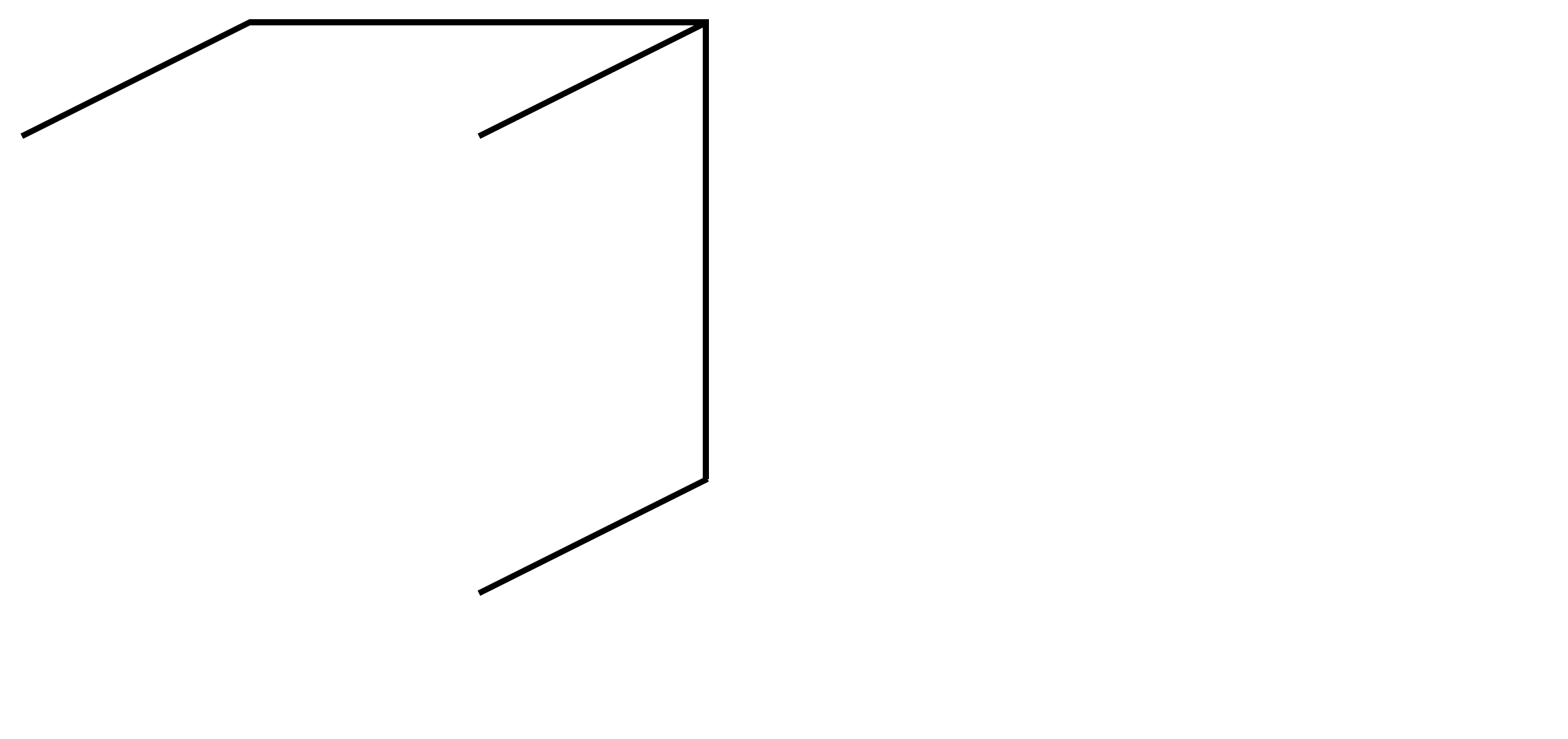}}%
    \put(0.13424745,0.03144546){\color[rgb]{0,0,0}\makebox(0,0)[lb]{\smash{$K$}}}%
    \put(0,0){\includegraphics[width=\unitlength,page=2]{face_neighbor_hex_facetree_tex.pdf}}%
    \put(0.79739285,0.03144546){\color[rgb]{0,0,0}\makebox(0,0)[lb]{\smash{$G$}}}%
    \put(0.18748731,0.25018946){\color[rgb]{0,0,0}\makebox(0,0)[lb]{\smash{$E$}}}%
    \put(0,0){\includegraphics[width=\unitlength,page=3]{face_neighbor_hex_facetree_tex.pdf}}%
    \put(0.67788548,0.28645259){\color[rgb]{0,0,0}\makebox(0,0)[lb]{\smash{$F$}}}%
  \end{picture}%
\endgroup%
 \caption[Constructing the face element $F$ to an element $E$ at a tree face
$G$.]
{Constructing the face element $F$ to an element $E$ at a tree face
$G$. We can interpret the face of the 3D tree $K$ as a 2D tree $G$.  The face
$F$ of $E$ is an element in this tree.}
\figlabel{fig:construct-face}
\end{figure}

\begin{remark}
  Since we construct a lower-dimensional element as the face of a
  higher-dimensional one, there are two conditions that need to be satisfied for
  the implementations of the two element shapes involved.
\begin{enumerate}
 \item The refinement pattern of a face of the higher dimensional elements must
       conform to the lower dimensional refinement pattern.
 \item The maximum possible refinement level of higher dimensional elements must
       not exceed the one of the lower dimensional elements.
\end{enumerate}
If one or both of these conditions are not fulfilled, then there exist
faces of the higher dimensional elements for which an interpretation as a lower
dimensional element is not possible.
For Morton-type SFCs, these two conditions are naturally fulfilled.

\end{remark}
\begin{remark}
 For the simplicial and hexahedral Morton SFC with maximum refinement level
 $\mathscr{L}$, the anchor node coordinates of an element of level $\ell$ are
 integer multiples of $2^{\mathscr L - \ell}$.
 Suppose the maximum level of hexahedral elements is $\mathscr{L}_1$ and the
 maximum level of a face boundary quadrilateral element is $\mathscr{L}_2 \geq
 \mathscr{L}_1$, then we will have to multiply a hexahedral coordinate with
 ${2^{\mathscr{L}_2-\mathscr{L}_1}}$ to transform it into a
 quadrilateral coordinate.
For simplicity, we reduce our presentation to the case that
all element shapes have the same
maximum possible refinement level and omit the scaling factor.
\end{remark}

The implementation for the Morton index is straightforward; see
Table~\ref{tab:quadhexfaces}.

\begin{table}
\center
\begin{tabular}[t]{|c|c|}\hline
\multicolumn{2}{|c|}{Quadrilateral}\\\hline
 $f$  & $F.x$ \\ \hline
 $0$, $1$ & $Q.y$ \\ \hline
 $2$, $3$ & $Q.x$ \\ \hline
\end{tabular}\hspace{2ex}%
\begin{tabular}[t]{|c|c||c|c|}\hline
\multicolumn{4}{|c|}{Hexahedron}\\\hline
 $f$ & $(F.x, F.y)$  & $f$ & $(F.x, F.y)$\\ \hline
 $0$ & $(Q.y, Q.z)$  & $3$ & $(Q.x, Q.z)$ \\ \hline
 $1$ & $(Q.y, Q.z)$  & $4$ & $(Q.x, Q.y)$ \\ \hline
 $2$ & $(Q.x, Q.z)$  & $5$ & $(Q.x, Q.y)$ \\ \hline
\end{tabular}
\caption
[\texttt{t8\_element\_boundary\_face} for quadrilaterals and hexahedra.]
{\texttt{t8\_element\_boundary\_face} for quadrilaterals and hexahedra.
  Left: For a quadrilateral $Q$ with anchor node $(Q.x, Q.y)$ and a face
$f$, the corresponding anchor node coordinate $F.x$ of the face line element.
Right: For a hexahedron $Q$ with anchor node $(Q.x, Q.y, Q.z)$ and a face
$f$, the corresponding anchor node coordinates $(F.x, F.y)$ of the face
quadrilateral element.
In either case, computing the coordinates is equivalent to a projection.}
\figlabel{tab:quadhexfaces}
\end{table}

For simplices with the TM-index, we note that we
shall restrict ourselves to
those combinations of element and face number that
occur on the tree boundary.
In particular, all possible faces are subfaces of the faces of the root simplex
$S_0$.

Triangles of type $1$ never lie on the root tree boundary, hence we only need
to consider type $0$ triangles.
The result solely depends on the face number $f$.

A tetrahedron that lies on the root tree boundary has a type different
from $3$. In order to compute the boundary face, we distinguish
two cases.  Let $g$ be the face of the root tetrahedron $S_0$ corresponding to
the boundary face $f$ of $T$.
\begin{enumerate}
\item
$g = 0$ or $g = 1$.
These faces of $S_0$ lie in the $(x = 0)$-plane or the $(x=z)$-plane of the
coordinate system, and $(F.x, F.y) = (T.z, T.y)$.
\item
$g = 2$ and $g = 3$. These faces lie in the $(y = 0)$-plane or the
$(y=z)$-plane, and the anchor node of $F$ is given by $(F.x, F.y) = (T.x,T.z)$.
\end{enumerate}

We show all possible results in Table~\ref{tab:facecat}.

\begin{table}
\center
 \begin{tabular}[t]{|c|c|c|} \hline
 \multicolumn{3}{|c|}{Triangle}\\\hline
 \mytabvspace $\type{(T)}$ & $f$ & $F.x$ \\ \hline
        0  &  $0$ & $T.y$  \\
           &  $1$ & $T.x$  \\
           &  $2$ & $T.x$  \\\hline
 \end{tabular}\hspace{3ex}%
 \begin{tabular}[t]{|c|c|c|c|c|} \hline
 \multicolumn{5}{|c|}{Tetrahedron}\\\hline
 \mytabvspace $\type{(T)}$ & $f$ & case & $\type{(F)}$ & $(F.x,F.y)$ \\ \hline
   0  &  $0$   & 1 & 0 & $(T.z,T.y)$ \\
      &  $1$   & 1 & 0 & $(T.z,T.y)$ \\
      &  $2$   & 2 & 0 & $(T.x,T.z)$ \\
      &  $3$   & 2 & 0 & $(T.x,T.z)$ \\\hline
   1  &  $0$   & 1 & 1 & $(T.z,T.y)$ \\\hline
   2  &  $2$   & 1 & 1 & $(T.z,T.y)$ \\\hline
   3  &  $-$   & $-$ & $-$ & $-$     \\\hline
   4  &  $1$   & 2 & 1 & $(T.x,T.z)$ \\\hline
   5  &  $3$   & 2 & 1 & $(T.x,T.z)$ \\\hline
\end{tabular}
\caption%
  {\texttt{t8\_element\_boundary\_face ($T$, $f$)} for triangles and tetrahedra.
   Left: The $x$ coordinate of the anchor node of the boundary line $F$ at face
   $f$ of a triangle $T$ in terms of $T$'s coordinates. Right: two cases occur,
   which we list together with
   the type of the boundary triangle $F$ at a face $f$ of tetrahedron $T$ and
   the anchor node coordinates $(F.x, F.y)$.}%
\figlabel{tab:facecat}%
\end{table}%

\subsection{(iii) Constructing $F'$ from $F$}
If we know the tree face index $g$, we can look up the corresponding face index
$g'$ of the face in $K'$ from the coarse mesh connectivity~\cite{BursteddeHolke17}.

In order to transform the coordinates of $F$ to obtain $F'$ we need
to understand how the vertices of the face $g$ connect to the vertices of
the face $g'$.
Each face's vertices form a subset of the vertices of the trees as in
Figure~\ref{fig:vertexface}.
Let
$\set{v_0,\dots,v_{n-1}}$ and $\set{v'_0,\dots,v'_{n-1}}$ be these vertices for
$g$ and $g'$ in ascending order, thus $v_i<v_{i+1}$ and $v'_i < v'_{i+1}$.
The face-to-face connection of the two trees determines a permutation
$\sigma\in S_n$ such that vertex $v_i$ connects to vertex $v'_{\sigma(i)}$.
In theory, there are $n!$ possible permutations.
However, not all of them occur.
\begin{definition}
 Since we exclude trees with negative volume, there is exactly one way
 to connect two trees across the faces $g$ and $g'$ in such a way that the
 vertices $v_0$ and $v'_0$ are connected.
 We call the corresponding permutation $\sigma_0$.

\end{definition}

We obtain all other possible permutations $\sigma$ by rotating the face $g'$.
This rotation is encoded in the orientation information of the coarse mesh.

\begin{definition}[From {\cite[Definition 2.2]{BursteddeHolke17}}]
 The \emph{orientation} of a face connection is the index $j$ such
 that $v_0$ connects with $v'_j$. Thus,
 \begin{equation}
   \orient (g, g', \sigma) = \sigma(0).
 \end{equation}
\end{definition}

\begin{remark}
  If we look at the same face connection, but change the order of $g$ and $g'$,
  the permutation $\sigma$ becomes $\sigma^{-1}$.
  In 3D $\sigma(0)$ is in general not equal to $\sigma^{-1}(0)$ and thus the
  orientation depends on the order of the faces $g$ and $g'$ (if unequal).
  In order to make the orientation unique, we use the following convention:
  If $K$ and $K'$ have the same shape then the smaller face is considered as $g$.
  If $K$ and $K'$ have different shapes, we consider $g$ as the face of the smaller shape,
  regarding the order: hexahedron $<$ prism $<$ pyramid; %
  tetrahedron $<$ prism $<$ pyramid.  For more details, we refer the reader
  to~\cite{BursteddeHolke17}.
\end{remark}

From the initial permutation $\sigma_0$ and the orientation we can
reconstruct $\sigma$.
$\sigma_0$ is determined by the shapes of $K$ and $K'$ and the face indices $g$
and $g'$. In fact, since the orientation encodes the possible rotations, the
only data we need to know is the sign of $\sigma_0$.

\begin{definition}
\label{def:facesign}
  Let $K$ and $K'$ be two trees of shapes $t$ and $t'$,
  and let $g$, $g'$ faces of $K$ and $K'$ of the same element shape.
  We define the \emph{sign} of $g$ and $g'$ as the sign of the
  permutation $\sigma_0$,
  \begin{equation}
    \sign_{t,t'}(g,g') := \sign(\sigma_0).
  \end{equation}

\end{definition}

\begin{remark}
This definition does not depend on the order of the faces $g$ and $g'$, since
  \begin{equation}
    \sign_{t',t}(g',g) = \sign(\sigma_0^{-1}) = \sign(\sigma_0) = \sign_{t,t'}(g,g').
  \end{equation}
\end{remark}

Using the orientation, the sign, and the face index $g'$, we transform the
coordinates of $F$ to obtain the corresponding face $F'$ as a subface of the
face $G'$ of $K'$. For this task we introduce the low-level function

\lowleveloneline{$F' \leftarrow$ t8\_element\_transform\_face}
          {(face element $F$, orientation $o$, sign $s$)}{.}

\begin{remark}
\label{rem:facetrafo}
The transformation $o = i$, $s = -1$ is the same
as first using $o = 0$, $s = -1$ and then $o = i$, $s = 1$.
Thus, we only need to implement all cases with $s = 1$ and one additional case
$o = 0$, $s = -1$.
\end{remark}

\begin{remark}
 The sign is always 0 for the boundary of line elements.
 If the faces are lines there are two possible face-to-face connections and
 these are already uniquely determined by the orientation of the connection.
 Thus, for 1D and 2D trees (lines, quadrilaterals, and triangles)
 it is not necessary to use the sign.
\end{remark}

  For hexahedra with the Morton index we compute the sign of two faces via the
  tables $\mathcal
  R, \mathcal Q, \mathcal P$ from~\cite[Table~3]{BursteddeWilcoxGhattas11} as
  \begin{equation}
   \label{eq:facetrafop4est}%
    \sign_{\mathrm{hex},\mathrm{hex}}(g,g') =
    \sign\left(i\mapsto \mathcal P\left(\mathcal Q( \mathcal R(g,g'),0), i\right)\right)
    = \neg\mathcal R(g,g').
  \end{equation}
  The permutation in the middle is exactly the permutation $\sigma_0$.
  The argument $0$ of $\mathcal Q$ is the orientation of a face-to-face
  connection, but the result is independent of it, and we could have chosen
  any other value.
We display the sign for all possible tree-to-tree connections of tetrahedra,
hexahedra, and prisms in Table~\ref{tab:facesigns}.

\begin{table}
  \center
  \begin{tabular}[t]{|cc|rrrr|}
    \hline
     \multicolumn{6}{|c|}{$K$ and $K'$ tetrahedra}\\\hline
    & & \multicolumn{4}{c|}{$g$}\\
    &  & 0 & 1 & 2 & 3\\\hline
      \multirow{4}{*}{$g'$}&
    0 &$-1$ &$ 1$ &$-1$ &$ 1$\\
   &1 &$ 1$ &$-1$ &$ 1$ &$-1$\\
   &2 &$-1$ &$ 1$ &$-1$ &$ 1$\\
   &3 &$ 1$ &$-1$ &$ 1$ &$-1$\\\hline
  \end{tabular}
  \begin{tabular}[t]{|cc|rrrrrr|}
    \hline
     \multicolumn{8}{|c|}{$K$ hexahedron, $K'$ prism}\\\hline
    & & \multicolumn{6}{c|}{$g$}\\
    &  & 0 & 1 & 2 & 3 &4 & 5 \\\hline
      \multirow{4}{*}{$g'$}&
    0 &$ 1$&$-1$&$-1$&$ 1$&$ 1$&$-1$\\
   &1 &$-1$&$ 1$&$ 1$&$-1$&$-1$&$ 1$\\
   &2 &$ 1$&$-1$&$-1$&$ 1$&$ 1$&$-1$\\ \hline
  \end{tabular}\\[1ex]
  \begin{tabular}[t]{|cc|rrrr|}
    \hline
     \multicolumn{6}{|c|}{$K$ tetrahedron, $K'$ prism}\\\hline
    & & \multicolumn{4}{c|}{$g$}\\
    &  & 0 & 1 & 2 & 3 \\\hline
      \multirow{2}{*}{$g'$}&
    3 &$-1$&$ 1$&$-1$&$ 1$\\
   &4 &$ 1$&$-1$&$ 1$&$-1$\\ \hline
  \end{tabular}
  \begin{tabular}[t]{|cc|rrrrr|}
    \hline
     \multicolumn{7}{|c|}{$K$ and $K'$ prisms}\\\hline
    & & \multicolumn{5}{c|}{$g$}\\
    &  & 0 & 1 & 2 & 3 & 4 \\\hline
      \multirow{4}{*}{$g'$}&
    0 & $-1$ &  $1$ & $-1$ &  $-$ &  $-$ \\
   &1 &  $1$ & $-1$ &  $1$ &  $-$ &  $-$ \\
   &2 & $-1$ &  $1$ & $-1$ &  $-$ &  $-$ \\
   &3 &  $-$ &  $-$ &  $-$ & $-1$ &  $1$ \\
   &4 &  $-$ &  $-$ &  $-$ &  $1$ & $-1$ \\\hline
  \end{tabular}
\caption%
{Value of $\sign_{t,t'}(g,g')$ from Definition~\ref{def:facesign} for four
 possible tree-to-tree connections.
 We obtain these values from Figure~\ref{fig:vertexface}.
 For two hexahedra, we refer to~\eqnref{facetrafop4est}.
}
\figlabel{tab:facesigns}%
\end{table}%

\begin{table}
\center
 \begin{tabular}[t]{|c|c|}\hline
  \multicolumn{2}{|c|}{Line}\\\hline
  \mytabvspace  $o$ & $\begin{pmatrix} F'.x\end{pmatrix}$\\[0.7ex] \hline
  \mytabvspace
   0  & $\begin{pmatrix}F.x\end{pmatrix}$ \\[0.5ex]
   1  &
  $\begin{pmatrix}2^\mathcal{L} - F.x - h\end{pmatrix}$ \\ \hline
 \end{tabular}
 \begin{tabular}[t]{|c|c||c|c|}\hline
  \multicolumn{4}{|c|}{Quadrilateral}\\\hline
  \myhugetabvspace  $o$ & $\begin{pmatrix} F'.x\\F'.y\end{pmatrix}$ &
  $o$ & $\begin{pmatrix} F'.x\\F'.y\end{pmatrix}$\\[2.0ex] \hline
   \myhugetabvspace
   0  & $\begin{pmatrix}F.x\\ F.y\end{pmatrix}$ &
   2  &
  $\begin{pmatrix} F.y \\ 2^\mathcal{L}-F.x-h\end{pmatrix}$\\[2.8ex]
   1  &
  $\begin{pmatrix}2^\mathcal{L} - F.y - h \\ F.x\end{pmatrix}$ &
   3  &
  $\begin{pmatrix} 2^\mathcal{L}-F.x-h\\2^\mathcal{L}-F.y-h \end{pmatrix}$\\ \hline
 \end{tabular}\\[2ex]
 \begin{tabular}{|c|c|c||c|c|c|}\hline
  \multicolumn{6}{|c|}{Triangle}\\\hline
  \myhugetabvspace $\type(F)$ & $o$ & $\begin{pmatrix} F'.x\\F'.y\end{pmatrix}$
 & $\type(F)$ & $o$ & $\begin{pmatrix} F'.x\\F'.y\end{pmatrix}$\\[2.0ex] \hline
   \myhugetabvspace
   0 &  0  & $\begin{pmatrix}F.x\\ F.y\end{pmatrix}$ &
   1 &  0  & $\begin{pmatrix}F.x\\ F.y\end{pmatrix}$ \\[2.8ex]
     &  1  &
  $\begin{pmatrix}2^\mathcal{L} - F.y - h \\ F.x - F.y\end{pmatrix}$ &
     &  1  &
  $\begin{pmatrix}2^\mathcal{L} - F.y - h \\ F.x - F.y - h\end{pmatrix}$ \\[2.8ex]
     &  2  &
  $\begin{pmatrix}2^\mathcal{L} - F.x + F.y - h\\ 2^\mathcal{L}-F.x-h\end{pmatrix}$ &
     &  2  &
  $\begin{pmatrix}2^\mathcal{L} - F.x + F.y\\ 2^\mathcal{L}-F.x-h\end{pmatrix}$
     \\ \hline
 \end{tabular}
\caption%
{Result of \texttt{t8\_transform\_face ($F$, $o$, $s = 1$)} for lines (top
left), quadrilaterals (top right) and triangles (bottom) with sign $1$.
For values with $s = -1$ see Table~\ref{tab:triangle-transform-b}
and Remark~\ref{rem:facetrafo}.}%
\figlabel{tab:triangle-transform}%
\end{table}%

For the classical and tetrahedral Morton indices we need to compute the anchor
node of $F'$ from the anchor node of the input face $F$.
We show the computation with $o = i$, $s = 1$ for lines, triangles and
quadrilaterals in Table~\ref{tab:triangle-transform}.
Since we transform faces, it is not necessary to discuss the routine for
3-dimensional face element shapes.
We describe the formulas for $o = 0$, $s = -1$ for triangles and quadrilaterals
in Table~\ref{tab:triangle-transform-b}.
As we mention in Remark~\ref{rem:facetrafo}, we can compute all $o$ and $s$
combinations from these two tables.
For quadrilaterals and hexahedra, \texttt{t8\_element\_transform\_face} is
equivalent to the internal coordinate transformation in
\texttt{p4est\_transform\_face}
due to~\eqref{eq:facetrafop4est}.

\begin{table}
  \center
  \begin{tabular}[t]{|c|c|}\hline
  \multicolumn{2}{|c|}{Triangle}\\\hline
  \myhugetabvspace $\type(F)$ & $\begin{pmatrix} F'.x\\F'.y\end{pmatrix}$\\[2.0ex] \hline
   \myhugetabvspace
   0 &  $\begin{pmatrix}F.x\\ F.x - F.y\end{pmatrix}$
     \\[2.8ex]
   1 &
  $\begin{pmatrix} F.x \\ F.x - F.y - h\end{pmatrix}$
    \\ \hline
 \end{tabular}
  \begin{tabular}[t]{|c|}\hline
  \multicolumn{1}{|c|}{Quadrilateral}\\\hline
  \myhugetabvspace $\begin{pmatrix} F'.x\\F'.y\end{pmatrix}$\\[2.0ex] \hline
   \myhugetabvspace
    $\begin{pmatrix}F.y\\ F.x\end{pmatrix}$
    \\ \hline
 \end{tabular}
  \caption%
{Result of
 \texttt{t8\_transform\_face ($F$, $o = 0$, $s = -1$)} for triangles (left) and
  quadrilaterals (right).  We compute any arbitrary combination of values for
  $o$ with $s = -1$ by first applying \texttt{t8\_transform\_face ($F$, $0$, $-1$)}
  and then \texttt{t8\_transform\_face ($F$, $o$, $1$)} from
  Table~\ref{tab:triangle-transform}.}%
  \label{tab:triangle-transform-b}%
\end{table}%

\subsection{(iv) Constructing $E'$ from $F'$}
We now have $E$, $F$, $F'$, $K$ and $K'$ and can construct the neighbor element
$E'$.
For this we introduce the function

\lowleveloneline{$E' \leftarrow$ t8\_element\_extrude\_face}
{(face element $F'$, tree $K'$, face index $g'$)}
  {.}
\noindent
This function has as input a face element and a root tree
face index and as output the element within the root tree that has as a boundary
face the given face element.
How to compute the element from this data depends on the element shape and
the root tree face.
For quadrilaterals, triangles, hexahedra, and tetrahedra with the (T-)Morton index
we show the formulas to compute the anchor node coordinates of $E'$ in
Table~\ref{tab:face-extrudeNEW}.

\begin{table}
\begin{minipage}[t]{0.45\textwidth}
\center
 \begin{tabular}[t]{|c|c||c|c|}\hline
  \multicolumn{4}{|c|}{2D -- coordinates}\\\hline
   \myhugetabvspace
  $g'$ & $\begin{pmatrix} E'.x\\ E'.y\end{pmatrix}$ &
  $g'$ & $\begin{pmatrix} E'.x\\E'.y\end{pmatrix}$\\[2.0ex]\hhline{|====|}
  \multicolumn{4}{|c|}{Quadrilateral from line}\\\hline
   \myhugetabvspace
   0  & $\begin{pmatrix}0\\ F'.x\end{pmatrix}$ &
   2  &
  $\begin{pmatrix} F'.x \\ 0\end{pmatrix}$\\[2.8ex]
   1  &
  $\begin{pmatrix}2^\mathcal{L} - h \\ F'.x\end{pmatrix}$ &
   3  &
  $\begin{pmatrix} F'.x \\2^\mathcal{L} - h \end{pmatrix}$\\ \hline

  \multicolumn{4}{|c|}{Triangle from line}\\\hline
   \myhugetabvspace
   0  & $\begin{pmatrix} 2^\mathcal L - h\\ F'.x\end{pmatrix}$ &
   2  &
  $\begin{pmatrix} F'.x \\ 0\end{pmatrix}$\\[2.8ex]
   1  &
  $\begin{pmatrix}F'.x \\ F'.x\end{pmatrix}$ &
     & \\ \hline
 \end{tabular}\\[2ex]
 \begin{tabular}[t]{|c|c|c|}\hline
  \multicolumn{3}{|c|}{3D -- types}\\\hline
    $g'$ & $\type(F')$ & $\type(E')$ \\ \hline
  \multicolumn{3}{|c|}{Tetrahedron from triangle}\\ \hline
   0 & 0 & 0 \\
     & 1 & 1 \\\hline
   1 & 0 & 0 \\
     & 1 & 2 \\\hline
   2 & 0 & 0 \\
     & 1 & 4 \\\hline
   3 & 0 & 0 \\
     & 1 & 5 \\\hline
  \multicolumn{3}{|c|}{Prism from triangle or quad}\\\hline
   0 & $-$ & 0 \\\hline
   1 & $-$ & 0 \\\hline
   2 & $-$ & 0 \\\hline
   3 & 0 & 0 \\
     & 1 & 1 \\\hline
   4 & 0 & 0 \\
     & 1 & 1 \\\hline
 \end{tabular}
\end{minipage}\hfill%
\begin{minipage}[t]{0.45\textwidth}
 \begin{tabular}[t]{|c|c||c|c|}\hline
  \multicolumn{4}{|c|}{3D - coordinates}\\\hline
   \myhugetabvspace
    $g'$ & $\begin{pmatrix} E'.x\\ E'.y \\ E'.z\end{pmatrix}$ &
    $g'$ & $\begin{pmatrix} E'.x\\ E'.y \\ E'.z\end{pmatrix}$\\[2.0ex] \hline
  \multicolumn{4}{|c|}{Hexahedron from quad}\\\hline
   \myhugetabvspace
   0  & $\begin{pmatrix}0\\ F'.x \\ F'.y\end{pmatrix}$ &
   3  &
  $\begin{pmatrix} F'.x \\ 2^\mathcal L - h \\ F'.y \end{pmatrix}$\\[3.6ex]
   1  &
  $\begin{pmatrix} 2^\mathcal L - h \\ F'.x \\ F'.y \end{pmatrix}$ &
   4  &
  $\begin{pmatrix} F'.x \\ F'.y \\ 0 \end{pmatrix}$\\[3.6ex]
   2  &
  $\begin{pmatrix} F'.x \\ 0 \\ F'.y\end{pmatrix}$ &
   5  &
  $\begin{pmatrix} F'.x \\ F'.y \\ 2^\mathcal{L} - h \end{pmatrix}$\\[3.6ex]
\hline
  \multicolumn{4}{|c|}{Tetrahedron from triangle}\\\hline
   {\scalebox{1.2}{\myhugetabvspace}}
   0  & $\begin{pmatrix} 2^\mathcal L - h\\ F'.y \\ F'.x\end{pmatrix}$ &
   2  &
  $\begin{pmatrix} F'.x \\ 0 \\ F'.y\end{pmatrix}$\\[3.6ex]
   1  &
  $\begin{pmatrix} F'.x \\ F'.y \\ F'.x\end{pmatrix}$ &
   3  &
  $\begin{pmatrix} F'.x \\ 0 \\ F'.y \end{pmatrix}$\\\hline
  \multicolumn{4}{|c|}{Prism from triangle or quad}\\\hline
   {\scalebox{1.2}{\myhugetabvspace}}
   0  & $\begin{pmatrix} 2^\mathcal L - h\\ F'.x \\ F'.y\end{pmatrix}$ &
   3  &
  $\begin{pmatrix} F'.x \\ F'.y \\ 0\end{pmatrix}$\\[3.6ex]
   1  &
  $\begin{pmatrix} F'.x \\ F'.x \\ F'.y\end{pmatrix}$ &
   4  &
  $\begin{pmatrix} F'.x \\ F'.y \\ 2^\mathcal L - h\ \end{pmatrix}$\\
   2  &
  $\begin{pmatrix} F'.x \\ 0 \\ F'.y\end{pmatrix}$ &
      & \\\hline
 \end{tabular}
\end{minipage}
\caption
{The computation of $E'=$ \texttt{t8\_element\_extrude\_face ($F'$, $T'$, $g'$)}.
Depending on the anchor node coordinates of $F'$ and the tree face index $g'$,
we determine the anchor node of the extruded element $E'$.
For tetrahedra and prisms, we additionally need to compute the type of $E'$,
which depends on $g'$ and the type of the triangle $F'$ (bottom left).
In the case of a triangle, the type of $E'$ is always $0$, since type $1$
triangles cannot lie on a tree boundary. Hence, we do not show a table for
this case.
$h$ refers to the length of the element $E'$ (resp.\ $F'$) and is computed as
$2^{\mathcal L -\ell}$, where $\ell$ is the refinement level of $E'$ and $F'$.}%
\figlabel{tab:face-extrudeNEW}%
\end{table}%

\subsection{Refined face-neighbors}
\label{sec:ghost-halfneighbors}
\seclab{Halfsizeghostv1}

In order to implement the ghost algorithm for balanced forests from
\cite{BursteddeWilcoxGhattas11}, we need to compute refined face-neighbors of an
element.
That is, given an element $E$ and a face $f$, construct the
neighbors
of $E$ across $f$ of refinement level $\ell(E) + 1$.
Most often these neighbors are half-size in each dimension, but if we are
dealing for example with a 1:$3^d$ Peano refinement, the neighbors are one
third the size.
Without loss of generality, we will be using the term half-size.

We construct half face-neighbors in three steps:
\begin{enumerate}[(i)]
 \item Construct the children $C_f$ of $E$ that have a face in $f$,
       in SFC order.
 \item For each child $C_f[i]$ compute the face index $f_i$ of the face
       that is a child of $f$ and a face of $C_f[i]$.
       (For the cubical Morton curve, the face indices $f_i$ are ascending
       by construction, but this is not generally true for hybrid refinements.)
 \item For each child $C_f[i]$ compute its face-neighbor across $f_i$.
       If we cross a tree boundary, these children may not be in order of the
       neighbor tree's SFC, which is relevant for application codes.
       Within the same tree we can use an optimized code path.
\end{enumerate}

The third step has been developed above.  Steps (i) and (ii) are performed by
two new low-level algorithms:

\lowlevel{$C[] \leftarrow$
          t8\_element\_children\_at\_face (element $E$, face index $f$)}{
Returns an array of children of $E$ that share a face with $f$.}
\vspace{-\baselineskip}

\lowleveloneline{$f' \leftarrow$ t8\_element\_child\_face}
                 {(element $E$, child index $i$, face index $f$)}
         {\newline
Given an element $E$, a child index $i$ of $E$, and a face index $f$ of some
face $F$ of $E$, compute the index $f'$ of the child's face that is a subface
of $F$.\\
It is required that the child lies on the face $F$.}

A typical implementation of \texttt{t8\_element\_children\_at\_face}
would look up the child indices of these children in a table and then
construct the children with these indices.
The child indices can be obtained from the refinement pattern. For the
quadrilateral Morton index, for example, the child indices at face $f = 0$ are
$0$ and $2$. For a hexahedron the child indices at face $f = 3$ are $2$, $3$,
$6$, and $7$.
For the TM index these indices additionally depend on the type of the
simplex.  We list all cases in Table~\ref{tab:ghost-childrenatface}.

\begin{table}
\scalebox{0.97}{
 \begin{tabular}[t]{|c|c|c|c|}
 \hline
 \multicolumn{4}{|c|}{Triangle}\\\hline
 &\multicolumn{3}{c|}{$f$}\\\hline
 $\type(T)$  & 0 & 1 & 2 \\ \hline
  0 & 1,3 & 0,3 & 0,1 \\
  1 & 2,3 & 0,3 & 0,2 \\\hline
 \end{tabular}
 \begin{tabular}[t]{|c|c|c|c|c|}
 \hline
 \multicolumn{5}{|c|}{Tetrahedron}\\\hline
 &\multicolumn{4}{c|}{$f$}\\\hline
 $\type(T)$  & 0 & 1 & 2 & 3 \\ \hline
  0 & 1, 4, 5, 7 & 0, 4, 6, 7 & 0, 1, 2, 7 & 0, 1, 3, 4 \\
  1 & 1, 4, 5, 7 & 0, 5, 6, 7 & 0, 1, 3, 7 & 0, 1, 2, 5 \\
  2 & 3, 4, 5, 7 & 0, 4, 6, 7 & 0, 1, 3, 7 & 0, 2, 3, 4 \\
  3 & 1, 5, 6, 7 & 0, 4, 6, 7 & 0, 1, 3, 7 & 0, 1, 2, 6 \\
  4 & 3, 5, 6, 7 & 0, 4, 5, 7 & 0, 1, 3, 7 & 0, 2, 3, 5 \\
  5 & 3, 5, 6, 7 & 0, 4, 6, 7 & 0, 2, 3, 7 & 0, 1, 3, 6 \\ \hline
 \end{tabular}\\
}
\center
\begin{tabular}[t]{|c|c|c|c|c|c|}
\hline
\multicolumn{6}{|c|}{Prism}\\\hline
&\multicolumn{5}{c|}{$f$}\\\hline
$\type(P)$  & 0 & 1 & 2 & 3 & 4\\ \hline
 0 & 1, 3, 5, 7 & 0, 3, 4, 7 & 0, 1, 4, 5 & 0, 1, 2, 3  & 4, 5, 6, 7\\
 1 & 2, 3, 6, 7 & 0, 3, 4, 7 & 0, 2, 4, 6 & 0, 1, 2, 3  & 4, 5, 6, 7\\ \hline
\end{tabular}
\caption[The child indices of all children of an element touching a given face.]
{The child indices of all children of an element touching a given face
for triangles, tetrahedra and prisms.
These indices are needed for \texttt{t8\_element\_children\_at\_face}.}
\figlabel{tab:ghost-childrenatface}
\end{table}

The low-level algorithm \texttt{t8\_element\_child\_face} can also be described
via lookup tables. Its input is a parent element $E$, a face index $f$ and a
child index $i$, such that the child $E_i$ of $E$ has a subface of the face
$f$. In other words, $E_i$ is part of the output of
\texttt{t8\_element\_children\_at\_face}. The return value of
\texttt{t8\_element\_child\_face} is the face index $f_i$ of the face of
$E[i]$ that is the subface of $f$.

For the classical Morton index, the algorithm is the identity on $f$, since the faces
of child quadrilaterals/hexahedra are labeled in the same manner as those of the
parent element.
For the TM index for triangles, the algorithm is also the identity, since only
triangle children of the same type as the parent can touch a face of the parent
and for same type triangles the faces are labeled in the same manner.

For tetrahedra with the TM-index, the algorithm is the identity on those
children that have the same type as the parent.
However, for each face $f$ of a tetrahedron $T$,
there exists a child of $T$ that has the middle face child of $f$
as a face. This child does not have the same type as $T$.
For this child the corresponding face value is computed as $0$ if $f=0$, $2$ if
$f = 1$, $1$ if $f=2$, or $3$ if $f=3$.

\section{Finding owner processes of elements}
\label{sec:ghost-findowner}
\seclab{faceownerghostv2}

For any ghost algorithm, after we have successfully constructed an element's
full-size or refined face-neighbor, we need to identify the owner process of
this neighbor.
We can use this information to shorten the list of potential neighbor
processes, eventually arriving at the tightest possible set to communicate
with.
\begin{definition}
Let $E$ be an element in a (partitioned) forest. A process $p$ is an
\emph{owner} of $E$ if there exists a leaf $L$ in the forest such that
\begin{enumerate}
 \item $L$ is in the partition of $p$, and
 \item $L$ is an ancestor or a descendant of $E$.
\end{enumerate}
\end{definition}
Unique ownership is thus guaranteed for leaf elements and their descendants,
but not for every ancestor element of a tree.
In any case, each element has at least one owner.

In his section, we describe how to find all owner processes of a given element
and how to find those processes that own leaf elements sharing a given face
with an element.
We focus on particularly efficient, recursive designs of this functionality.

\subsection{Owners of a forest node}

We begin with the algorithm \texttt{t8\_forest\_owner} that determines all
owner processes of a given forest element.

\begin{definition}
The \emph{first/last descendant} of an element $E$ is the descendant
of $E$ of maximum refinement level with smallest/largest SFC index.
\end{definition}

Since first/last descendants cannot be refined further, they are either a leaf or descendants
of a leaf. Hence, they have a unique owner process.
See also Figure~\ref{fig:owner-ex1} for an illustration.
We denote these owners by $p_\rfirst(E)$ and $p_\rlast(E)$.
Since a forest is always partitioned along the SFC in ascending order, it must hold for each
owner process $p$ of $E$ that
\begin{equation}
  \label{eq:ghost-owner}
p_\rfirst(E) \leq p \leq p_\rlast(E).
\end{equation}
Conversely, if a process $p$ fulfills inequality~\ref{eq:ghost-owner} and its
partition is not empty, then it must be an owner of $E$.
Furthermore, we conclude that an element has a unique owner if and only if
$p_\rfirst(E) = p_\rlast(E)$.

\begin{figure}
  \center
  \def\svgwidth{0.5\textwidth}
\begingroup%
  \makeatletter%
  \providecommand\color[2][]{%
    \errmessage{(Inkscape) Color is used for the text in Inkscape, but the package 'color.sty' is not loaded}%
    \renewcommand\color[2][]{}%
  }%
  \providecommand\transparent[1]{%
    \errmessage{(Inkscape) Transparency is used (non-zero) for the text in Inkscape, but the package 'transparent.sty' is not loaded}%
    \renewcommand\transparent[1]{}%
  }%
  \providecommand\rotatebox[2]{#2}%
  \ifx\svgwidth\undefined%
    \setlength{\unitlength}{93.19533691bp}%
    \ifx\svgscale\undefined%
      \relax%
    \else%
      \setlength{\unitlength}{\unitlength * \real{\svgscale}}%
    \fi%
  \else%
    \setlength{\unitlength}{\svgwidth}%
  \fi%
  \global\let\svgwidth\undefined%
  \global\let\svgscale\undefined%
  \makeatother%
  \begin{picture}(1,0.94514086)%
    \put(0,0){\includegraphics[width=\unitlength,page=1]{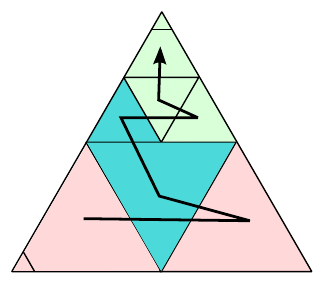}}%
    \put(0.1029275,0.02584887){\color[rgb]{0,0,0}\rotatebox{-0.16875751}{\makebox(0,0)[lb]{\smash{first descendant}}}}%
    \put(0.04890653,0.86824351){\color[rgb]{0,0,0}\makebox(0,0)[lb]{\smash{last descendant}}}%
    \put(0,0){\includegraphics[width=\unitlength,page=2]{owners_ex1_tex.pdf}}%
  \end{picture}%
\endgroup%
   \caption[An element $E$ and its leaf elements.]
  {An element $E$ and its leaf elements. We depict its first descendant
  (bottom left) and last descendant (top). Their owners are unique and we
  denote them by $p_\rfirst(E)$ (pink) and $p_\rlast(E)$
  (light green). For all other owners $p$---in this case, the process owning
  the blue leaves---we have $p_\rfirst(E) \leq p \leq p_\rlast(E)$.
  We note that, depending on the chosen space filling curve, the first and last
  descendants of an element need not be placed at its corners.}%
  \figlabel{fig:owner-ex1}%
\end{figure}%

Each process can compute the SFC index of the first descendant of its first
local element. From these SFC indices we build an array of size $P$, which
is the same on each process.
We can then
determine the owner process of a first or last descendant by performing a
binary search in this array if we combine it with the array of tree offsets.
This is the same approach as in \cite{BursteddeWilcoxGhattas11}.

Hence, we can compute all owner processes of an element by constructing its first and last descendant
and computing their owners.
When we know that an element has a unique owner---for example when it is a leaf
element---it suffices to construct its first descendant and compute its owner.

Since the search time depends on $P$, we will in practice work with a subwindow
onto the array that is searched, which is narrowed down by top-down tree
traversals.
Thus, we aim to use all prior knowledge to shorten the length of any search.

\subsection{Owners at a face}
\label{sec:ownersatface}
For the \texttt{Ghost\_v2} algorithm that works on an unbalanced forest---as
described in \cite{IsaacBursteddeWilcoxEtAl15} for cubical elements---we will
have to identify all owners of leaves at a face of a face-neighbor element of a
given element.
Since \pforest's algorithm \texttt{find\_range\_boundaries}
\cite{IsaacBursteddeWilcoxEtAl15}
is highly efficient, but specific to the Morton SFC and hypercubes,
we introduce the
algorithm \texttt{t8\_owners\_at\_face}.
Given an element $E$ and a face $f$,
\texttt{t8\_owners\_at\_face} determines the set $P_E$ of all processes that
have leaf elements that are descendants of $E$ and share a face with $f$.  It
is a recursive algorithm that we now describe in detail.

\begin{definition}
The first/last \emph{face descendant} of an element $E$ at a face $f$ is the
descendant of $E$ of maximum refinement level that shares a subface with $f$
and has smallest/largest SFC index.
\end{definition}
We denote the owner processes of an element's first and last
face descendants by $p_\rfirst(E,f)$ and $p_\rlast(E,f)$.
If these are equal to the same process $q$,
we can return $q$ as the single owner at that face.

As opposed to the owners of an element, not all nonempty processes in the range
from $p_\rfirst(E,f)$ to $p_\rlast(E,f)$ are necessarily owners
of leaves at the face of $E$; see for example face $f=0$ in
Figure~\ref{fig:ownersatface}. Here, $p_\rfirst(E,0) = 0$, $p_\rlast(E,0) = 2$,
and the owners at the face are processes 0 and 2 despite process $1$ being nonempty.

It is thus not sufficient to determine all nonempty processes between
$p_\rfirst(E,f)$ and $p_\rlast(E,f)$.
Therefore, if $p_\rfirst(E,f) < p_\rlast(E,f) - 1$, we enter a
recursion for each child of $E$ that lies on the face $f$.
This recursion is guaranteed to return immediately if the input element has
only descendants owned by a single process, which happens at the latest when
the input element is a leaf.  We will terminate the recursion earlier for
elements whose descendants at the face $f$ are all owned by a single
process, or by two processes whose ranks differ by 1.
In practice, this procedure prunes the search tree very quickly.

We outline the algorithm in Algorithm~\ref{alg:ownersatface} and illustrate an
example in Figure~\ref{fig:ownersatface}.

\begin{algorithm}
  \DontPrintSemicolon
  \caption{\texttt{t8\_owners\_at\_face}
           (forest $\forest F$, element $E$, face index $f$)}
  \label{alg:ownersatface}
  \algoresult{The set $P_E$ of all processes that own leaf elements that are descendants
  of $E$ and have a face that is a subface of $f$}\;
  $P_E\gets\emptyset$\;
  $\firstd \gets$ \texttt{t8\_element\_first\_desc\_face ($E$, $f$)}
    \Comment{First and last}
  $\lastd \gets$ \texttt{t8\_element\_last\_desc\_face ($E$, $f$)}
    \Comment{descendant of $E$ at $f$}
  $p_\rfirst \gets$ \texttt{t8\_forest\_owner ($\forest F$, $\firstd$)}
    \Comment{The owners of $\firstd$ and $\lastd$}
  $p_\rlast \gets$ \texttt{t8\_forest\_owner ($\forest F$, $\lastd$)}\;
  \algoeifcom {\IfComment{Only $p_\rfirst$ and $p_\rlast$ are}}
  {$p_\rfirst \in \set{p_\rlast, p_\rlast - 1} $ }
  {
    \Return $\set{p_\rfirst, p_\rlast}$ \Comment{owners of leaves at $f$}
  }(\IfComment{There may be other owners.  Enter the recursion})
  { %
    $C_f[]\gets$ \texttt{t8\_element\_children\_at\_face ($E$, $f$)}\;
    \algofor {$0\leq i < $\texttt{t8\_element\_num\_face\_children ($E$, $f$)}}
    {
      $j \gets$ \texttt{child\_index} $(C_f[i])$\Comment{Child number relative to $E$}
      $f' \gets$ \texttt{t8\_element\_child\_face ($E$, $j$, $f$)}\;
      $P_E \gets P_E \abst\cup$ \texttt{t8\_owners\_at\_face ($\forest F$, $C_f[i]$, $f'$)}
    }
    \Return $P_E$
  }
\end{algorithm}

\begin{figure}
  \center
  \def\svgwidth{0.55\textwidth}
\begingroup%
  \makeatletter%
  \providecommand\color[2][]{%
    \errmessage{(Inkscape) Color is used for the text in Inkscape, but the package 'color.sty' is not loaded}%
    \renewcommand\color[2][]{}%
  }%
  \providecommand\transparent[1]{%
    \errmessage{(Inkscape) Transparency is used (non-zero) for the text in Inkscape, but the package 'transparent.sty' is not loaded}%
    \renewcommand\transparent[1]{}%
  }%
  \providecommand\rotatebox[2]{#2}%
  \ifx\svgwidth\undefined%
    \setlength{\unitlength}{93.19533691bp}%
    \ifx\svgscale\undefined%
      \relax%
    \else%
      \setlength{\unitlength}{\unitlength * \real{\svgscale}}%
    \fi%
  \else%
    \setlength{\unitlength}{\svgwidth}%
  \fi%
  \global\let\svgwidth\undefined%
  \global\let\svgscale\undefined%
  \makeatother%
  \begin{picture}(1,0.94514086)%
    \put(0,0){\includegraphics[width=\unitlength]{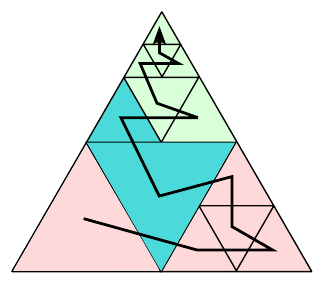}}%
    \put(0.17896196,0.17939129){\color[rgb]{0,0,0}\makebox(0,0)[lb]{\smash{$p=0$}}}%
    \put(0.41928747,0.27725086){\color[rgb]{0,0,0}\makebox(0,0)[lb]{\smash{$p=1$}}}%
    \put(0.53913259,0.52962814){\color[rgb]{0,0,0}\makebox(0,0)[lb]{\smash{$p=2$}}}%
    \put(0.77331875,0.51579107){\color[rgb]{0,0,0}\makebox(0,0)[lb]{\smash{$f = 0$}}}%
    \put(0.07587091,0.51258001){\color[rgb]{0,0,0}\makebox(0,0)[lb]{\smash{$f = 1$}}}%
    \put(0.42474783,0.04064462){\color[rgb]{0,0,0}\makebox(0,0)[lb]{\smash{$f = 2$}}}%
  \end{picture}%
\endgroup%
\hfill
  \def\svgwidth{0.28\textwidth}
\begingroup%
  \makeatletter%
  \providecommand\color[2][]{%
    \errmessage{(Inkscape) Color is used for the text in Inkscape, but the package 'color.sty' is not loaded}%
    \renewcommand\color[2][]{}%
  }%
  \providecommand\transparent[1]{%
    \errmessage{(Inkscape) Transparency is used (non-zero) for the text in Inkscape, but the package 'transparent.sty' is not loaded}%
    \renewcommand\transparent[1]{}%
  }%
  \providecommand\rotatebox[2]{#2}%
  \ifx\svgwidth\undefined%
    \setlength{\unitlength}{294.81169434bp}%
    \ifx\svgscale\undefined%
      \relax%
    \else%
      \setlength{\unitlength}{\unitlength * \real{\svgscale}}%
    \fi%
  \else%
    \setlength{\unitlength}{\svgwidth}%
  \fi%
  \global\let\svgwidth\undefined%
  \global\let\svgscale\undefined%
  \makeatother%
  \begin{picture}(1,1.85104584)%
    \put(0,0){\includegraphics[width=\unitlength,page=1]{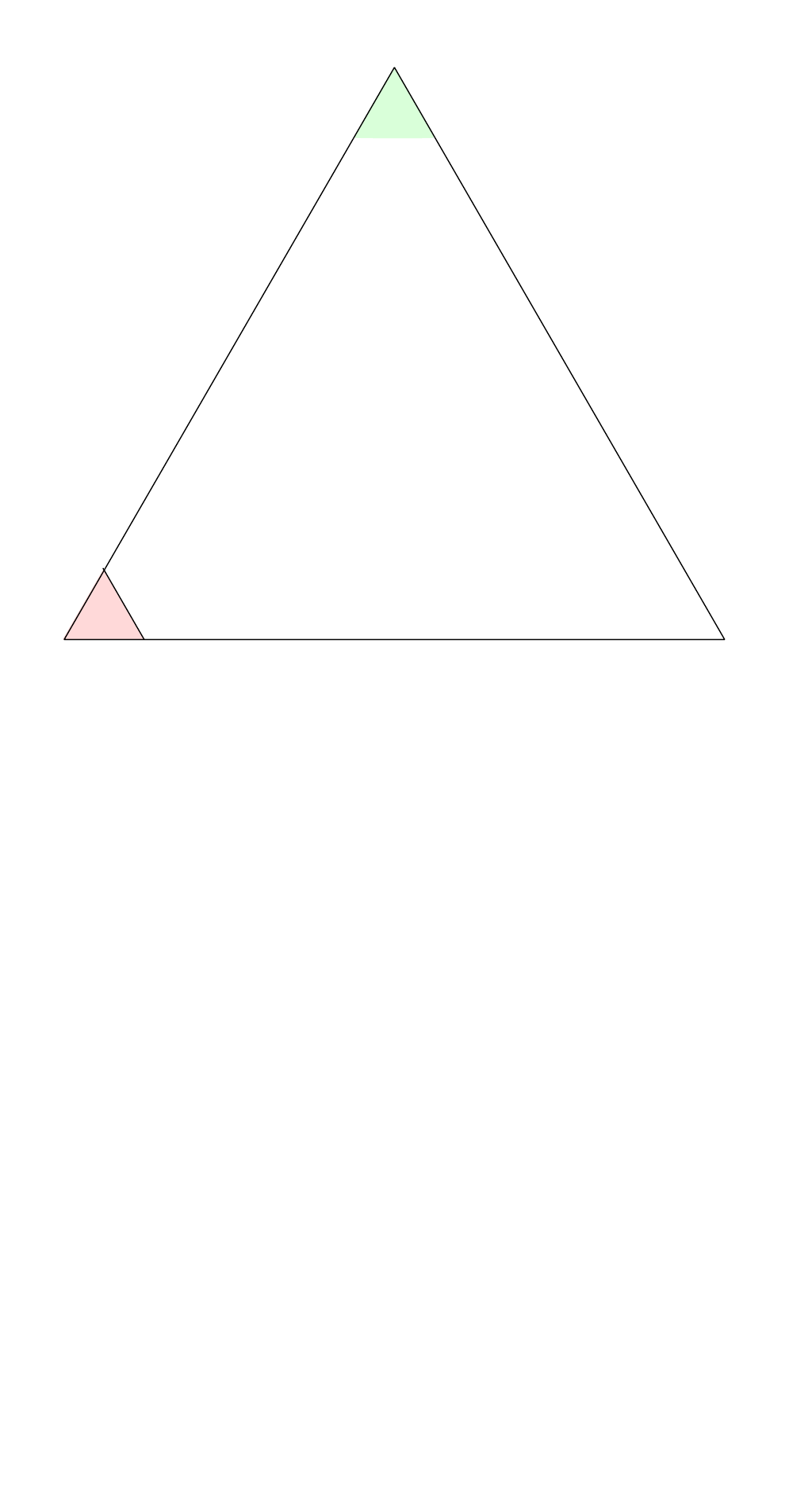}}%
    \put(0.1861849,1.11045593){\color[rgb]{0,0,0}\makebox(0,0)[lb]{\smash{0}}}%
    \put(0,0){\includegraphics[width=\unitlength,page=2]{owners_at_face_ex1_detail_b_tex.pdf}}%
    \put(0.46557029,1.59564253){\color[rgb]{0,0,0}\makebox(0,0)[lb]{\smash{2}}}%
    \put(0.05497321,1.49788517){\color[rgb]{0,0,0}\makebox(0,0)[lb]{\smash{$f = 1$}}}%
    \put(0,0){\includegraphics[width=\unitlength,page=3]{owners_at_face_ex1_detail_b_tex.pdf}}%
    \put(0.15098151,0.96949184){\color[rgb]{0,0,0}\makebox(0,0)[lb]{\smash{\small{first face descendant}}}}%
    \put(0,0){\includegraphics[width=\unitlength,page=4]{owners_at_face_ex1_detail_b_tex.pdf}}%
    \put(0.53683924,1.74521062){\color[rgb]{0,0,0}\makebox(0,0)[lb]{\smash{\small{last face}}}}%
    \put(0,0){\includegraphics[width=\unitlength,page=5]{owners_at_face_ex1_detail_b_tex.pdf}}%
    \put(0.19431916,0.18348732){\color[rgb]{0,0,0}\makebox(0,0)[lb]{\smash{0}}}%
    \put(0.27029986,0.3373873){\color[rgb]{0,0,0}\makebox(0,0)[lb]{\smash{0}}}%
    \put(0.38969799,0.56610424){\color[rgb]{0,0,0}\makebox(0,0)[lb]{\smash{1}}}%
    \put(0.46606641,0.6948064){\color[rgb]{0,0,0}\makebox(0,0)[lb]{\smash{2}}}%
    \put(0.56248599,1.6669476){\color[rgb]{0,0,0}\makebox(0,0)[lb]{\smash{\small{descendant}}}}%
  \end{picture}%
\endgroup%
   \caption[An example for \texttt{t8\_owners\_at\_face}]{An example for Algorithm~\ref{alg:ownersatface}, \texttt{t8\_forest\_owners\_at\_face}.
   Left: A triangle element $E$ with the TM-index as SFC whose descendants
   are owned by three different processes: $0$ (red), $1$ (blue), and
   $2$ (green). The owners at the faces are $\set{0, 2}$ at face $0$,
   $\set{0, 1, 2}$ at face $1$, and $\set{0}$ at face $2$.
   Right: The iterations of \texttt{t8\_forest\_owners\_at\_face} at face $f =
   1$.  At first the first and last descendant of $E$ at $f$ are
   constructed. We compute their owner processes $0$ and $2$, and since their
   difference is greater one, we continue the recursion. In the second
   iteration the algorithm is called once for the lower left child and once for
   the upper child of $E$. We determine their first and last descendants at the
   respective subface of $f$.  For the lower left child, the recursion stops since
   both face descendants are owned by process $0$. For the upper child the owner
   processes are $1$ and $2$ and since there are no other possible owner processes
   in between, we stop the recursion as well.
  }
  \figlabel{fig:ownersatface}
\end{figure}

We optimize our implementation of \texttt{t8\_owners\_at\_face}
by taking into account that
the first and last owners $p_f$ and $p_l$ at the current recursion step form
lower and upper bounds for the first and last owners in any upcoming recursion
step.
Thus, we restrict the binary searches in \texttt{t8\_forest\_owner} to the
interval $[p_\rfirst, p_\rlast]$ instead of $[0, P-1]$.

We also exploit that the first descendant of an element $E$ at a face $f$ is
at the same time the first face descendant of $E$'s first child at $f$.
The same holds for the last descendant and the last child at $f$.
Thus, we reuse the first/last face descendants and owners of $E$ when we enter
the recursion with the first/last child at $f$.

\section{The ghost algorithms}
\seclab{theghostalgorithms}

In this section we review existing ghost algorithms, which we call v1 and v2
and extend to hybrid meshes, and propose an updated version that is at
least as scalable as its predecessors, albeit being currently restricted to
face ghosts.
In practice, we execute it on over 1e12 elements, which is a novel achievement
for a true AMR code.

\subsection{Variants of existing ghost algorithms}
\seclab{exististingghost}

For the classic \ghostb algorithm \cite{BursteddeWilcoxGhattas11} we assume
that the forest is balanced and hence we know that all
face-neighbor leaf elements of $E$ have a refinement level between $\ell(E)
-1$ and $\ell(E) + 1$. Therefore, all neighbor elements of $E$ with level
$\ell(E) + 1$ must have a unique owner process.
Thus, to identify the neighbor processes of $E$ across a face $F$ it suffices
to construct the face-neighbors of $E$ across $F$ of level $\ell(E) + 1$ and
determine their owner processes.
To enable this functionality available in \tetcode, we construct these
face-neighbors via the function \texttt{t8\_forest\_half\_face\_neighbors}; see
Section ~\ref{sec:ghost-halfneighbors}.
We list the \ghostb algorithm in Algorithm~\ref{alg:ghostb}, which collects
the sets $R_p^q$ on each process.
Here and in the following, we omit the subsequent communication phase to
exchange $R_p^q$ with all processes $q$ for which it is non-empty, since this
involves only standard symmetric pairs of senders and receivers.

For \texttt{Ghost\_v2} \cite{IsaacBursteddeWilcoxEtAl15} we drop the assumption
of a balanced forest.
In consequence, there are no longer any restrictions on the sizes of
face-neighbor leaves of $E$.
To compute $E$'s remote processes across $F$, we first construct the
corresponding face-neighbor $E'$ of the same-level as $E$.
For this element we know that it is either a descendant of a
forest leaf (including the case that it is a leaf itself),
which then has a unique owner, or an ancestor of
multiple forest leaves, which could all have different owners.
We need to compute only the owners of those descendant/ancestor forest
leaves of $E'$ that touch the face $F$. We achieve this with
Algorithm~\ref{alg:ownersatface} \texttt{t8\_forest\_owners\_at\_face} that we
describe in Section~\ref{sec:ownersatface}.
We show the \texttt{Ghost\_v2} Algorithm in \ref{alg:ghost}.  It uses
the function \texttt{dual\_face} (not listed), which, given an element $E$ and
a face index $f$, returns the face index $f'$ from the neighboring element.

\begin{algorithm}
\DontPrintSemicolon
\caption{\ghostb (\texttt{forest $\forest F$})\newline\mbox{}\hfill
           (for balanced forests only and simplified to face-only neighbors)%
}%
\label{alg:ghostb}
\algofor{$K\in \forest F$\texttt{.trees}}
 { %
 \algofor{$E \in K$\texttt{.elements}}
  { %
   \algofor {$0\leq f <$ \texttt{t8\_element\_num\_faces($E$)}}
    { %
      $E'[]\gets$ \texttt{t8\_forest\_half\_face\_neighbors ($\forest F$, $E$, $f$)}\;
      \algofor {$0\leq i <$ \texttt{t8\_element\_num\_face\_children($E$, $f$)}}
      {   %
      $q \gets$ \texttt{t8\_forest\_owner ($\forest F$, $E'[i]$)}\;
        \algoif {$q \neq p$}
        {
          $R_p^q = R_p^q \cup \set {E}$\;
        }
      }
    }
  }
 }
\end{algorithm}

\begin{algorithm}
\DontPrintSemicolon
\caption{\texttt{Ghost\_v2} (\texttt{forest $\forest F$})}
\label{alg:ghost}
\algofor{$K\in \forest F$\texttt{.trees}}
 { %
 \algofor{$E \in K$\texttt{.elements}}
  { %
   \algofor {$0\leq f < $ \texttt{t8\_element\_num\_faces ($E$)}}
    { %
      $E'\gets$ \texttt{t8\_forest\_face\_neighbor ($\forest F$, $E$, $f$)}\;
      $f'\gets$ \texttt{dual\_face($E$, $E'$, $f$)}\;
      $P_{E'} \gets$ \texttt{t8\_forest\_owners\_at\_face ($\forest F$, $E'$, $f'$)}\;
      \algofor {$q\in P_{E'}$}
      {   %
        \algoif {$q \neq p$}
        {
          $R_p^q = R_p^q \cup \set {E}$\;
        }
      }
    }
  }
 }
\end{algorithm}

\subsection{Optimizing the runtime of \texttt{Ghost}}
\seclab{forestsearchghostv3}

The \ghostb and \texttt{Ghost\_v2} algorithms that we have presented above in
naive, stripped down versions both iterate
over all local leaf elements to identify the partition boundary leaves on the process.
For each leaf we generate all (half) face-neighbors and compute their
owners. Thus, their runtime is proportional to the number of all local leaves.
However, for most meshes only a portion of the leaf elements actually are
partition boundary elements, depending on the surface-to-volume ratio of the process's
partition. Since the surface of a volume grows quadratically while the volume itself
grows cubically, the number of partition boundary leaves can become arbitrarily small in
comparison to the number of all leaves.
Ideally, the runtime of \texttt{Ghost} should be proportional to the
number of partition boundary elements.

Our goal is therefore to improve the runtime of the algorithms by excluding
non-boundary leaves from the iteration.
In \pforest these locally surrounded leaves have been excluded from the iteration by checking
for each quadrilateral/hexahedron whether its $3\times 3$ neighborhood
\cite{SundarSampathBiros08, BursteddeWilcoxGhattas11},
namely all same-level face-(edge-/vertex-)neighbors, are process local.
This extra test makes the \texttt{Ghost\_v2} algorithm fast and scalable.
However, since this approach uses particular geometrical properties of the
quadrilateral/hexahedral shapes and of the Morton SFC, it is not practical for
our hybrid, element-shape independent approach.

To exclude the locally surrounded leaves in \tetcode, we replace the leaf iteration with a
top-town traversal, repurposing a recursive approach originally proposed for
searches \cite{IsaacBursteddeWilcoxEtAl15}.
Starting with a tree's root element, we
check whether it may have partition boundary leaf descendants, and if so, we create the
children of the element and continue recursively.
If we reach a leaf element,
we check whether it is a partition boundary element---and if so for which processes---in
the way described in the previous section.
This approach allows us to terminate the recursion as soon as we
reach a locally surrounded element, thus
saving the iteration over all descendant leaves of that element.

\subsubsection{The recursive top-down search}

In \cite{IsaacBursteddeWilcoxEtAl15} the authors present the general recursive
\texttt{search} algorithm for octree AMR, which easily extends to arbitrary
tree-based AMR. The setting is that we search a leaf or a set of leaves in a
forest that satisfy given conditions. One numerical example for such a search
arises in semi-Lagrangian advection solvers
\cite{Albrecht16,MirzadehGuittetBursteddeEtAl16}. To interpolate the values of
an advected function $\phi_t$ at time
$t$, each grid point $x_i$ is tracked back in time to its previous position
$\hat x_i$ at $t-\Delta t$. This point $\hat x_i$ lies in a leaf element $E_i$
of the forest and an element-local interpolation with the values of
$\phi_{t-1}$ is used to determine the value $\phi_t(x_i)$.
Thus, in each time step, we have to search the forest for the leaf elements
$\set{E_i}$ given the points $\set{\hat x_i}$.

In our case, we apply \texttt{search} to the problem of identifying all local
leaf elements at a process's partition boundary.
The \texttt{search} algorithm has been shown to be especially efficient when
looking for multiple matching leaves at once \cite{IsaacBursteddeWilcoxEtAl15},
which is the case in our setting.
It has been extended to search remote partitions as well \cite{Burstedde19},
but we are not yet using this additional functionality
in our variant \texttt{t8\_forest\_search}.

As presented in \cite{IsaacBursteddeWilcoxEtAl15} the idea of search is to
perform a recursive top-down traversal for each tree by
starting with the root element of that tree and recursively creating its children
until we end up with a leaf element. On each intermediate element we call a
user-provided callback function which returns true only if the search should
continue with this element. If the callback returns false, the recursion for
this element stops and its children are excluded from the search.
If the search has reached a leaf element, the callback also performs the
desired operations if the leaf matches the search.

For our ghost algorithm the callback returns false for locally surrounded
elements, thus
excluding possibly large areas from
the search and hence speeding up the computation.
Once a leaf element is reached, we check whether it is a partition boundary
element or not.
Thus, we iterate over the leaf's faces and compute the owners at the respective
neighbor faces as in the inner for-loop of Algorithm~\ref{alg:ghost}.

We show our version of \texttt{search} in Algorithm~\ref{alg:search}.
It is a simplified version of Algorithm~3.1 in \cite{IsaacBursteddeWilcoxEtAl15}
without queries, since we do not need these for \texttt{Ghost}.
We also use the function \texttt{split\_array} from
\cite{IsaacBursteddeWilcoxEtAl15}. This function takes as input an element $E$
and an array $L$ of (process local) leaf elements in $E$, sorted in SFC order.
\texttt{split\_array} returns a set of arrays $\set{M[i]}$, such that for the
$i$-th child $E_i$ of $E$ the array $M[i]$ contains exactly the leaves in $L$ that
are also leaves of $E_i$. Thus, $L = \dot\bigcup_i M[i]$.

To search in the complete forest, we iterate over all trees, and for each
we compute the finest element $E$ such that all local tree leaves are still
descendants of $E$
(we skip any tree that has no local leaves).
We compute $E$ as the nearest common ancestor of the first
and last leaf element of the tree.
With $E$ and the leaf elements of the tree, we call the
\texttt{element\_recursion}; see Algorithm~\ref{alg:forest_search}.

\begin{algorithm}
  \DontPrintSemicolon
  \caption{\texttt{element\_recursion} (element $E$, leaves $L$, callback
           \texttt{Match})%
}
  \label{alg:search}
  \algorithmicrequire{ The leaves in $L$ must be descendants of $E$ and ascending}\;
  isLeaf $\gets L = \set{E}$ \Comment{Determine whether $E$ is a leaf element}
  \algoif{\texttt{Match($E$, isLeaf)}\algoand \textbf{not} \texttt{isLeaf}}
  {
    $M[] \gets$ \texttt{split\_array} ($L$, $E$)%
    \Comment{$M[i]$ are no-copy views onto $L$}
    $C[] \gets$ \texttt{t8\_element\_children} ($E$)\;
    \algofor {$0\leq i <$ \texttt{t8\_element\_num\_children} ($E$)}
    {
      \algoif {$M[i] \ne \emptyset$} {%
      \texttt{element\_recursion} ($C[i]$, $M[i]$, \texttt{Match})\;
      }%
    }
  }
\end{algorithm}

\begin{algorithm}
  \DontPrintSemicolon
  \caption{\texttt{t8\_forest\_search} (forest $\forest F$, callback \texttt{Match})}
  \label{alg:forest_search}
  \algofor {$K \in \forest F.\mathrm{trees}$, $K$ contains local leaves}
  {
    $E_1 \gets$ \texttt{first\_tree\_element} ($\forest F$, $K$)
    \Comment{First and last local}
    $E_2 \gets$ \texttt{last\_tree\_element} ($\forest F$, $K$)
    \Comment{leaf in the tree}
    $E \gets$ \texttt{t8\_element\_nearest\_common\_ancestor} ($E_1$, $E_2$)\;
    $L \gets$ \texttt{tree\_leaves} ($\forest F$, $K$)
    \Comment{No-copy view of tree leaves}
    \texttt{element\_recursion} ($E$, $L$, \texttt{Match})
    \Comment{Top-down traversal}
  }
\end{algorithm}

\subsubsection{The optimized \texttt{Ghost} algorithm}

The algorithm \texttt{t8\_forest\_search} requires a callback function, in our
case \ghostmatch (Algorithm~\ref{alg:ghost_match}), which works as follows.
If the element $E$ passed to \ghostmatch is not a leaf element, we check
whether the element and all of its possible face-neighbors are owned by the
current process.
For the element's owners, we do not call the function
\texttt{t8\_forest\_owner}, but instead save runtime by computing the first and
last process that own leaves of the element and checking whether they are
equal.
For these computations we construct $E$'s first and last descendant.
Analogously, for the owners at the neighbor faces we compute the first and last
owner processes.  If for $E$ the first and last process is $p$ and at each
face-neighbor the first and last owner at the corresponding face is also $p$,
$E$ is a locally surrounded element and cannot have any partition boundary leaves as descendants. Thus,
we return 0 and the search does not continue for the descendants of $E$.

If $E$ is a leaf element, then it may or may not be a partition boundary element.
We thus compute all owner processes for all face-neighbors using
\texttt{t8\_forest\_owners\_\-at\_\-face} and add $E$ as a partition boundary
element to all of these that are not $p$.

\begin{algorithm}
  \caption{\ghostmatch (element $E$, bool \texttt{isLeaf})}
  \label{alg:ghost_match}
  \DontPrintSemicolon
  \algoresult{If $E$ is a leaf, compute the owners of the face-neighbors
and add to the sets $R_p^q$.
If not, then terminate if $E$ is a locally surrounded element
}\;

  \algoeifcom {\IfComment{$E$ is a leaf. Compute the owners at}}{\texttt{isLeaf}}
  {
  \algoforcom{\IfComment{its faces%
}}{$0\leq f <$ \texttt{t8\_element\_num\_faces ($E$)}}
    { %
      $E' \gets$ \texttt{t8\_forest\_face\_neighbor ($\forest F$, $E$, $f$)}\;
      $f' \gets$ \texttt{dual\_face($E$, $E'$, $f$)}\;
      $P_{E'} \gets $\texttt{t8\_forest\_owners\_at\_face ($\forest F$, $E'$, $f'$)}\;
      \algofor {$q\in P_{E'}$}
      {   %
        \algoif {$q \neq p$}
        {
          $R_p^q = R_p^q \cup \set {E}$\;
        }
      }
    }
  }(\IfComment{$E$ is not a leaf})
  {
   $p_\rfirst(E) \gets$ \texttt{t8\_element\_first\_owner} ($E$)\;
   $p_\rlast(E) \gets$ \texttt{t8\_element\_last\_owner} ($E$)\label{algline:ghostmatchrec}\;
    \algofor {$0\leq f <$ \texttt{t8\_element\_num\_faces ($E$)}
    \label{algline:ghostmatchfor}}
    { %
      $E' \gets$ \texttt{t8\_forest\_face\_neighbor ($\forest F$, $E$, $f$)}\;
      $f' \gets$ \texttt{dual\_face($E$, $E'$, $f$)}\;
      $p_\rfirst(E', f') \gets$ \texttt{t8\_first\_owner\_at\_face ($\forest F$, $E'$, $f'$)}\;
      $p_\rlast(E', f') \gets$ \texttt{t8\_last\_owner\_at\_face ($\forest F$, $E'$, $f'$)}\;
      \algoif{$ p_\rfirst(E',f') \neq p$\algor $p_\rlast(E',f') \neq p$}{
        \Return 1
\Comment{Not all face-neighbor leaves owned by $p$}
      }
    }
    \algoif{$p_\rfirst(E) = p_\rlast(E) = p$}
    {
      \Return 0
      \Comment{Terminate recursion}
    }
  }
  \Return 1
  \Comment{Continue recursion}
\end{algorithm}

\begin{algorithm}
\caption{\texttt{Ghost} (forest $\forest F$)}
\label{alg:ghost_v3}
\DontPrintSemicolon
\algoresult{The ghost layer of $\forest F$ is constructed}\;
\texttt{t8\_forest\_search} ($\forest F$, \ghostmatch)
\end{algorithm}

\subsubsection{Further implementation details}

For each child $C$ of an element $E$ the ranks $p_\rfirst(E),
p_\rlast(E), p_\rfirst(E,f)$, and $p_\rlast(E,f)$ serve as
lower and upper bounds for the corresponding ranks for $C$.
Thus, in our implementation of \ghostmatch in \tetcode, we store these ranks
for each recursion level reducing the search range for the binary owner search
for $C$ from $[0, P-1]$ to $[p_\rfirst(E), p_\rlast(E)]$, and
to $[p_\rfirst(E,f), p_\rlast(E,f)]$ for the faces.
To compute these bounds it is necessary to always enter the \texttt{for}-loop
in Line~\ref{algline:ghostmatchfor}, even though we do not spell this out in
Algorithm~\ref{alg:ghost_match}.

\section{Numerical results}
\seclab{numericalstudies}

In this section we discuss our numerical results in different
runtime studies on the \juqueen~\cite{Juqueen} and the \juwels~\cite{Juwels}
supercomputers at the FZ J\"ulich.
\juqueen is an IBM BlueGene/Q system consisting of 28,675 compute nodes, each
with 16 IBM PowerPC-A2 cores at 1,6 GHz.
Each compute node has 16GB RAM.
\juqueen was in operation until May 2018.
\juqueen's successor \juwels is a Bull Sequana X1000 system consisting of 2,271
compute nodes, each node with 96 GB RAM and two 24-core Intel Xeon SC 8168 CPUs
running at 2,7 GHz.
The latter system has faster processors by a large factor, which
shows clearly in our experiments.
We use one MPI rank per core throughout.

\subsection{Comparing the different ghost versions}

To verify that the additional complexity of implementing the top-down search
is worth the effort, we perform runtime tests of the different ghost algorithms.

We use two meshes on a unit cube geometry.
The first consists of a single hexahedron tree and the second of six
tethrahedron trees with a common diagonal as shown in
Figure~\ref{fig:sechstetras}.
For each mesh we run two types of tests, one with a uniform mesh and one with
an adaptive mesh, where we refine every
third element (in SFC order) recursively in $k$ rounds from some base level
$\ell$ to level $\ell + k$; see Figure~\ref{fig:ghost-comparetest}.

We use 64 compute nodes of \juqueen and display our results in
Table~\ref{tab:ghost-comparetest}.
As expected, the iterative versions
scale linearly with the number of elements.
In contrast, our proposed algorithm scales
with the number of ghost elements, which grows less quickly compared
to the number of elements.
We conclude that we indeed skip most of the elements that do not lie
on the partition boundary of a process partition.
The improved version shows overall a significantly better performance and is up
to a factor of 23.7 faster (adaptive tetrahedra, level $8$) than the iterative
version. For smaller or degraded meshes where the number of ghosts is on the
same order as the number of leaf elements, the improved version shows no
disadvantage compared to the iterative version. This underlines that we do not
lose runtime to the \texttt{Search} overhead, even if each element is a
partition boundary element. For small meshes all algorithms show negligible runtimes on
the order of milliseconds.

We conclude that our \ghostn algorithm based on the top-down search is the ideal
choice among the three versions tested and will use this version
for all experiments presented in the following.

\begin{figure}
\center
\includegraphics[width=0.49\textwidth]{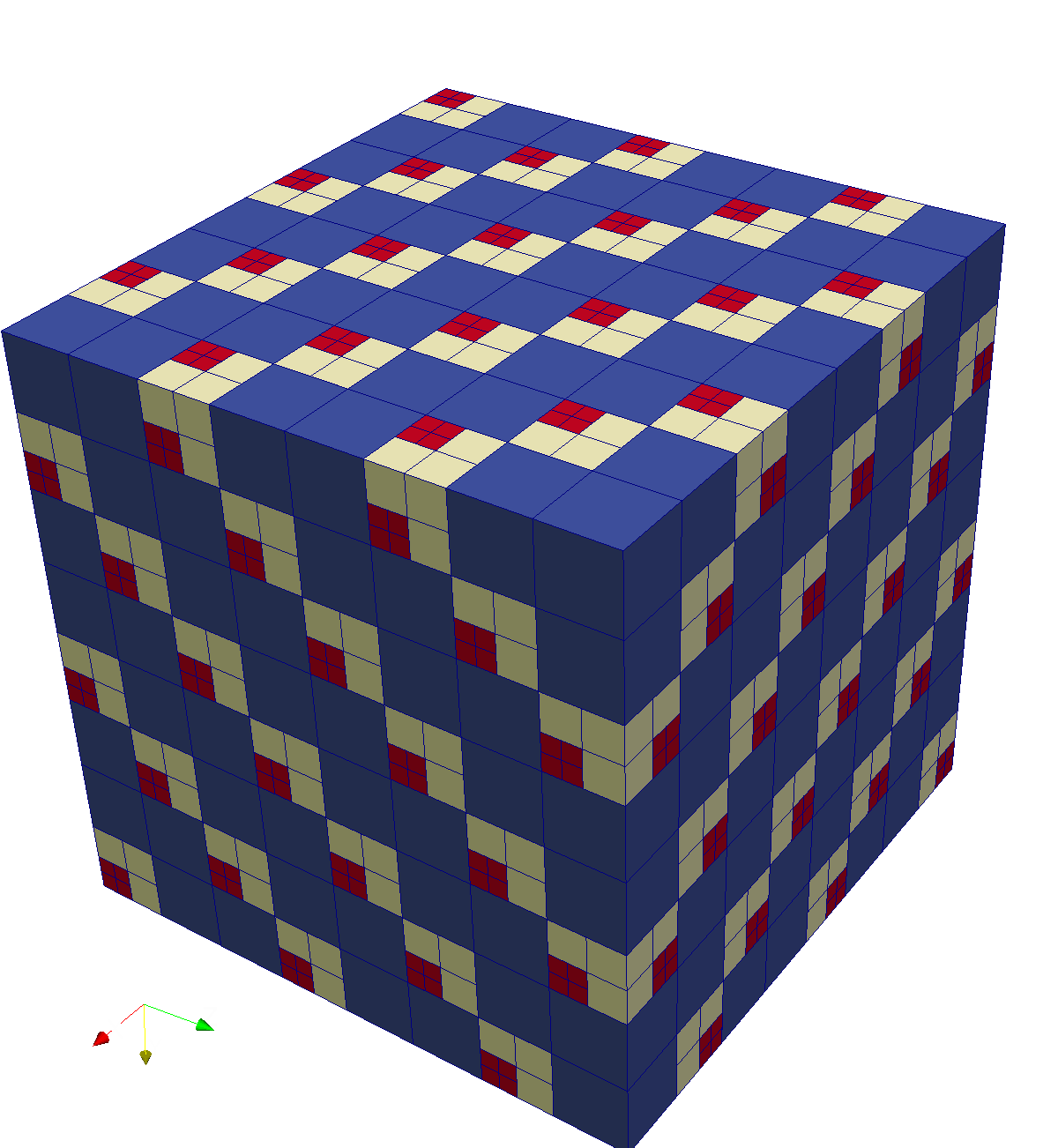}
\includegraphics[width=0.49\textwidth]{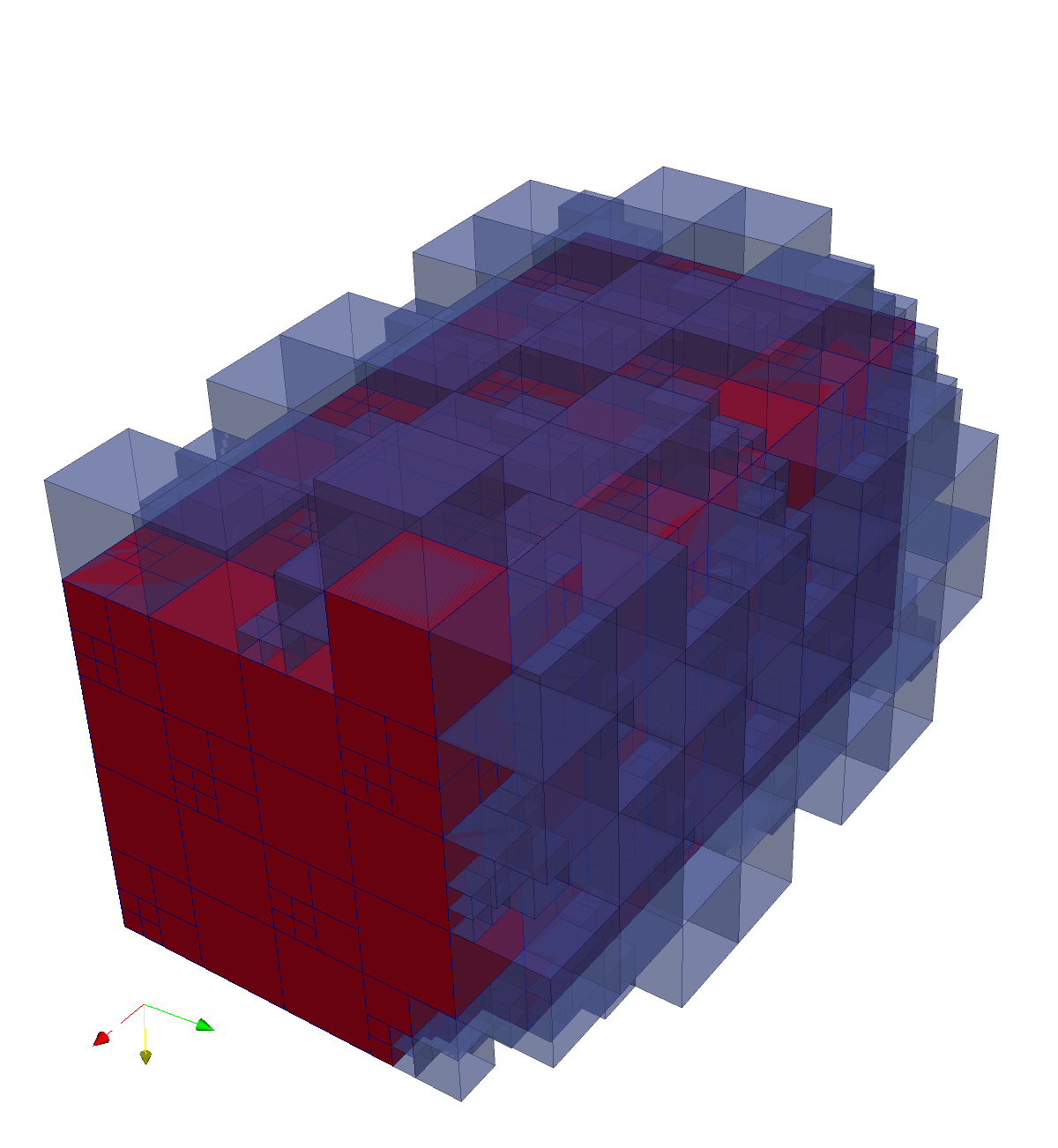}
\caption[Comparing the different implementations of \texttt{Ghost}.]
{We compare the different implementations of \texttt{Ghost} by
testing them on a unit cube geometry with 1024 MPI ranks of \juqueen.
Left: an adaptive mesh with minimum level $\ell = 3$ for one hexahedron tree.
We refine every third element in SFC order and repeat the process a second time
with the refined elements to reach level $5$.
Right: for an example computation on 4 MPI ranks, we show the local leaf
elements of the process with MPI rank 1 (red) and its ghost elements (blue,
transparent).}
\figlabel{fig:ghost-comparetest}
\end{figure}

\begin{table}
\center
 \begin{tabular}{|c||r|r|r||r|r|r|}\hline
  \multicolumn{7}{|c|}{tetrahedra}\\ \hline
  & \multicolumn{3}{c||}{uniform}
  & \multicolumn{3}{c|}{adaptive}\\ \hhline{|=||===||===|}
  $\ell$ & 9 & 8 & 4 & 8--10 & 7--9 & 3--5 \\ \hline
  elements/proc & 786,432 & 98,304 & 24 &1,015,808 & 126,976 & 31 \\ \hline
  ghosts/proc   & 32,704  &  8,160 & 30 &   31,604 &   8,137 & 56 \\ \hhline{|=||===||===|}
  Ghost\_v1 [s]    & 172.3  &  21.64 & 7.99\e-3 & -&-&-\\ \hline
  Ghost\_v2 [s]    & 129.6  &  16.19 & 5.93\e-3 & 167.94 & 20.88 & 8.10\e-3 \\ \hline
  Ghost [s]        &  7.41  &   1.75 & 5.01\e-3 &   7.08 &  1.69 & 8.12\e-3 \\ \hline
 \end{tabular}\\[1ex]

 \begin{tabular}{|c||r|r|r||r|r|r|}\hline
  \multicolumn{7}{|c|}{hexahedra}\\ \hline
  & \multicolumn{3}{c||}{uniform}
  & \multicolumn{3}{c|}{adaptive}\\ \hhline{|=||===||===|}
  $\ell$ & 9 & 8 & 4 & 8--10 & 7--9 & 4--6 \\ \hline
  elements/proc & 131,072 & 16,384 & 4 & 169,301 & 21,162 & 41\\ \hline
  ghosts/proc   & 8,192   &  2,048 & 8 &   7,681 &  1,913 & 30\\ \hhline{|=||===||===|}
  Ghost\_v1 [s]    & 29.51 & 3.742 & 2.87\e-3 & -&-&-\\ \hline
  Ghost\_v2 [s]    & 18.25 & 2.302 & 2.32\e-3 & 23.79 & 2.964 & 8.01\e-3 \\ \hline
  Ghost [s]        &  3.14 & 0.711 & 2.90\e-3 &  2.81 & 0.649 & 8.12\e-3 \\ \hline
 \end{tabular}
\caption[Runtime tests for the three different ghost algorithms.]
{Runtimes for the three different ghost algorithms on \juqueen.
We run the tests on 1,024 MPI ranks.
For tetrahedra and hexahedra we test a uniform level $\ell$
mesh and a mesh that adapts every third element of a uniform level $\ell$ mesh
up to level $\ell + 2$; \confer Figure~\ref{fig:ghost-comparetest}.
Since the adaptive forests are not balanced, they are not supported by \ghostb.
We observe that our new \texttt{Ghost} is superior to the other versions by
a factor of up to 23 and scales with the number of ghost elements, not
the number of leaf elements.
We note that for the hexahedral case, \texttt{Ghost\_v2} would be sped up
similarly by a ($3 \times 3)$-test, as is demonstrated by the near-ideal
version implemented in \pforest \cite{IsaacBursteddeWilcoxEtAl15}.
However, this optimization is not available for non-cubic shapes.}%
\figlabel{tab:ghost-comparetest}
\end{table}

\subsection{A single-shape test case}
\label{sec:testcase}

In this test we use a similar setting to the tests in~\cite{BursteddeHolke17}
for coarse mesh partitioning.
We start with a uniform forest of level $\ell$ and refine it in a band along
an interface defined by a plane to level $\ell + k$.
We then establish a 2:1 balance among the elements (using a ripple propagation
algorithm not discussed here) and repartition the mesh
using the \texttt{Partition} algorithm.
Afterwards, we create a layer of ghost elements with \texttt{Ghost}.
The interface moves through the domain in time in direction of the plane's
normal vector.
In each time step we adapt the mesh, such that we coarsen elements outside
of the band to level $\ell$ and refine within the band to level $\ell + k$.
Then we repeat balance, partition, and \texttt{Ghost}.
As opposed to the test in~\cite{BursteddeHolke17}, we take the unit cube as
our coarse mesh geometry. We run the test once with a hexahedral mesh
consisting of one tree and once with a tetrahedral mesh of six trees forming a
unit cube, similar to the previous section (see also
Figure~\ref{fig:baghoex-1}).

We choose the normal vector
$\frac{3}{2}\begin{pmatrix}1, &1,& \frac{1}{2}\end{pmatrix}^t$ and $\frac{1}{4}$
for the width of our refinement band. We move the refinement band with speed
$v = \frac{1}{64}$ and scale the time step $\Delta t$ with the refinement level as
\begin{equation}
\label{eq:balance-deltat}
 \Delta t (\ell) = \frac{0.8}{2^\ell v} .
\end{equation}
The constant $0.8$ can be seen as width of the band of level $\ell$ elements that will be refined
to level $k$ in the next time step.
We start the band at position $x_0 (\ell) = 0.56 - 2.5\Delta t(\ell)$
and simulate up to 5 time steps.
The strong and weak scaling results collected in the following are obtained
with the \texttt{t8\_time\_forest\_partition} example of \tetcode
version~0.3\cite{tetcodeweb19}.

\begin{figure}
\center
\includegraphics[width=0.45\textwidth]{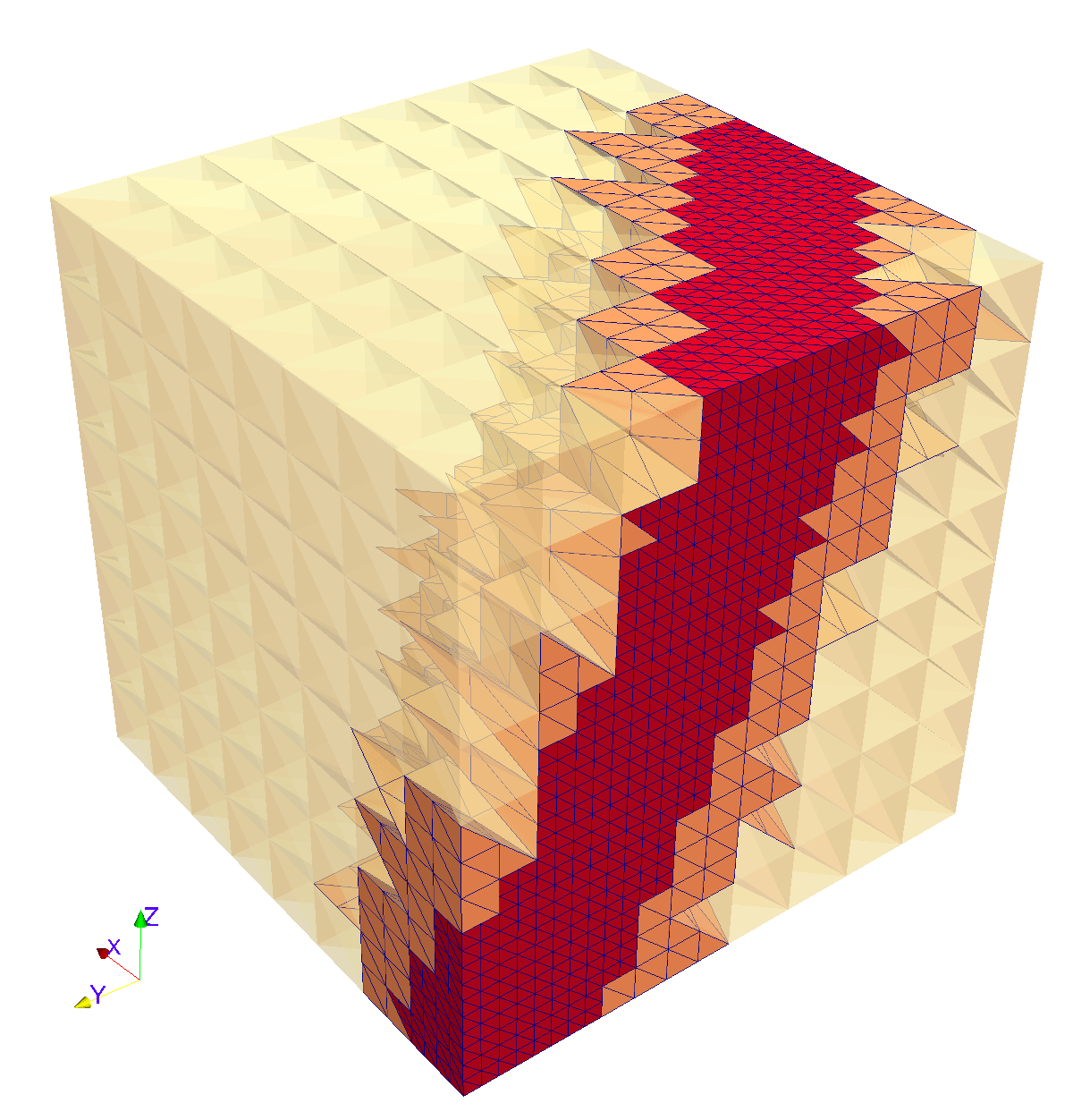}\hfill
\includegraphics[width=0.45\textwidth]{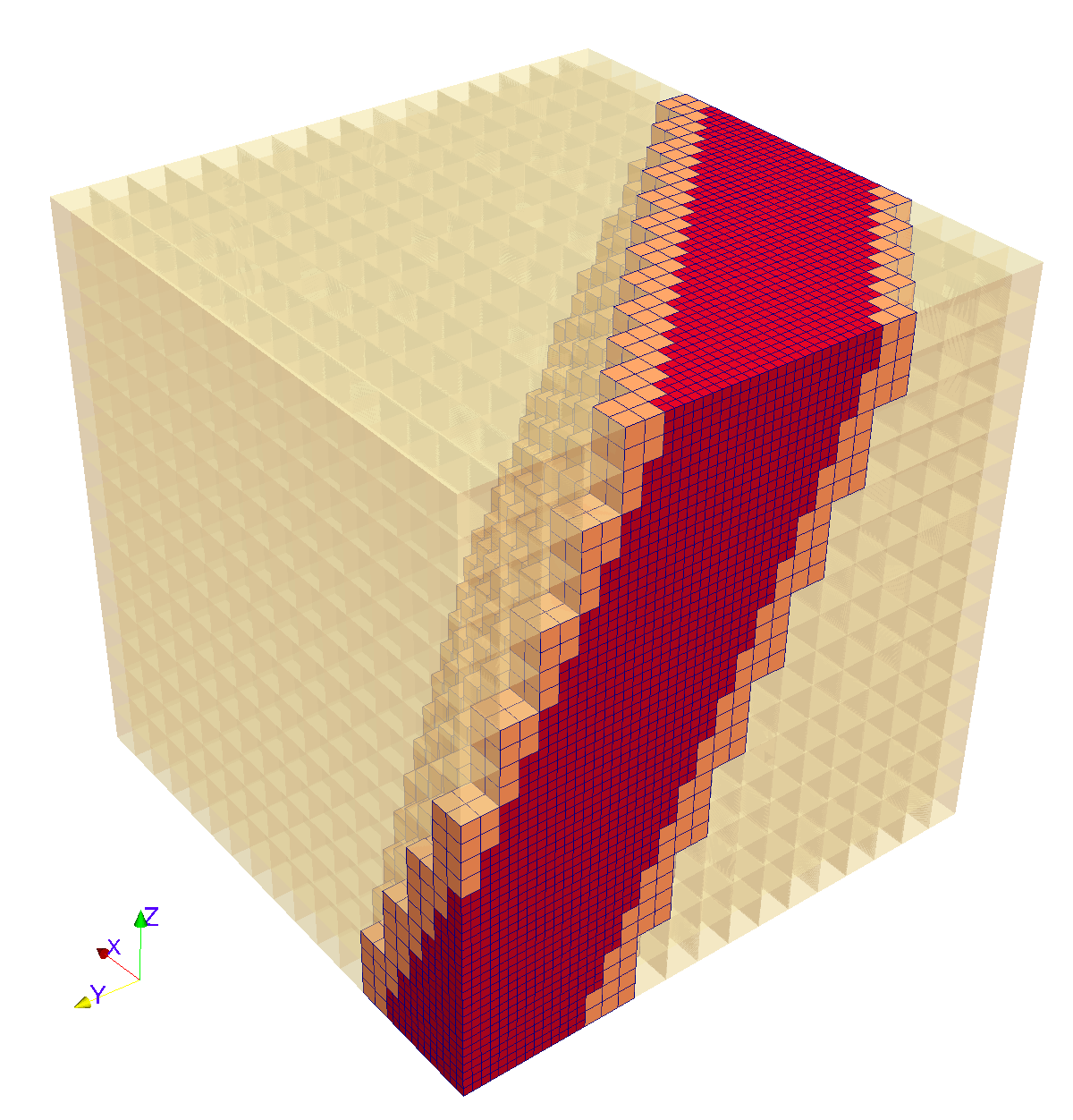}
\caption
  {On \juqueen,
  we test \texttt{Ghost} on a unit cube geometry
  consisting of six tetrahedral trees (left) or one hexahedral tree (right).
  Starting with a uniform level $\ell$, we refine the forest in a band around a
  plane to level $\ell + k$. We then 2:1 balance the forest and create the ghost
  layer. In the next time step, the band moves in the direction of the plane's
  normal vector and we repeat the steps, coarsening previously fine forest
  elements if they now reside outside of the band. We show the forest after
  \texttt{Balance} at time step
  $t = 2\Delta t(\ell)$ for two different configurations.
  Left: Tetrahedral elements with $\ell = 3$, $k = 2$. In total
  we have 56,566 tetrahedral elements.  Right: Hexahedral elements with $\ell =
  4$, $k = 2$, summing up to 78,100 hexahedral elements in total. The color
  represents the
  refinement level. We draw level $\ell$ elements opaque.}
\figlabel{fig:baghoex-1}
\end{figure}

\subsubsection{Strong scaling}

We run a strong scaling test with tetrahedral elements and
refinement parameters $\ell = 8$, $k = 2$ on 8,192 to 131,072 MPI ranks,
increasing the process count by a factor of 2 in each step. We list the
runtimes at time $t = 4\Delta t$ for \texttt{Ghost} in
Table~\ref{tab:baghoex-1-strong} and plot them together with those of
\texttt{Partition} in Figure~\ref{fig:baghoex-1-strong}.

As wee have already seen in Table~\ref{tab:ghost-comparetest}, the runtime of
\texttt{Ghost} depends linearly on the number of ghost elements per process.
Consider two runs with $P_1$ and $P_2$ processes, respectively, and let $G_1$
and $G_2$ denote the numbers of ghost elements
per process, then the parallel efficiency of the second run in relation to the
first run is
\begin{equation}
\label{eq:effgho}
 e_\mathrm{Ghost} = \frac{T_1 G_2}{T_2 G_1}.
\end{equation}

Our results demonstrate that we achieve ideal strong scaling efficiency for
\texttt{Ghost}.

\begin{table}
\center
\begin{tabular}{|r|r|r|r|r|}\hline
\multicolumn{5}{|c|}{Tetrahedral case with $\ell = 8$, $k = 2$, $C=0.8$ at $t=4\Delta t$}\\ \hline
 $P$ & $E/P$ & $G/P$ & Time [s] & Par.\ Eff.\ \\ \hline
 8,192  & 234,178 & 17,946  & 3.25  & 100.0\% \\
16,384  & 117,089 & 11,311  & 2.12  &  96.6\% \\
32,768  &  58,545 &  7,184  & 1.27  & 102.4\% \\
65,536  &  29,272 &  4,560  & 0.79  & 104.5\% \\
131,072 &  14,636 &  2,859  & 0.52  &  99.5\% \\  \hline
\end{tabular}
\caption[Strong scaling with tetrahedral elements.] {
The results for strong scaling of \texttt{Ghost}
with tetrahedral elements, $\ell = 8$, and $k=2$.
We show the runtimes at time $t=4\Delta t$.
The mesh consists of approximately 1.91\e9 tetrahedra.
In addition to the runtimes, we show the number of elements per process $E/P$,
and ghosts per process $G/P$.
The last column contains the parallel efficiency according to \eqref{eq:effgho}
in reference to the smallest run with 8,192 processes.
}
\figlabel{tab:baghoex-1-strong}
\end{table}

\begin{figure}
\center
\includegraphics[width=0.49\textwidth]{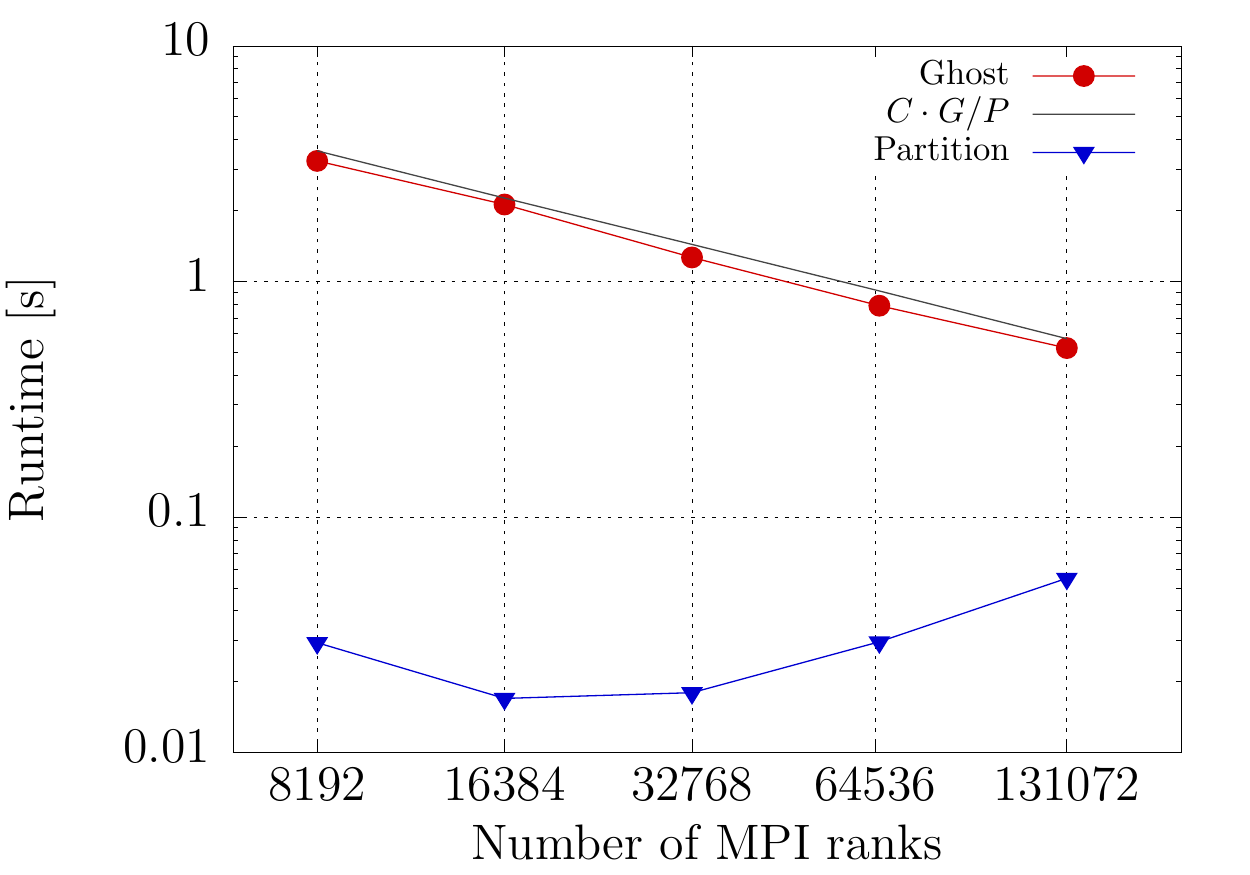}
\caption[Strong scaling with tetrahedral elements.]
{Strong scaling with tetrahedral elements.
We plot the
runtimes of \texttt{Ghost} and \texttt{Partition} for
  the test case from Section~\ref{sec:testcase} with $\ell = 8$, $k = 2$ at
  time step $t = 4\Delta t$. The forest mesh consists of approximately 1.91\e9
  tetrahedra.
Ideally, \texttt{Ghost} scales with the number of ghost elements per process, $G/P$.
This number is indicated by the black line.
As we observe in the plot and in Table~\ref{tab:baghoex-1-strong}, we achieve
perfect scaling for \texttt{Ghost}.
 The runtime of \texttt{Partition} is below 0.1 seconds even for the largest
run.}%
\figlabel{fig:baghoex-1-strong}%
\end{figure}

\subsubsection{Weak scaling}
For weak scaling we increase the global number of elements while also increasing
the process count, keeping the local number of elements nearly constant.
Since with each refinement level $\ell$ the number of global elements grows by a factor
of 8, we multiply the process count with 8 as well.
We test the following configurations, again with $k=2$:
\begin{itemize}
  \item Tetrahedral elements with 8,192 processes, 65,536 processes, and 458,752 processes,
  with refinement levels $\ell = 8$, $\ell = 9$, $\ell = 10$. This amounts to
  about 235k elements per process.
  Thus, the largest run has about $108\e9$ elements.
  \item Tetrahedral elements with 2,048 processes, 16,384 processes, and
131,072 processes, with refinement levels $\ell = 8$, $\ell = 9$, $\ell = 10$.
Here we have about 155k elements per process, summing up to
 $20.3\e9$ elements on 131,072 processes.
  \item Hexahedral elements with the same process counts and levels
  $\ell = 9$, $\ell = 10$ and $\ell = 11$
  ($162\e9$ elements in total).
\end{itemize}

Note that 458,752 is actually 7 times 65,536. We choose it since it is the
maximum possible process count on \juqueen with 16 processes per node, using all
28,672 compute nodes. The number of elements per process is thus about 14\%
greater than on the other process counts in the configuration. However,
\eqref{eq:effgho} still applies for computing the parallel efficiency.

The largest test case that we run is for hexahedra on 458,752 processes with $\ell = 11$
 and 162\e9 elements.

\begin{table}
\center
\begin{tabular}{|r|r|r|r|r|r|}\hline
\multicolumn{6}{|c|}{Tetrahedral case with $k = 2$, $C=0.8$ at $t=4\Delta t$}\\ \hline
  $P$ & $\ell$ & $E/P$ & $G/P$ & Time [s] & Par.\ Eff.\ \\ \hline
 8,192  & 8 & 234,178 & 17,946 & 3.25  & 100.0\% \\
65,536  & 9 & 233,512 & 18,282 & 3.76  &  88.2\% \\
458,752 &10 & 266,494 & 20,252 & 3.79  &  96.8\% \\  \hhline{|======|}
  2,048 & 7 & 117,630 & 10,999 & 1.99 & 100.0\% \\
 16,384 & 8 & 117,089 & 11,311 & 2.12 &  96.5\% \\
131,072 & 9 & 116,756 & 11,478 & 2.18 &  95.2\% \\ \hline
\end{tabular}\\[2ex]
\begin{tabular}{|r|r|r|r|r|r|}\hline
\multicolumn{6}{|c|}{Hexahedral case with $k = 2$, $C=0.8$ at $t=2\Delta t$}\\ \hline
  $P$ & $\ell$ & $E/P$ & $G/P$  & Time [s] & Par.\ Eff.\ \\ \hline
 8,192  & 9 & 309,877 & 34,600  & 6.79 & 100.0\% \\
65,536  &10 & 310,163 & 35,136  & 6.85 & 100.7\% \\
458,752 &11 & 354,746 & 38,833  & 7.86 &  96.9\% \\  \hhline{|======|}
  2,048  & 8 & 156,178 & 21,536 & 4.18 & 100.0\% \\
 16,384  & 9 & 155,702 & 22,036 & 4.25 & 100.6\% \\
131,072  &10 & 155,460 & 22,284 & 4.36 &  98.9\% \\ \hline
\end{tabular}
  \caption[Weak scaling with tetrahedral and hexahedral elements.]
{Weak scaling for \texttt{Ghost} with tetrahedral (top) and
  hexahedral (bottom) elements.
  We increase the base level by one, keeping $k = 2$, and multiply the process
  count by eight to maintain the same number of local elements per process.
  Notice that the highest process count of 458,752 is only seven times 65,536
  resulting in $\approx 14\%$ more local elements.
  Overall, the runtime of \texttt{Ghost} is lower for tetrahedra than for hexahedra
  due to the lower number of faces per element.
  Similar to the strong scaling tests, the parallel efficiency is nearly ideal
  (see also Figure~\ref{fig:baghoex-1-weak}).
  The maximum global numbers of elements are $108\e9$ for tetrahedra
  and $162\e9$ for hexahedra.}
\figlabel{tab:baghoex-1-weak}
\end{table}

\begin{figure}
  \center
\includegraphics[width=0.49\textwidth]{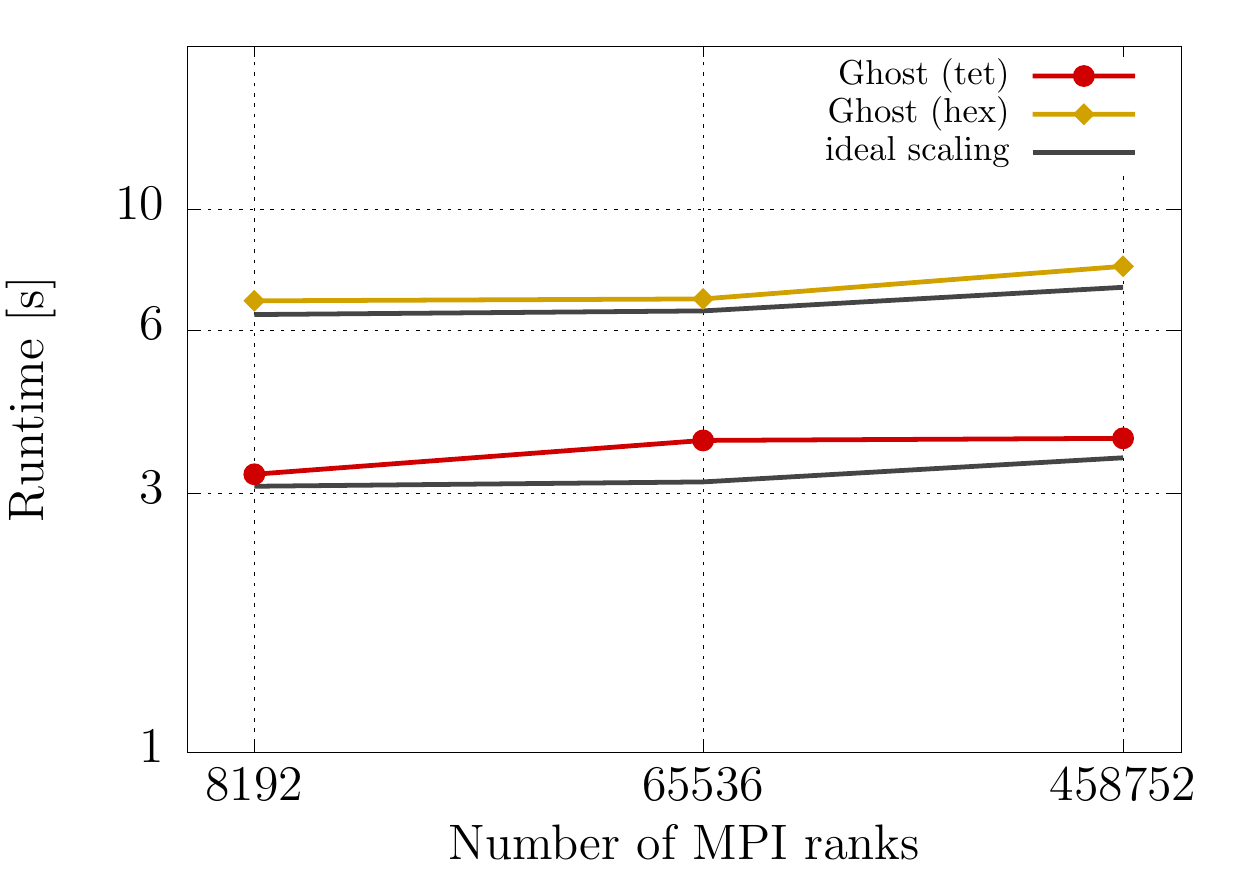}
\includegraphics[width=0.49\textwidth]{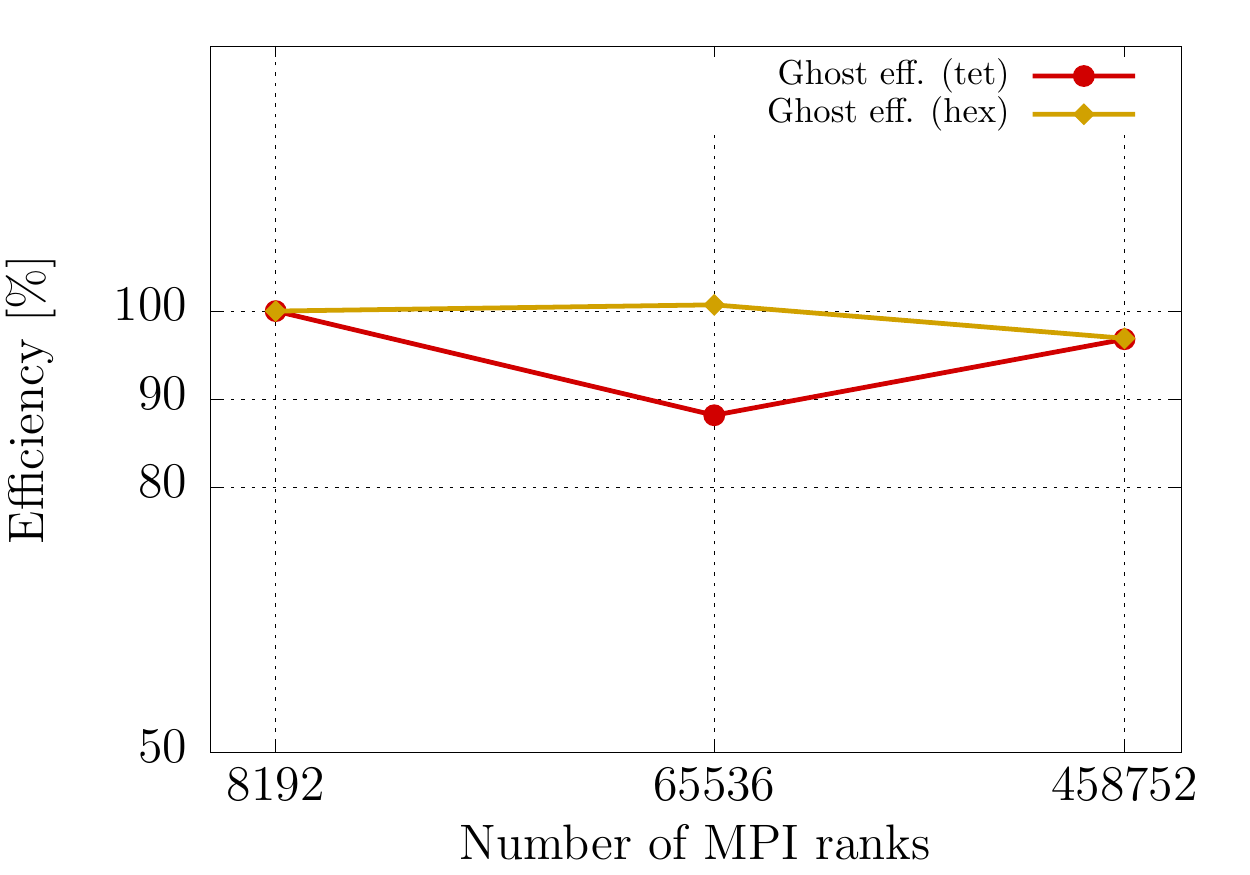}
  \caption[Weak scaling with tetrahedral and hexahedral elements.]
 {%
  Weak scaling results for tetrahedra with refinement levels 8,
  9, and 10, and for hexahedra with refinement levels 9, 10, and 11,
  $k = 2$,
  on \juqueen.
  This amounts to 233k elements per process for tetrahedra
  and 310k elements per process for hexahedra.
  This number differs slightly for the 458,752 process runs, since 458,752
  is only seven times 65,536, while we increase the number of mesh elements
  by the factor 8.
  On the left-hand side we plot the runtimes of \texttt{Ghost}
  with the ideal scaling in black.
  On the right-hand side we plot the parallel efficiency in $\%$.
  We display all values in Table~\ref{tab:baghoex-1-weak}.}
  \figlabel{fig:baghoex-1-weak}
\end{figure}

We show these results in Table~\ref{tab:baghoex-1-weak} and
Figure~\ref{fig:baghoex-1-weak}.
We notice that \texttt{Ghost} for tetrahedra is faster than \texttt{Ghost} for hexahedra.
The reason is the smaller number of ghosts due to less faces per element.
In all tests we observe excellent strong and weak scaling with
efficiencies in the order of 95\%.

\subsection{A hybrid test case}

\begin{figure}
 \center
 \includegraphics[height=0.45\textwidth]{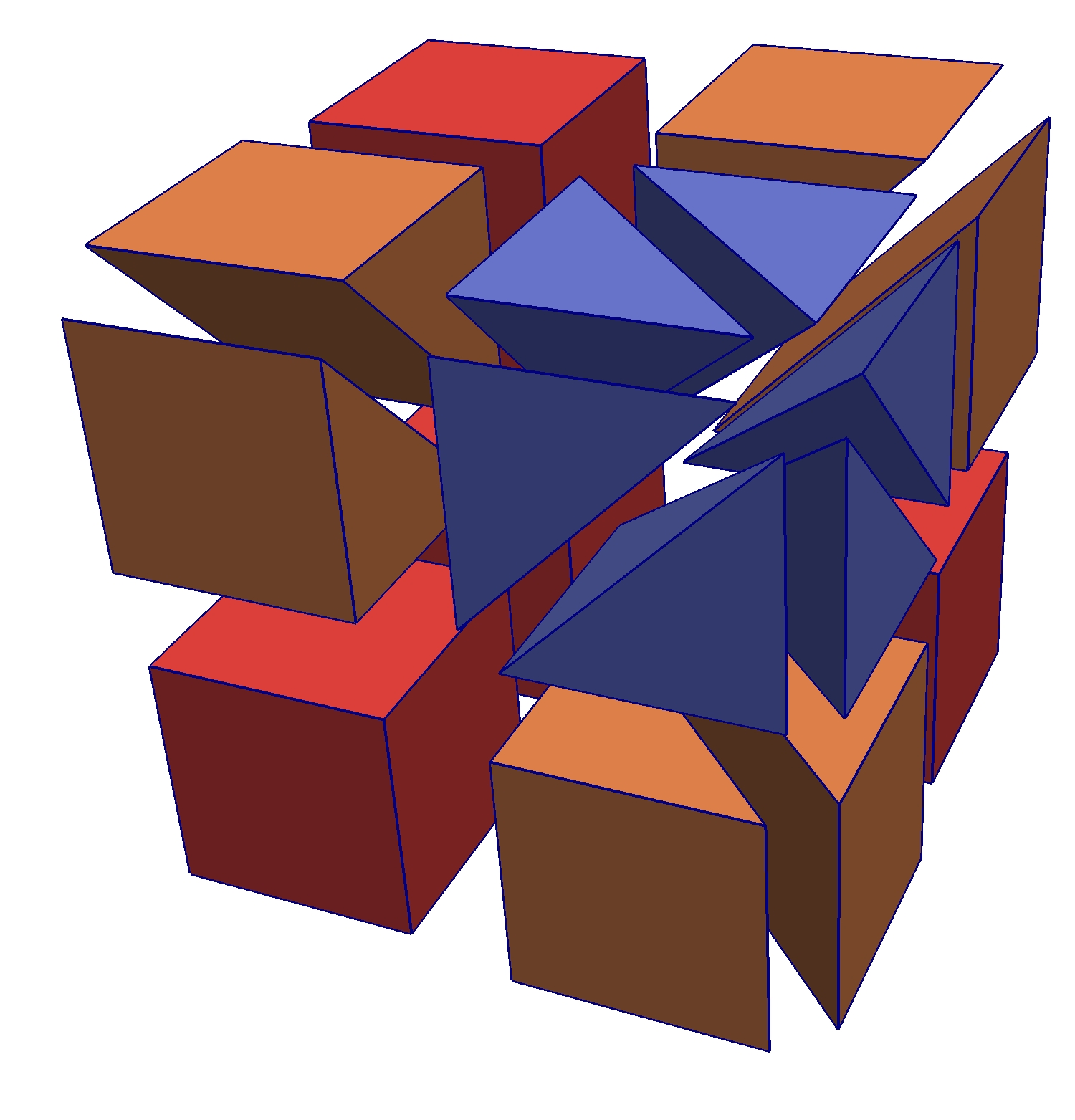}
 \hfill
 \includegraphics[height=0.45\textwidth]{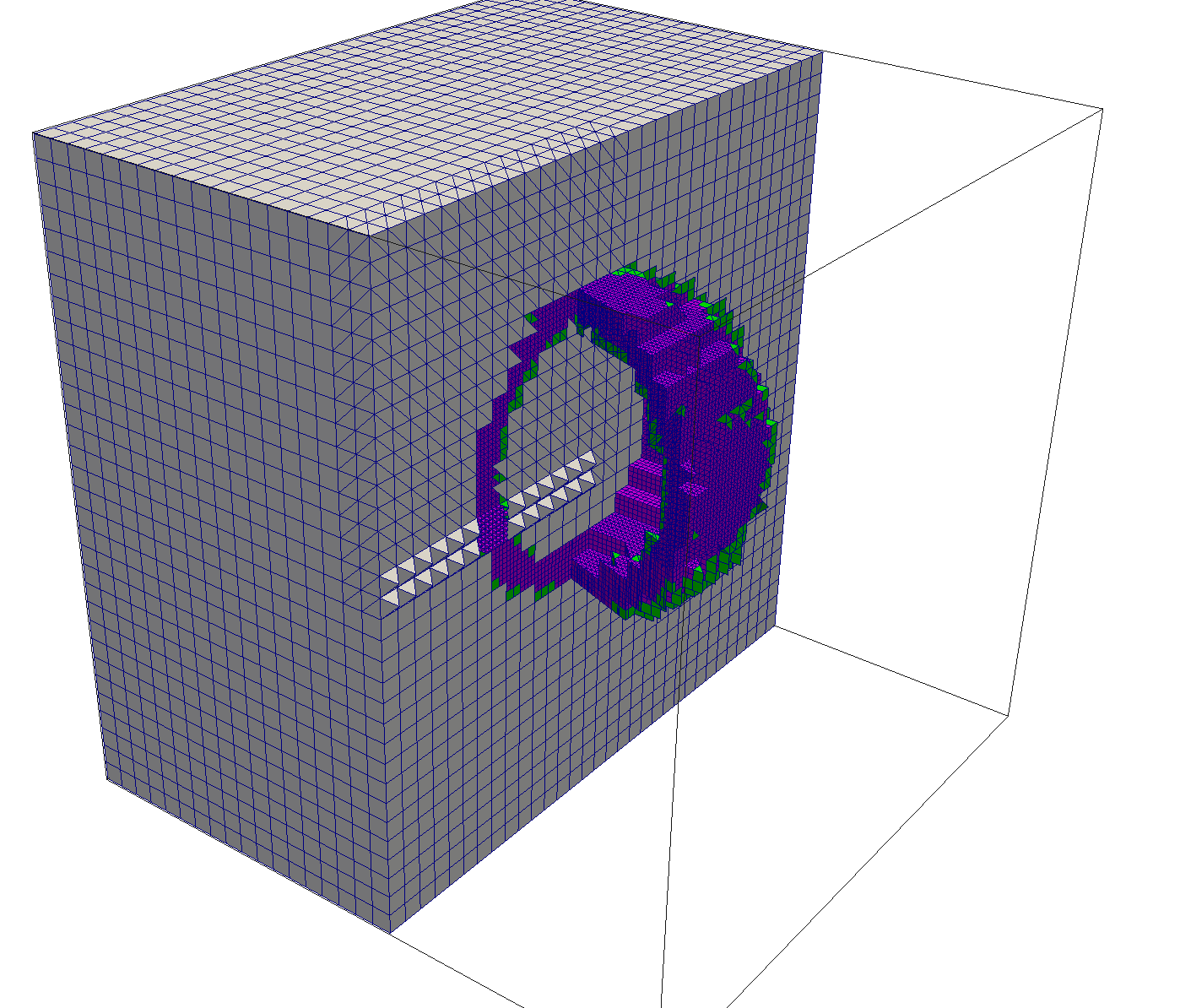}
 \caption{For tests on \juwels we choose a hybrid cube mesh as geometry.
Left: This coarse mesh consists of six tetrahedra, six prisms and four
hexahedra.
We arrange the trees such that every element in any refined mesh
aligns at least two of its faces parallel to a coordinate plane.
Right: We uniformly refine to a base level $\ell$ and then
add $k = 2$ rounds of adaptive refinement near a spherical shell.
This mesh is not 2:1 balanced. The colors indicate different
refinement levels; purple for level $\ell+2$, green for
level $\ell+1$ and gray for level $\ell$.}
\label{fig:hybridgeom}
\end{figure}

In our last test case we test hybrid meshes on the \juwels supercomputer.
For the coarse mesh we model a cube with four hexahedra, six prisms and six
tetrahedra. We refine this geometry uniformly to level $\ell$ and then
adaptively refine along a spherical shell to level $\ell + 2$; see
Figure~\ref{fig:hybridgeom}. In this test case we do not establish a
2:1 balance in the mesh, demonstrating the capability of our algorithm
to handle unbalanced forests; cf.\ \cite{IsaacBursteddeWilcoxEtAl15}.
The results for weak and strong scaling are listed in
Tables~\ref{tab:hybridweakscale} and~\ref{tab:hybridstrongscale}, respectively.
Notice that due to the increased memory per process on \juwels we are able to
create larger amounts of elements per process than on \juqueen,
while the run time per element and process is much less.

\begin{table}
\center
\begin{tabular}{|r|r|r|r|r|r|}\hline
  $P$ & $\ell$ & $E/P$ & $G/P$ & Time [s] & Par.\ Eff.\\ \hline
  192    &  8 & 1,601,702 & 70,467 & 0.384 & -- \\ %
  1,536  &  9 & 1,601,702 & 74,586 & 0.363 & 112\% \\ %
  12,288 & 10 & 1,601,702 & 76,514 & 0.355 & 117\% \\ %
  98,304 & 11 & 1,601,702 & 77,454 & 0.369 & 114\% \\  \hline %
\end{tabular}
\caption{Weak scaling for the hybrid mesh with hexahedra, prisms and tetrahedra
on \juwels;
see also Figure~\ref{fig:hybridgeom}. We increase the base level from $\ell = 8$ to
$\ell = 11$ and the process count from 192 to 98,304.
The largest mesh has approximately 157.5\e9 elements.
As before with non-hybrid meshes we observe excellent parallel efficiency.}
\label{tab:hybridweakscale}
\end{table}

\begin{table}
\center
\begin{tabular}{|r|r|r|r|r|r|}\hline
  $P$ & $\ell$ & $E/P$ & $G/P$ & Time [s] & Par.\ Eff. \\ \hline
12,288 & 11 & 12,813,617 & 305,290 & 1.428 &  --\\      %
24,576 & 11 &  6,406,808 & 194,919 & 0.912 & 99.2\% \\  %
49,152 & 11 &  3,203,404 & 121,987 & 0.550 & 103\%  \\  %
98,304 & 11 &  1,601,702 &  77,454 & 0.369 & 97.4\%  \\  \hline   %
\end{tabular}
\caption{Strong scaling for the hybrid mesh with hexahedra, prisms and tetrahedra
on \juwels;
see also Figure~\ref{fig:hybridgeom}.
We fix the base level to $\ell = 11$
and increase the process count from 12,288 to 98,304.
The mesh has approximately 157.5\e9 elements.
Absolute processing rates are above 200k local ghosts elements per second.
We achieve nearly perfect parallel efficiency at sub-second full-system
runtimes.}%
\label{tab:hybridstrongscale}
\end{table}

\begin{table}
 \center
\begin{tabular}{|r|r|r|r|r|}\hline
  $P$ &\#Elements  & \texttt{Ghost} [s] & \texttt{Partition} [s]\\ \hline
 49,152 & 1,099,511,627,776 & 2.08 & 0.73\\
 \hline
\end{tabular}
\caption{The largest mesh that we create is uniform at level 12 and
1.1\e 12 elements.}
\label{tab:hybrid1e12}
\end{table}

As a final experiment we create a uniform level 12 mesh with more than $1\e12$
elements.
Even for this exceptionally large mesh the runtime of our optimized
\texttt{Ghost} algorithm is only 2.08 seconds (Table~\ref{tab:hybrid1e12}).

\section{Conclusion}

In this paper, we present a parallel ghost layer assembly for adaptive forest
meshes.
It is general with respect to the shape of the trees, which may be cuboidal,
simplicial, or of any other shape that yields a conforming coarse mesh.
Furthermore, it is general in terms of non-conforming adaptivity by recursive
refinement and works with and without a 2:1 balance property.
The runtimes we show are proportional to the number of local ghost elements,
which is the optimal rate to be expected, and scale to over 1e12 elements
total.
Absolute runtimes are low at less than 5~\microsec per local ghost element on
the Xeon-based \juwels supercomputer.

We have limited the exposition to face-only connectivity, which suffices to
implement flux- and mortar-based numerical methods, for example of finite
volume, spectral element, or discontinuous Galerkin type.
The extension of the algorithm to vertex and 3D edge connectivity is still
open.
Based on our experience with assembling the fully connected ghost layer for
adaptive hexahedral meshes, we judge this extension to be feasible.
In addition to implementing the necessary low-level functions to compute vertex
and 3D edge neighbors within a tree,
the major task will be to extend the tree-to-tree neighbor computation
accordingly.
To do so, the method that we describe in this paper, namely constructing the lower
dimensional element, transforming it into its neighbor and then extruding it to
the neighbor element, can be applied for arbitrary codimension as well.
The challenge
compared to face-neighbors is that at a single inter-tree
vertex/edge can connect an arbitrary number of trees.
Encoding and identifying these neighbor trees is a task that requires an
appropriate extension of the coarse mesh connectivity.
Once this is accomplished, the neighbor elements can be constructed by using
the techniques described in this paper.
The primary use for the presented algorithm is the support of element-based
numerical methods, which can be realized in multiple ways.
Since the ghost layer is an ordered linear structure, it can be binary searched
directly by an application, for example to count and allocate the communication
buffers for flux-based methods.
Alternatively, it can be used in a joined top-down traversal of the mesh to
collect globally consistent face and node numbers for element-based methods
and store them in a lookup table, which then serves as the interface to the
numerical application.
Other uses include CFL-limited particle tracking and semi-Lagrangian methods,
which require quick owner search of points that leave the local partition.

In summary, we hope to have provided both an abstract technique and a usable
software module that is indispensible to many numerical applications, and which
hides the complexity of the algorithmic details behind a minimal interface.

\bibliographystyle{siamplain}
\bibliography{./ccgo_new,./group,./donna}

\end{document}